\documentclass[12pt]{article}
\usepackage{fullpage}
\usepackage{graphics}
\usepackage{amsmath,amssymb,bm}

\allowdisplaybreaks[1]
\def\be#1\ee{\begin{eqnarray}#1\end{eqnarray}}
\def\bal#1\eal{\begin{align}#1\end{align}}
\def\bat#1\eat{\begin{alignat}{2}#1\end{alignat}}
\def\bmu#1\emu{\begin{multline}#1\end{multline}}
\def\bga#1\ega{\begin{gather}#1\end{gather}}
\newcommand{\ba}{\begin{array}}
\newcommand{\ea}{\end{array}}
\newcommand{\n}{\notag}

\newcommand{\abs}[1]{\lvert#1\rvert}
\newcommand{\supfi}[1]{{}^{\,#1\!}}
\newcommand{\subfi}[1]{{}_{\,#1}}
\renewcommand{\d}{\partial}
\renewcommand{\bf}{\mathbf}
\renewcommand{\cal}{\mathcal}

\newcommand{\ds}{\displaystyle}

\title{\textbf{Bound states in Yukawa theory}}

\author{Axel Weber\thanks{Electronic address: 
\texttt{axel@itzel.ifm.umich.mx}} \\
\normalsize \em Instituto de F\'{\i}sica y Matem\'aticas, \\[-1mm] 
\normalsize \em Universidad Michoacana de San Nicol\'as de Hidalgo, \\[-1mm]
\normalsize \em Edificio C-3, Ciudad Universitaria, A. Postal 2-82, \\[-1mm]
\normalsize \em 58040 Morelia, Michoac\'an, Mexico
\and
Norbert E. Ligterink\thanks{Electronic address:
\texttt{n.e.ligterink@utwente.nl}} \\
\normalsize \em Control Engineering, \\[-1mm]
\normalsize \em Faculty of Electrical Engineering, Mathematics,
and Computer Science, \\[-1mm]
\normalsize \em University of Twente, \\[-1mm]
\normalsize \em P.O. Box 217, 7500 AE Enschede, Netherlands}

\date{\normalsize June 13, 2005}

\begin{document}

\maketitle

\begin{abstract}
A generalization of the Gell-Mann--Low Theorem is applied to bound state
calculations in Yukawa theory. The resulting effective Schr\"odinger equation 
is solved numerically for two-fermion bound states with the exchange of a
massless boson. The complete low-lying bound state spectrum is obtained for
different ratios of the constituent masses. No abnormal solutions are found. 
We show the consistency of the non-relativistic and one-body limits and 
discuss the special cases of identical fermions and fermion-antifermion 
states. To our knowledge, this is the first consistent calculation
of bound states in pure Yukawa theory (without UV cutoff).
\end{abstract}


\section{Introduction}

During the more than 50 years that have passed since Bethe and Salpeter
formulated their famous equation \cite{BS51,GL51}, the calculation of
relativistic bound states has proved to be one of the truely hard
problems in quantum field theory. For most of the numerous proposals
for its solution, a model theory of two scalar bosons interacting via
the exchange of a third scalar has served as a first testing ground.
In the case of the Bethe-Salpeter equation itself, it was shown by Wick
and Cutkosky that this model theory has an analytical solution in the
popular ``ladder approximation'' to the equation, in case that the
exchanged boson is massless \cite{WC54}.
It is hence quite surprising that for one of the most natural 
generalizations of this model where two spin 1/2 fermions interact through 
the exchange of a scalar boson, not a single consistent formalism has been 
devised for its solution to date.

In this introduction, we will give a brief review of the most important 
intents to calculate relativistic two-fermion bound states in this model,
i.e., in Yukawa theory. It was clear even before Wick and Cutkosky's
solution of the purely scalar model that the case of fermionic
constituents means a lot more to the Bethe-Salpeter equation than just a 
technical complication due to the inclusion of spin degrees of freedom 
\cite{Nak69}: the fact that the kernel of the equation in the ladder
approximation (after a Wick rotation) is not of Fredholm type in this 
case represents a fundamental difficulty for analytic as much as for 
numerical investigations. In the case of equal constituent masses and 
zero boson mass, it was shown by Goldstein \cite{Gol53} (later corrected 
and improved upon by Green \cite{Gre55}), originally for the subspace of 
pseudoscalar bound states only, that the ladder approximation gives a 
\emph{continuum} of coupling constants corresponding to a ``tightly bound'' 
state with zero mass, i.e., such that the binding energy completely 
compensates the masses of the constituents, while one expects a series of 
discrete values if a discrete spectrum of energy eigenvalues is to exist. 
The definite version of the argument is reviewed in Ref.\ \cite{Set88} 
(see the references there). In the case of Yukawa theory, this is essentially 
all that is known about the solutions of the Bethe-Salpeter equation. 

The exchange of a scalar boson between spin 1/2 fermions has come to
play an important role in the description of the nucleon-nucleon
interaction in the context of one-boson exchange (OBE) models. 
Although the scalar boson exchange is used as an effective and approximate
description of an actual two-pion exchange in this case, it is believed
to give the dominant attractive contribution to the intermediate-range
potential. In the numerous numerical calculations for the deuteron
system within OBE models, the scalar boson exchange is always accompanied 
by the exchange of other mesons, most importantly pions and rho and omega
mesons. For this type of calculations, the Bethe-Salpeter equation as well 
as several of its three-dimensional reductions have been employed, notably 
the Blankenbecler-Sugar-Logunov-Tavkhelidze (BSLT) equation \cite{BSL63}
and the Gross or spectator equation \cite{Gro69}. The latter equations 
avoid problems with abnormal solutions, the one-body limit, and others that 
arise in the Bethe-Salpeter equation (in the ladder approximation) 
\cite{Nak69, Gro82}, see, however, Ref.\ \cite{Pas97}. Among the vast 
literature on OBE models, we mention the work of Tjon and collaborators 
\cite{ZT80} who use the Bethe-Salpeter equation, the work of Gross and 
collaborators \cite{Gro74} who base their analysis on the Gross equation, 
and the quite complete ``Bonn model'' by Holinde, Machleidt, et al.\ reviewed
in Ref.\ \cite{Mac87}, which employs the Bloch-Horowitz scheme \cite{BH58}
for the bound state calculations, a formalism that is \emph{not} based on 
the Bethe-Salpeter equation. 

In all these approaches, a momentum cutoff 
is introduced in order to make the equations well-defined. It is interpreted 
phenomenologically as representing the spatial extension of the nucleons and 
mesons. Although we have not found a corresponding calculation in
the published literature, it appears plausible in view of the singular 
nature of the Bethe-Salpeter kernel in the ladder approximation that the
introduction of a momentum cutoff would also be necessary in the case of a 
pure Yukawa interaction, and that the corresponding numerical results
depend on the value of the cutoff. Note, however, the work of Gari and 
collaborators \cite{KG83} who dispense with a momentum cutoff and rather 
determine the nucleon and meson structures self-consistently through loop 
corrections within the same nucleon-meson model. It is perhaps not a 
coincidence that their approach employs the Okubo transformation technique
\cite{Oku54}, which is unrelated to the Bethe-Salpeter equation but very 
similar to the formalism proposed in the present paper.

More recently, calculations of bound states in Yukawa theory have been
attempted in light-front quantization, both in a Tamm-Dancoff approximation 
\cite{GW93} and in the so-called covariant light-front dynamics \cite{MCK03}. 
Just as in the equal-time quantized theory, an additional momentum cutoff
has to be introduced for the one-boson exchange, and the solutions for the
bound states turn out to depend on this cutoff. It was pointed out by van 
Iersel and Bakker \cite{IB04} that the light front ladder approximation is 
incomplete as long as instantaneous terms are not taken into account. One 
possibility to include such terms was presented in Ref.\ \cite{SFC01} and 
may help to solve the renormalization problem in the future.

In this paper, we propose to use a generalization \cite{Web00} of the 
Gell-Mann--Low theorem \cite{GL51} for the calculation of relativistic 
bound states. After the obligatory application to the purely scalar model 
in a previous publication \cite{WL02}, we now turn to Yukawa theory.
Surprisingly, the application of the generalized Gell-Mann--Low theorem 
turns out to be straightforward and, except for the necessary technical 
complications due to the spin degrees of freedom, completely analogous
to the scalar case. No inconsistencies whatsoever have been found, and
no necessity for the introduction of a cutoff (for the interaction
between the fermions) arises. For the numerical calculation of the
bound state spectrum, we will focus on the case of a massless exchanged 
boson, for the following reasons: first, the massless case is the most 
singular one, and if the formalism works in this case, it is expected to 
be applicable to the case of a massive exchanged boson as well. Second, 
the solutions in the non-relativistic limit are known analytically in this
case (Coulomb potential) and our numerical solutions can be compared
against them. In particular, one would like to find the degeneracies 
characteristic of the non-relativistic limit. Finally, we plan to 
extract the lowest-order fine and hyperfine structure from our effective
Hamiltonian in the near future, and this can be done analytically only
in the massless (Coulomb) case.

The organization of the paper is as follows: in the next section, we
introduce the generalization of the Gell-Mann--Low theorem on which
our approach is based, and comment on aspects relevant to the present
work. We also present, in the same section, the effective Hamiltonians
in the zero-, one-, and two-fermion sectors which result from the
application of the generalized Gell-Mann--Low theorem to lowest non-trivial
order. They describe the renormalization of the vacuum energy, the fermion
mass renormalization, and the effective potential for the two-fermion
dynamics, respectively. In Section 3, we perform a series of non-trivial
analytical checks on the results obtained in Section 2. In particular,
we investigate the non-relativistic and one-body limits of the
effective Schr\"odinger equation, replace fermions by antifermions,
and consider the case of identical constituents. All the properties of
our effective Hamiltonian description turn out to be in accord with physical 
expectations. We present numerical bound state solutions in Section 4,
for fine structure constants between zero and one and arbitrary ratios of
the constituent masses. The eight lowest-lying states are calculated in
each case, including states with non-zero relative orbital angular
momentum and mixtures of spin singlet and triplet states. The 
non-relativistic, one-body, and equal-mass limits are discussed in detail.
Section 5 contains our conclusions. There are also four appendices.
Appendices A and B are concerned with the appearing loop corrections and 
their regularization in different schemes from a Hamiltonian, not manifestly 
covariant perspective. They prepare the ground for future calculations at 
higher orders in the expansion of the effective Hamiltonian, where
regularization and renormalization are expected to become central issues.
Appendices C and D present several formulas which are used in the
analytical separation of angular and spin variables in the effective 
Schr\"odinger equation.

\section{The Bloch-Wilson Hamiltonian}

The physical idea behind the Bloch-Wilson or effective Hamiltonian is very
similar to the Born-Oppenheimer approximation: the integration over the
``light'' degrees of freedom generates the dynamics for the ``heavy''
degrees of freedom (the constituents in our case), where the ``light'' 
degrees of freedom are given by the so-called interacting ``virtual clouds'' 
around the constituents. The virtual clouds are, in turn, created by the 
constituents as described here through an adiabatic process. Technically, a 
generalization of the Gell-Mann--Low theorem is used which we will briefly 
describe in the following (for further details, see Ref.\ \cite{Web00}).

The adiabatic evolution operator $U_\epsilon$ provides a map from an 
eigenstate of the free Hamiltonian to an exact eigenstate of the full 
interaction Hamiltonian, \emph{if its application to the free eigenstate, 
after a suitable normalization, is well defined}. This is the content of 
the original theorem by Gell-Mann and Low \cite{GL51}. To this end, the
full Hamiltonian $H$ is split into a free part $H_0$ and an interacting 
part $H_1$. The adiabatic damping is implemented by replacing $H = 
H_0 + H_1$ through
\be
H(t) = H_0 + e^{-\epsilon \abs{t}} H_1 \:.
\ee
An eigenstate $| \phi \rangle$ of $H_0 = \lim_{t \to -\infty} H (t)$ is 
then evolved according to the time-dependent Schr\"odinger equation with 
Hamiltonian $H(t)$ from $t \to - \infty$ to the state $U_\epsilon 
| \phi \rangle$ at $t = 0$. According to the adiabatic theorem, in the limit 
$\epsilon \to 0$ the state $U_\epsilon | \phi \rangle$ is an eigenstate of 
$H (t=0) = H$. A possible infinite phase (see, for example, \cite{FW71}) is 
taken care of by imposing the normalization condition
\be
\langle \phi | \psi \rangle = 1 \:, \label{normGL}
\ee
so that
\be
| \psi \rangle = \frac{U_\epsilon | \phi \rangle}{\langle \phi | U_\epsilon 
| \phi \rangle} \:.
\ee

As a generalization of this theorem, it can be shown \cite{Web00} that the 
adiabatic evolution operator also maps $H_0$-invariant subspaces $\Omega_0$,
$H_0 \Omega_0 \subset \Omega_0$, to $H$-invariant subspaces $\Omega$, 
$H \Omega \subset \Omega$, \emph{if its application to the $H_0$-invariant 
subspace, after a suitable normalization, is well defined}. When a subspace 
is invariant under a hermitian operator, this operator can be diagonalized in 
the subspace, in other words, the subspace is a direct sum of eigenspaces of 
the operator, in this case the Hamiltonian. Since the only important property 
of the adiabatic evolution operator in this context is its $H$-invariant 
image, there is still an infinity of possible ``normalizations'' of the 
operator that map between the same subspaces $\Omega_0$ and $\Omega$.
Here we choose the normalization condition
\be
P_0 U_{\text{BW}} = \bf{1}_{\Omega_0} \:, \label{normUBW}
\ee
where $P_0$ is the orthogonal projector to $\Omega_0$ and $\bf{1}_{\Omega_0}$
the identity operator in $\Omega_0$. $U_{\text{BW}}$ is the ``Bloch-Wilson
operator'', the normalized version of the adiabatic evolution operator
given explicitly by
\be
U_{\text{BW}} = U_\epsilon (P_0 U_\epsilon)^{-1} \:.
\ee
Eq.\ \eqref{normUBW} naturally generalizes the normalization condition
\eqref{normGL} in the original Gell-Mann--Low theorem.
The subspace $\Omega_0$ has to be 
small enough in order that it is possible in practice to diagonalize the 
Hamiltonian in $\Omega = U_{\text{BW}} \Omega_0$, at least numerically. 
On the other hand, it has to be 
large enough for the Bloch-Wilson operator to be well defined. This is in 
principle a subtle issue and may depend on the normalization condition chosen. 
We will limit ourselves here to show that everything is well defined for a 
natural choice of the subspace (for the field theory and the perturbative 
order considered here, see the discussion below). In the
case of a two-constituent bound state, a natural choice for the
$H_0$-invariant subspace $\Omega_0$ is the space of all states of the two
constituents as free particles. This will be mapped by the normalized
adiabatic evolution operator to the $H$-invariant subspace $\Omega$,
a direct sum of eigenstates of the full
Hamiltonian, which is expected to coincide with the space of all
physical two-particle states, scattering as well as bound states.

A further simplification is obtained by realizing that the
Bloch-Wilson operator effects a similarity transformation between
the subspaces $\Omega_0$ and $\Omega$. Since the free subspace 
$\Omega_0$ is usually much easier to work in, it is convenient to 
similarity transform the full Hamiltonian back to this subspace, thereby
defining an effective or Bloch-Wilson Hamiltonian $H_{\text{BW}}$ in
$\Omega_0$,
\be
H_{\text{BW}} = P_0 H U_{\text{BW}} \:,
\ee
where we have taken into account that $P_0 |_\Omega$ is the inverse map to 
$U_{\text{BW}}$ as a consequence of the normalization \eqref{normUBW}. In the 
above example, the effective Hamiltonian acts on the subspace of all states of
the free constituents. Its eigenvalues are exactly equal to the eigenvalues
of the full Hamiltonian (in the corresponding subspace $\Omega$), and there 
is a one-to-one relation between the corresponding eigenstates. Mathematically,
the time-independent Schr\"odinger equation for the effective Hamiltonian
takes the form of a non-relativistic Schr\"odinger equation for the two
constituents, with the relativistic expression for the kinetic energies
and a complicated (generally non-local and non-hermitian) interaction
term. With the normalization chosen, the eigenstates $| \phi \rangle$ of 
the effective Hamiltonian have the immediate meaning of wavefunctions for 
the constituents: 
\be
| \phi \rangle = P_0 (U_{\text{BW}} | \phi \rangle) \:,
\ee
hence $| \phi \rangle$ is the two-particle component of the complicated
exact eigenstate of $H$, $U_{\text{BW}} | \phi \rangle$.

Obviously, the Bloch-Wilson operator $U_{\text{BW}}$ cannot be determined 
exactly in any interesting practical application. However, the adiabatic
evolution operator $U_\epsilon$, and hence $U_{\text{BW}} = 
U_\epsilon (P_0 U_\epsilon)^{-1}$, have a well-known perturbative expansion, 
the Dyson series:
\be
U_{\epsilon} = \sum_{n=0}^\infty \frac{(-i)^n}{n!}
\int_{-\infty}^0 dt_1 \cdots \int_{-\infty}^0 dt_n \, e^{-\epsilon (|t_1| +
\ldots + |t_n|)} \, T [ H_1 (t_1) \cdots H_1 (t_n) ] \:,
\ee
where
\be
H_1 (t) = e^{i H_0 t} H_1 \, e^{-i H_0 t} \:.
\ee
Every order in this expansion determines an 
approximation to the effective Hamiltonian and consequently of its 
eigenvalues and eigenstates. To compare with, in the Bethe-Salpeter
approach an infinite part of the Dyson series has to be summed up to find an
approximation to the bound states. The sum is performed in practice 
implicitly by solving an integral equation (the homogeneous Bethe-Salpeter
equation) with a perturbative approximation to its kernel, and the 
counterpart in our Hamiltonian approach is the solution
of the (approximated) effective Schr\"odinger equation. All of the above
is exemplified in the application to a simple scalar model in \cite{WL02}.

Finally, let us comment on the sense in which the adiabatic evolution
operator, or rather its normalized version $U_{\text{BW}}$, is well-defined in 
$\Omega_0$. In the proof of the generalized Gell-Mann--Low theorem, as well
as in the proof of the original theorem, the adiabatic evolution operator 
is treated as a formal power series (in $H_1$) given by the Dyson series. 
Consequently, $U_{\text{BW}}$ is considered well-defined if every term of 
the corresponding series
is well-defined (finite) disregarding convergence properties of the
series as a whole. However, the UV divergencies of the usual covariant
perturbation theory cannot be escaped: they appear in the context of
the Bloch-Wilson Hamiltonian (in the most straightforward approach) as
divergencies of the three-dimensional momentum integrals for large
momenta. In the scalar model considered before it was shown that these
divergencies can be dealt with (to the lowest non-trivial order considered
there) rather easily, and, in fact, in the same way as in covariant
perturbation theory. On the other hand, divergencies of the IR type
are considered to potentially invalidate the existence of the 
Bloch-Wilson operator (for the specific subspace $\Omega_0$ 
chosen). In the time-independent version presented in \cite{Web00},
which is obtained by performing the time integrals in the Dyson series,
the IR divergencies arise from vanishing energy denominators. They can
usually be avoided by appropriately enlarging the subspace $\Omega_0$.
However, the usefulness of the entire approach rests on the possibility
of choosing a subspace $\Omega_0$ which is manageably small, ideally
the space of all states of the two constituents as free particles in the
example mentioned above. Here, as for the scalar model before, we take
a pragmatic approach: we will be content with the fact that no IR
divergence arises to the perturbative order and for the choice of
$\Omega_0$ considered.

We will now apply these ideas to Yukawa theory. To simplify matters, we
consider two different fermion species, $A$ and $B$, which interact
with a scalar boson. Hence the Hamiltonian of the theory consists of the
free Hamiltonians for the fermions $A$ and $B$ and the scalar boson, and
the (normal-ordered) interaction term
\be
H_1 = g \int d^3 x \bm{:} \left[ \bar{\psi}_A (\bf{x}) \psi_A (\bf{x}) + 
\bar{\psi}_B (\bf{x}) \psi_B (\bf{x}) \right] \varphi (\bf{x}) \bm{:} \:.
\label{defHint}
\ee
We are interested in bound states of one fermion $A$ and one fermion $B$.
Other bound states, of identical fermions or fermion and antifermion
(e.g., in a theory with only one fermion species), are simply
related to bound states of type $AB$ and can be treated by exactly the
same method. They will be discussed in some detail in the next section.
The application of the generalized Gell-Mann--Low theorem to lowest 
non-trivial order follows in very close analogy the purely scalar case
\cite{WL02}. In particular, for a subspace $\Omega_0$ with a fixed number
of (free) particles $P_0 H_1 P_0 = 0$, and the effective Hamiltonian 
becomes to lowest non-trivial order
\be
H_{BW} = H_0 P_0 - i \int_{-\infty}^0 dt \, e^{- \epsilon |t|} P_0
H_1 (0) H_1 (t) P_0 + {\cal O} (H_1^4) \label{HBW2nd}
\ee
(the limit $\epsilon \to 0$ being understood). For the rest of this
chapter, we will discuss in detail the cases where $\Omega_0$ is the
subspace of zero, one (fermionic), and two ($AB$) particles.
We will go through the essential steps briefly, present
the corresponding results, and take the opportunity to comment on several
issues that have not received due attention in \cite{WL02}. For illustrative
purposes, it may be helpful to have a look at the diagrams presented there
which are the same in the present case, only that spin labels have to
be added to the external fermion lines.

\subsection{Vacuum state and zero-point energy renormalization}

Beginning with the vacuum state, we consider the Bloch-Wilson Hamiltonian for
the subspace $\Omega_0 = \mathbb{C} | 0 \rangle$, or, equivalently, for the
orthogonal projector on $\Omega_0$, $P_0 = | 0 \rangle \langle 0 |$.
In this case, the application of the generalized version of the 
Gell-Mann--Low theorem is equivalent to the original Gell-Mann--Low theorem.
The result for the Bloch-Wilson Hamiltonian is $H_{\text{BW}} = E_V P_0$, 
where (by Wick's theorem)
\bmu
E_V - E_0 \\
= i g^2 \int_{-\infty}^0 dt \, e^{- \epsilon \abs{t}} 
\int d^3 x \, d^3 x' \Delta_F (0 - t, \bf{x} - \bf{x}') \, \text{tr} \left[
S_F^A (0 - t, \bf{x} - \bf{x}') S_F^A (t - 0, \bf{x}' - \bf{x}) \right] \\
+ i g^2 \int_{-\infty}^0 dt \, e^{- \epsilon \abs{t}} 
\int d^3 x \, d^3 x' \Delta_F (0 - t, \bf{x} - \bf{x}') \, \text{tr} \left[
S_F^B (0 - t, \bf{x} - \bf{x}') S_F^B (t - 0, \bf{x}' - \bf{x}) \right] \:. 
\label{vaccor}
\emu
Here $E_0$ is the vacuum energy of the free theory and $S_F^{A,B}$ and
$\Delta_F$ are the fermionic (for particles $A,B$) and the bosonic 
Feynman propagators. In our conventions,
\bal
S_F^A (t, \bf{x}) &= i \int \frac{d^4 p}{(2 \pi)^4} \,
\frac{p \cdot \gamma + m_A}{p^2 - m_A^2 + i \epsilon} \, 
e^{- i (p_0 t - \bf{p} \cdot \bf{x})} \label{Fpropcov} \\
&= \int \frac{d^3 p}{(2 \pi)^3 \, 2 E_{\bf{p}}^A} \left[ \theta (t)
\left( p \cdot \gamma + m_A \right) e^{- i (E_{\bf{p}}^A t - \bf{p} \cdot
\bf{x})} - \theta (-t) \left( p \cdot \gamma - m_A \right) 
e^{ i (E_{\bf{p}}^A t - \bf{p} \cdot \bf{x})} \right] \label{Fpropnc}
\eal
in the covariant and non-covariant representation, respectively. 
We are using the following shorthands for the relativistic kinetic 
energies:
\be
E_{\bf{p}}^{A,B} = \sqrt{m_{A,B}^2 + \bf{p}^2} \;, \quad
\omega_{\bf{p}} = \sqrt{\mu^2 + \bf{p}^2}
\ee
($\mu$ is the boson mass). The bosonic propagator $\Delta_F (t, \bf{x})$ 
is simply given by the $m_A$-coefficient of $S_F^A (t, \bf{x})$ (and the 
substitution $m_A \to \mu$). In Eq.\ \eqref{vaccor},
note the opposite sign as compared to the bosonic case \cite{WL02}.

We can use time translation invariance of the integrands on the r.h.s.\ of
Eq.\ \eqref{vaccor} to move the vertices from times $t < 0$ and $t=0$ to
$t=0$ and $-t > 0$. Mathematically, this corresponds to replacing $t$ by
$-t$ and exchanging $\bf{x}$ and $\bf{x}'$, just as we did for the
one-particle states in the scalar case \cite{WL02}. If we, furthermore,
use the invariance of the bosonic propagator under the change of the
direction of propagation and apply these manipulations to half the r.h.s.\
of Eq.\ \eqref{vaccor}, we arrive (in the limit $\epsilon \to 0$) at the 
usual expression for the one-loop correction to the vacuum energy in 
covariant perturbation theory \cite{PS95},
\bal
E_V - E_0 &= i \, \frac{g^2}{2} \int d^4 x \, d^4 x' \, \delta(x^0)
\Delta_F (x - x') \, \text{tr} \left[ S_F^A (x - x') S_F^A (x' - x)
\right] \n \\
&\phantom{=} + i \, \frac{g^2}{2} \int d^4 x \, d^4 x' \, \delta(x^0)
\Delta_F (x - x') \, \text{tr} \left[ S_F^B (x - x') S_F^B (x' - x)
\right] \:. \label{vaccov}
\eal
The appearance of $\delta(x^0)$ eliminates a factor
\be
2 \pi \delta (0) = \int_{- \infty}^\infty dt
\ee
(due to time translation invariance) in the result, and leaves us with
\be
(2 \pi)^3 \delta^{(3)} (0) = \int d^3 x
\ee
(due to spatial translation invariance), to be interpreted as the volume
of space.

We can immediately express Eqs.\ \eqref{vaccor} and 
\eqref{vaccov} in momentum space, by use of the non-covariant and covariant 
expressions for the Feynman propagators in momentum space, Eqs.\
\eqref{Fpropnc} and \eqref{Fpropcov}, respectively. The results are 
\cite{Est03} 
\bal
\lefteqn{E_V - E_0} \n \\
&= \left[ g^2 \int \frac{d^3 p}{(2 \pi)^3} \frac{d^3 p'}{(2 \pi)^3}
\frac{1}{2 E_{\bf{p}}^A \, 2 E_{\bf{p}'}^A \, 2 \omega_{\bf{p} + \bf{p}'}}
\frac{4 \left( -E_{\bf{p}}^A E_{\bf{p}'}^A + \bf{p} \cdot \bf{p}' + m_A^2
\right)}{E_{\bf{p}}^A + E_{\bf{p}'}^A + \omega_{\bf{p} + \bf{p}'}}
\right. \n \\
&\phantom{=} \left. {}+ g^2 \int \frac{d^3 p}{(2 \pi)^3} 
\frac{d^3 p'}{(2 \pi)^3}
\frac{1}{2 E_{\bf{p}}^B \, 2 E_{\bf{p}'}^B \, 2 \omega_{\bf{p} + \bf{p}'}}
\frac{4 \left( -E_{\bf{p}}^B E_{\bf{p}'}^B + \bf{p} \cdot \bf{p}' + m_B^2
\right)}{E_{\bf{p}}^B + E_{\bf{p}'}^B + \omega_{\bf{p} + \bf{p}'}}
\right] (2 \pi)^3 \delta^{(3)} (0) \n \\
&= \left\{ \frac{g^2}{2} \int \frac{d^4 p}{(2 \pi)^4} \frac{d^4 p'}{(2 \pi)^4}
\frac{4 (p \cdot p' + m_A^2)}{[ (p - p')^2 - \mu^2 + i \epsilon]
[p^2 - m_A^2 + i \epsilon] [{p'}^2 - m_A^2 + i \epsilon]} \right. \n \\
&\phantom{=} \left. {}+ \frac{g^2}{2} \int \frac{d^4 p}{(2 \pi)^4} 
\frac{d^4 p'}{(2 \pi)^4} \frac{4 (p \cdot p' + m_B^2)}
{[ (p - p')^2 - \mu^2 + i \epsilon] [p^2 - m_B^2 + i \epsilon] 
[{p'}^2 - m_B^2 + i \epsilon]} \right\} (2 \pi)^3 \delta^{(3)} (0) \:.
\label{vacmom}
\eal
The equivalence between the two expressions can, of course, 
be established directly by integrating
over $p_0$ and $p'_0$ in the manifestly covariant expression. However,
there is a subtlety involved in the integration which may be of interest 
for future calculations at higher orders and will be described in detail
in Appendix \ref{appint}. The results \eqref{vacmom} are highly UV
divergent and will be treated as formal expressions only. It is, however,
important to remark that they are IR finite in the limit $\mu \to 0$
(we are here only interested in massive constituents $m_A, m_B \neq 0$).

\subsection{One-fermion states and mass renormalization}

We will now turn to the one-fermion states, considering for concreteness
fermions of type $A$. The $H_0$-invariant subspace considered is,
correspondingly, 
\be
\Omega_0 = \text{span} \left\{ | \bf{p}_A, r \rangle | \bf{p}_A \in 
\mathbb{R}^3, r=1,2 \right\} \:, 
\ee
where $| \bf{p}_A, r \rangle$ stands
for a state of one fermion $A$ with 3-momentum $\bf{p}_A$ and spin
orientation $r$ (we do not fix the basis to be used in spin space yet).
The one-fermion states are normalized in a non-covariant fashion,
\be
\langle \bf{p}_A, r | \bf{p}'_A, s \rangle = (2 \pi)^3 \delta ( \bf{p}_A
- \bf{p}'_A ) \, \delta_{rs} \:. \label{statenorm}
\ee

The application of the generalized Gell-Mann--Low theorem to lowest 
non-trivial order yields the following matrix elements of the
Bloch-Wilson Hamiltonian in $\Omega_0$,
\bmu
\langle \bf{p}_A, r | H_{\text{BW}} | \bf{p}'_A, s \rangle = 
\left( E_V + E_{\bf{p}_A}^A \right) (2 \pi)^3 \delta (\bf{p}_A - \bf{p}_A') \,
\delta_{rs} \\
- i g^2 \int_{-\infty}^0 dt \, e^{- \epsilon \abs{t}}
\int d^3 x \, d^3 x' \left[ \bar{\psi}_{\bf{p}_A, r}^A (0, \bf{x}) 
S_F^A (0 - t, \bf{x} - \bf{x}') \psi_{\bf{p}'_A, s}^A (t, \bf{x}') \right] 
\Delta_F (0 - t, \bf{x} - \bf{x}') \\
- i g^2 \int_{-\infty}^0 dt \, e^{- \epsilon \abs{t}}
\int d^3 x \, d^3 x' \left[ \bar{\psi}_{\bf{p}_A, r}^A (t, \bf{x}') 
S_F^A (t - 0, \bf{x}' - \bf{x}) \psi_{\bf{p}'_A, s}^A (0, \bf{x}) \right] 
\Delta_F (t - 0, \bf{x}' - \bf{x}) \:, \label{onepcor}
\emu
where the fermion wave functions $\psi_{\bf{p}, r}^A (t, \bf{x})$ are given by
\be
\psi_{\bf{p}, r}^A (t, \bf{x}) = \frac{u_A (\bf{p}, r)}
{\sqrt{2 E_{\bf{p}}^A}} \, e^{-i E_{\bf{p}}^A t + i \bf{p} \cdot \bf{x}}
\label{wavef}
\ee
with the Dirac spinors normalized to
\be
\bar{u}_A (\bf{p}, r) u_A (\bf{p}, s) = 2 m_A \delta_{rs} \:.
\ee

Using 3-momentum conservation at the vertices and, consequently,
$E_{\bf{p}_A}^A = E_{\bf{p}'_A}^A$, the vertices in
the last integral in Eq.\ \eqref{onepcor} can be translated from
times $t < 0$ and $t=0$ to $t=0$ and $-t > 0$, just as we did in the scalar
case \cite{WL02} and also for the corrections to the vacuum energy above.
Equation \eqref{onepcor} can then be written in the covariant form
\bmu
\langle \bf{p}_A, r | H_{\text{BW}} | \bf{p}'_A, s \rangle = 
\left( E_V + E_{\bf{p}_A}^A \right) (2 \pi)^3 \delta (\bf{p}_A - \bf{p}_A') \,
\delta_{rs} \\
- i g^2 \int d^4 x \, d^4 x' \, \delta(x^0) \left[ 
\bar{\psi}_{\bf{p}_A, r}^A (x) S_F^A (x - x') \psi_{\bf{p}'_A, s}^A (x') 
\right] \Delta_F (x - x') \:,  \label{onepcov}
\emu
where the integral over $d^3 x$ serves to implement 3-momentum 
conservation.

Finally, then, the matrix elements of $H_{\text{BW}}$ take the form
\bmu
\langle \bf{p}_A, r | H_{\text{BW}} | \bf{p}'_A, s \rangle = 
\left( E_V + E_{\bf{p}_A}^A \right) (2 \pi)^3 \delta (\bf{p}_A - \bf{p}_A') \,
\delta_{rs} \\
+ \frac{1}{2 E_{\bf{p}_A}^A} \left[ \bar{u}_A (\bf{p}_A, r)
G (\bf{p}_A) u_A (\bf{p}_A, s) \right]
(2 \pi)^3 \delta (\bf{p}_A - \bf{p}_A') \:, \label{onepform}
\emu
where $G (\bf{p}_A)$ is written in momentum space by use of the 
non-covariant and covariant expressions \eqref{Fpropnc} and \eqref{Fpropcov}
for the Feynman propagators in momentum space in Eqs.\ \eqref{onepcor} and 
\eqref{onepcov}, respectively, to give the following equivalent expressions 
\cite{Est03}
\bal
G (\bf{p}) &= - g^2 \int \frac{d^3 p'}{(2 \pi)^3} 
\frac{1}{2 E_{\bf{p}'}^A \, 2 \omega_{\bf{p} - \bf{p}'}}
\frac{E_{\bf{p}'}^A \gamma_0 - \bf{p}' \cdot \bm{\gamma} + m_A}
{E_{\bf{p}'}^A + \omega_{\bf{p} - \bf{p}'} - E_{\bf{p}}^A} \n \\ 
&\phantom{=} {}- g^2 \int \frac{d^3 p'}{(2 \pi)^3} 
\frac{1}{2 E_{\bf{p}'}^A \, 2 \omega_{\bf{p} - \bf{p}'}}
\frac{-E_{\bf{p}'}^A \gamma_0 - \bf{p}' \cdot \bm{\gamma} + m_A}
{E_{\bf{p}'}^A + \omega_{\bf{p} - \bf{p}'} + E_{\bf{p}}^A} \n \\
&= \left. i g^2 \int \frac{d^4 p'}{(2 \pi)^4} \frac{p' \cdot \gamma + m_A}
{[ (p - p')^2 - \mu^2 + i \epsilon ][ {p'}^2 - m_A^2 + i \epsilon ]}
\right|_{p_0 = E_{\bf{p}}^A} \:. \label{onepmom}
\eal
The equivalence between the two expression is shown directly in Appendix
\ref{appint} by performing the integration over $p'_0$ in the manifestly
covariant form.

We will now show that the contribution to $H_{\text{BW}}$ in the second line
of Eq.\ \eqref{onepform} can be absorbed into a renormalization of the
mass $m_A$. To this end, consider the covariant expression for $G (\bf{p})$
in Eq.\ \eqref{onepmom}: from covariance arguments, one has that 
$G (\bf{p})$ is of the form
\be
G (\bf{p}) = \left[ G_1 (p^2) p \cdot \gamma + G_0 (p^2) m_A
\right]_{p_0 = E_{\bf{p}}^A} \:, \label{Gform}
\ee
hence by use of the Dirac equation
\bal
\bar{u}_A (\bf{p}, r) G (\bf{p}) u_A (\bf{p}, s)
&= \bar{u}_A (\bf{p}, r) \left[ G_1 (m_A^2) m_A + G_0 (m_A^2) m_A \right]
u_A (\bf{p}, s) \n \\
&= 2 m_A^2 \left[ G_1 (m_A^2) + G_0 (m_A^2) \right] \delta_{rs} \:.
\label{dirac}
\eal
Consequently, we define
\be
\Delta m_A^2 = 2 m_A^2 \left[ G_1 (m_A^2) + G_0 (m_A^2) \right]
\label{DeltamAdef}
\ee
in order to write
\be
\langle \bf{p}_A, r | H_{\text{BW}} | \bf{p}'_A, s \rangle = 
\left( E_V + E_{\bf{p}_A}^A + \frac{\Delta m_A^2}{2 E_{\bf{p}_A}^A} \right) 
(2 \pi)^3 \delta (\bf{p}_A - \bf{p}_A') \, \delta_{rs} \:. \label{onepform2}
\ee

We are now in a position to perform the mass renormalization, 
\emph{completely within the Hamiltonian framework}. First, define the
renormalized or physical mass through
\be
\left[ E_V + E_{\bf{p}_A}^A + \frac{\Delta m_A^2}{2 E_{\bf{p}_A}^A}
\right]_{\bf{p}_A = 0} = E_V + M_A \:, \label{massren}
\ee
then 
\be
M_A = m_A + \frac{\Delta m_A^2}{2 m_A} + \cal{O} (g^4) \:.
\label{rencond}
\ee
We can use Eq.\ \eqref{rencond} to express $m_A$ in terms of $M_A$ in
Eq.\ \eqref{onepform2}, in particular in $E_{\bf{p}_A}^A$, taking into
account that $\Delta m_A^2$ is of order $g^2$. Working
consequently to order $g^2$, we arrive at
\be
E_V + E_{\bf{p}_A}^A + \frac{\Delta m_A^2}{2 E_{\bf{p}_A}^A}
= E_V + \sqrt{M_A^2 + \bf{p}_A^2} + \cal{O} (g^4) \:. 
\label{onepformren}
\ee
Remarkably, through the mass renormalization (to the order presently
considered) we have obtained an expression for the energy which is exactly 
covariant, in contradistinction to Eq.\ \eqref{onepform2}.

Now, the arguments to arrive at the manifestly covariant expression in
Eq.\ \eqref{onepmom} (see also Appendix \ref{appint}) and from there
to the crucial equation \eqref{Gform}, are formal in the sense that the
(unregularized) expression for $G (\bf{p})$ is UV divergent. We
show in Appendix \ref{appreg} that a careful derivation using different
regularization schemes leads to the same results. Observe, again, the
absence of IR divergencies in the limit $\mu \to 0$ in the expressions
\eqref{onepmom} (or in the corresponding explicit expressions presented in 
Appendix \ref{appreg}).

\subsection{Two-fermion states and effective potential}

Finally, in order to obtain the effective Schr\"odinger equation for $AB$
bound states, we consider the Bloch-Wilson Hamiltonian for the subspace
\be
\Omega_0 = \text{span} \left\{ | \bf{p}_A, r; \bf{p}_B, s \rangle |
\bf{p}_A, \bf{p}_B \in \mathbb{R}^3, r, s = 1, 2 \right\}
\ee
of all (free) states of one fermion $A$ and one fermion $B$, 
non-covariantly normalized as in Eq.\ \eqref{statenorm}. Then the matrix
elements of the Bloch-Wilson Hamiltonian to lowest non-trivial order turn 
out to be
\bal
\lefteqn{\langle \bf{p}_A, r; \bf{p}_B, s | H_{\text{BW}} 
| \bf{p}'_A, r'; \bf{p}'_B, s' \rangle} \n \\
&= \left( E_V + \sqrt{M_A^2 + \bf{p}_A^2} + \sqrt{M_B^2 + \bf{p}_B^2}
\right) (2 \pi)^3 \delta (\bf{p}_A - \bf{p}'_A) \delta_{r r'} (2 \pi)^3
\delta (\bf{p}_B - \bf{p}'_B) \delta_{s s'} \n \\
&\phantom{=} {}- i g^2 \int_{-\infty}^0 dt \, e^{- \epsilon \abs{t}}
\int d^3 x \, d^3 x' \left[ \bar{\psi}_{\bf{p}_B, s}^B (0, \bf{x}) 
\psi_{\bf{p}'_B, s'}^B (0, \bf{x}) \right]  \n \\
&\phantom{=} \hspace{6cm} {}\times \Delta_F (0 - t, \bf{x} - \bf{x}')
\left[ \bar{\psi}_{\bf{p}_A, r}^A (t, \bf{x}') 
\psi_{\bf{p}'_A, r'}^A (t, \bf{x}') \right] \n \\
&\phantom{=} {}- i g^2 \int_{-\infty}^0 dt \, e^{- \epsilon \abs{t}}
\int d^3 x \, d^3 x' \left[ \bar{\psi}_{\bf{p}_A, r}^A (0, \bf{x}) 
\psi_{\bf{p}'_A, r'}^A (0, \bf{x}) \right] \n \\
&\phantom{=} \hspace{6cm} {}\times \Delta_F (0 - t, \bf{x} - \bf{x}')
\left[ \bar{\psi}_{\bf{p}_B, s}^B (t, \bf{x}') 
\psi_{\bf{p}'_B, s'}^B (t, \bf{x}') \right] \:, \label{twopcor}
\eal
where $M_A$ and $M_B$ are the renormalized masses defined as in Eq.\
\eqref{massren}, and the wave functions $\psi^{A,B}_{\bf{p}, r} (t, \bf{x})$
have been introduced in Eq.\ \eqref{wavef}.

Due to the non-conservation of the perturbative energies, generally
$E_{\bf{p}_A}^A + E_{\bf{p}_B}^B \neq E_{\bf{p}'_A}^A + E_{\bf{p}'_B}^B$ 
for $\bf{p}_A + \bf{p}_B = \bf{p}'_A + \bf{p}'_B$, 
and the corresponding lack of time
translation invariance, the expressions with Feynman propagators in Eq.\
\eqref{twopcor} can\emph{not} be converted to on-shell Feynman diagrams.
It has not been much emphasized in \cite{WL02} that the mixing of states 
with different perturbative energies is essential for the formation
of bound states, because it is imperative for the localization of the
constituents in relative position space: consider, e.g., states with
total momentum $\bf{p}_A + \bf{p}_B = 0$, then a continuous superposition
of states with different relative momenta $\bf{p} = \bf{p}_A$ is
necessary to obtain a wavefunction of finite extension in relative position
space.

Equation \eqref{twopcor} leads to the following effective Schr\"odinger
equation,
\bmu
\left( \sqrt{M_A^2 + \bf{p}_A^2} + \sqrt{M_B^2 + \bf{p}_B^2} 
\right) \phi (\bf{p}_A, r; \bf{p}_B, s) \\
+ \sum_{r', s' = 1}^2 \int \frac{d^3 p'_A}{(2 \pi)^3}
\frac{d^3 p'_B}{(2 \pi)^3} \langle \bf{p}_A, r; \bf{p}_B, s | V 
| \bf{p}_A', r'; \bf{p}_B', s' \rangle 
\phi (\bf{p}_A', r'; \bf{p}_B', s') \\
= \left( E - E_V \right) \phi (\bf{p}_A, r; \bf{p}_B, s) \label{twopschr}
\emu
for the two-particle wave function in momentum space
\be
\phi (\bf{p}_A, r; \bf{p}_B, s) = \langle \bf{p}_A, r; \bf{p}_B, s |
\phi \rangle \:, \quad | \phi \rangle \in \Omega_0 \:.
\ee
Through the use of the non-covariant representations \eqref{Fpropnc} of 
the Feynman propagators in momentum space, the effective potential can be 
written as \cite{Est03}
\bmu
\langle \bf{p}_A, r; \bf{p}_B, s | V 
| \bf{p}_A', r'; \bf{p}_B', s' \rangle \\
= - \frac{g^2}{\sqrt{2 E_{\bf{p}_A}^A \, 2 E_{\bf{p}_B}^B \, 
2 E_{\bf{p}'_A}^A \, 2 E_{\bf{p}'_B}^B}} \,
\frac{1}{2 \omega_{\bf{p}_A - \bf{p}'_A}} \left( \frac{1}{E_{\bf{p}_A}^A +
\omega_{\bf{p}_A - \bf{p}'_A} - E_{\bf{p}'_A}^A} + \frac{1}{E_{\bf{p}_B}^B +
\omega_{\bf{p}_B - \bf{p}'_B} - E_{\bf{p}'_B}^B} \right) \\
\times \left[ \bar{u}_A (\bf{p}_A, r) \, u_A (\bf{p}'_A, r') 
\right] \left[ \bar{u}_B (\bf{p}_B, s) \, u_B (\bf{p}'_B, s') \right]
(2 \pi)^3 \delta (\bf{p}_A + \bf{p}_B - \bf{p}'_A - \bf{p}'_B) \:.
\label{twopeff}
\emu
The masses $m_{A,B}$ appearing in the kinetic energies $E_{\bf{p}}^{A,B}$ 
in the potential term can be replaced by their renormalized counterparts
$M_{A,B}$ to the present order in the perturbative expansion.
Eq.\ \eqref{twopeff} differs from its scalar analogue only through the
products of Dirac spinors (cf.\ Ref.\ \cite{WL02}).

The effective Hamiltonian commutes with the total 3-momentum operator
$\bf{P} = \bf{p}_A + \bf{p}_B$, hence we consider total momentum eigenstates
from now on. In particular, we specialize to the center-of-mass system
$\bf{P} = 0$ where the effective Schr\"odinger equation becomes
\bmu
\left( \sqrt{M_A^2 + \bf{p}^2} + \sqrt{M_B^2 + \bf{p}^2} \right)
\phi (\bf{p}; r, s) - g^2 \sum_{r', s' = 1}^2 \int \frac{d^3 p'}{(2 \pi)^3} 
\frac{1}{\sqrt{2 E_{\bf{p}}^A \, 2 E_{\bf{p}}^B \, 2 E_{\bf{p}'}^A \, 
2 E_{\bf{p}'}^B}} \\
\times \frac{1}{2 \omega_{\bf{p} - \bf{p}'}} \left( \frac{1}{E_{\bf{p}}^A +
\omega_{\bf{p} - \bf{p}'} - E_{\bf{p}'}^A} + \frac{1}{E_{\bf{p}}^B +
\omega_{\bf{p} - \bf{p}'} - E_{\bf{p}'}^B} \right) \\[2mm]
\times \left[ \bar{u}_A (\bf{p}, r) \, u_A (\bf{p}', r') 
\right] \left[ \bar{u}_B (- \bf{p}, s) \, u_B (- \bf{p}', s') \right]
\phi (\bf{p}'; r', s') = \left( E - E_V \right) \phi (\bf{p}; r, s) \:, 
\label{schroeff}
\emu
with the relative momentum $\bf{p} = \bf{p}_A = - \bf{p}_B$ and
\be
\phi (\bf{p}_A, r; \bf{p}_B, s) = \phi (\bf{p}_A; r, s) (2 \pi)^3
\delta (\bf{p}_A + \bf{p}_B) \:. \label{cmswavef}
\ee
Observe in 
the potential term in Eqs.\ \eqref{twopeff} and \eqref{schroeff} the square 
roots of the kinetic energies which are characteristic of the non-locality of 
the interaction, and the differences of energies in the denominators which
are due to the retardation of the interaction. The latter lead to a further
non-locality in the effective potential, and also introduce
non-hermiticity in the effective Hamiltonian.

For the discussion of the non-relativistic and the one-body limits in the
next section, and also for the numerical solution of the effective 
Schr\"odinger equation, it is convenient to cast Eq.\ \eqref{schroeff} 
into 2-spinorial form. To this end, the Dirac spinors are expressed in terms 
of Pauli spinors, most conveniently in the Dirac-Pauli representation,
\be
u_A (\bf{p}, r) = \sqrt{E^A_{\bf{p}} + M_A} 
\left( \ba{c} \chi_A (\bf{p}, r) \\[2mm]
\ds \frac{ \bf{p} \cdot \bm{\sigma} }{E^A_{\bf{p}} + M_A} \, 
\chi_A (\bf{p}, r) \ea \right) \:.
\ee
Here the Pauli spinors are normalized in the usual way,
\be
\chi_A^\dagger (\bf{p}, r) \chi_A (\bf{p}, s) = \delta_{rs} \:.
\ee
The Pauli spinors may or may not depend on the momentum $\bf{p}$. The
possibility of a momentum dependence is important if one wishes to employ
helicity eigenspinors. In the present work, however, we will not make use
of a momentum-dependent basis.

Eq.\ \eqref{schroeff} can now be rewritten in spinorial form as
\bmu
\left( \sqrt{M_A^2 + \bf{p}^2} + \sqrt{M_B^2 + \bf{p}^2} \right)
\phi (\bf{p}) \\
- g^2 \int \frac{d^3 p'}{(2 \pi)^3} \sqrt{
\frac{E^A_{\bf{p}} + M_A}{2 E^A_{\bf{p}}} \,
\frac{E^B_{\bf{p}} + M_B}{2 E^B_{\bf{p}}} \,
\frac{E^A_{\bf{p}'} + M_A}{2 E^A_{\bf{p}'}} \,
\frac{E^B_{\bf{p}'} + M_B}{2 E^B_{\bf{p}'}}} \\
\times \frac{1}{2 \omega_{\bf{p} - \bf{p}'}} \left( \frac{1}{E_{\bf{p}}^A +
\omega_{\bf{p} - \bf{p}'} - E_{\bf{p}'}^A} + \frac{1}{E_{\bf{p}}^B +
\omega_{\bf{p} - \bf{p}'} - E_{\bf{p}'}^B} \right) \\[2mm]
\times \left[ 
1 - \frac{\bf{p} \cdot \bm{\sigma}_A}{E^A_{\bf{p}} + M_A} \,
\frac{\bf{p}' \cdot \bm{\sigma}_A}{E^A_{\bf{p}'} + M_A} \right]
\left[ 1 - \frac{\bf{p} \cdot \bm{\sigma}_B}{E^B_{\bf{p}} + M_B} \,
\frac{\bf{p}' \cdot \bm{\sigma}_B}{E^B_{\bf{p}'} + M_B} \right]
\phi (\bf{p}') = \left( E - E_V \right) \phi (\bf{p}) \:, \label{spinrep}
\emu
where the spinorial wave function $\phi ({\bf p})$ is defined as
\be
\phi (\bf{p}) = \sum_{r,s} \phi( \bf{p}; r,s ) \left[ \chi_A (\bf{p}, r) 
\otimes \chi_B (-\bf{p}, s) \right] \:. \label{spinwavef}
\ee
As usual, $\bm{\sigma}_A$ is understood to act on $\chi_A(\bf{p}, r)$ only.

\section{Limiting cases, identical fermions and antifermions}

In the present section, we will perform a series of non-trivial analytical
checks on the effective Schr\"odinger equation. We will begin with the
non-relativistic limit: if the wave function $\phi (\bf{p}; r, s)$
is strongly suppressed for $\bf{p}^2 \gtrsim M_r^2$, where
\be
M_r = \frac{M_A M_B}{M_A + M_B}
\ee
is the (renormalized) reduced mass, we can approximate the effective
Schr\"odinger equation in the center-of-mass frame, Eq.\ \eqref{spinrep},
by
\be
\frac{\bf{p}^2}{2 M_r} \, \phi (\bf{p}) - \int \frac{d^3 p'}{(2 \pi)^3}
\frac{4 \pi \alpha}{\mu^2 + (\bf{p} - \bf{p}')^2} \, \phi (\bf{p}') =
\Big( E - E_V - M_A - M_B \Big) \phi (\bf{p}) \:, \label{NRmom}
\ee
where we have introduced the ``fine structure constant''
\be
\alpha = \frac{g^2}{4 \pi}
\ee
(for details on the approximation of the energy denominators, see Ref.\
\cite{WL02}). Observe in particular that the effective Hamiltonian acts
trivially on the spin degrees of freedom in this limit, as
expected from non-relativistic scattering processes where spin orientations
remain unchanged. As a consequence, Eq.\ \eqref{NRmom} has the same form as
in the case of scalar constituents \cite{WL02}.

The non-relativistic limit of Eq.\ \eqref{spinrep} is hence precisely 
what we expect on physical
grounds, a non-relativistic Schr\"odinger equation with the usual Yukawa 
potential (after Fourier transforming to position space). As discussed in 
detail in Ref.\ \cite{WL02}, the limit is attained for 
$\alpha \ll 1$ and $\mu \ll M_r$, both conditions being necessary. 

We will now consider the so-called one-body limit ``$M_B \to \infty$''
where $M_B^2 \gg M_A^2$ and the wave function $\phi (\bf{p}; r, s)$ is
negligibly small for $\bf{p}^2 \gtrsim M_B^2$. In this case, we can
approximate the effective Schr\"odinger equation \eqref{spinrep} by
\bmu
\sqrt{M_A^2 + \bf{p}^2} \, \phi (\bf{p})
- g^2 \int \frac{d^3 p'}{(2 \pi)^3} \sqrt{
\frac{E^A_{\bf{p}} + M_A}{2 E^A_{\bf{p}}} \,
\frac{E^A_{\bf{p}'} + M_A}{2 E^A_{\bf{p}'}}} \\
\times \frac{1}{2 \omega_{\bf{p} - \bf{p}'}} \left( \frac{1}{
\omega_{\bf{p} - \bf{p}'}} + \frac{1}{E_{\bf{p}}^A +
\omega_{\bf{p} - \bf{p}'} - E_{\bf{p}'}^A} \right)
\left[ 1 - \frac{\bf{p} \cdot \bm{\sigma}_A}{E^A_{\bf{p}} + M_A} \,
\frac{\bf{p}' \cdot \bm{\sigma}_A}{E^A_{\bf{p}'} + M_A} \right]
\phi (\bf{p}') \\[2mm]
= \left( E - E_V - M_B \right) \phi (\bf{p}) \:. \label{1Blimit}
\emu
The effective Hamiltonian acts trivially on the spin of particle $B$
in this limit. Eq.\ \eqref{1Blimit} has the form of a relativistic equation 
for fermion $A$ in an external potential, independent of the mass (except
for a constant shift in the energy) and the spin orientation of particle
$B$, in accord with physical expectations \cite{Gro82}.

We can, however, go one step further and compare Eq.\ \eqref{1Blimit} with
the relativistic equation for particle $A$ in the external potential
due to a fixed source which exchanges spinless bosons of mass $\mu$ with
particle $A$. The latter physical situation can also be described within 
the same general formalism, providing an internal consistency check for the
application of the generalized Gell-Mann--Low theorem.

To this end, we begin by defining the Hamiltonian $H'$ of the fixed source 
system, which consists of the Hamiltonians corresponding to free fermions 
$A$ and scalar bosons of mass $\mu$, and the interaction term
\be
H_1' = g \int d^3 x \bm{:} \bar{\psi}_A (\bf{x}) \psi_A (\bf{x}) 
\varphi (\bf{x}) \bm{:} + \: g \, \varphi (\bf{0}) \:. \label{defHfixed}
\ee
The form of the second contribution to the interaction Hamiltonian results 
from replacing the dynamical fermion field $B$ in Eq.\ \eqref{defHint} with 
a fixed source at $\bf{x} = \bf{0}$. To see that, we consider 
$\psi_B (\bf{x})$ as a classical Dirac field with probability density 
given by
\be
\rho (\bf{x}) = \psi^\dagger_B (\bf{x}) \psi_B (\bf{x}) \approx 
\bar{\psi}_B (\bf{x}) \psi_B (\bf{x}) \:,
\ee
the latter approximate equality holding when $\psi_B (\bf{x})$ describes
a particle (not an anti-particle) and the relevant momenta satisfy $\bf{p}^2
\ll M_B^2$. For a fermion $B$ localized at $\bf{x} = \bf{0}$ we then have
\be
\bar{\psi}_B (\bf{x}) \psi_B (\bf{x}) = \delta (\bf{x}) \:,
\ee
from which Eq.\ \eqref{defHfixed} follows.

The Bloch-Wilson Hamiltonian in the one-fermion sector is represented
diagrammatically to lowest non-trivial order [see Eq.\ \eqref{HBW2nd}]
in Ref.\ \cite{WL02} (adding spin labels to the external lines for the
present case). The corresponding algebraic expressions lead to a mass
renormalization for fermion $A$ which is identical to the one discussed
in the previous section, and the following matrix elements of the effective
Hamiltonian:
\bal
\lefteqn{\langle \bf{p}_A, r | H_{\text{BW}}' | \bf{p}'_A, r' \rangle} \n \\
&= \left( E_V' + \sqrt{M_A^2 + \bf{p}_A^2} \right) 
(2 \pi)^3 \delta (\bf{p}_A - \bf{p}'_A) \delta_{r r'} \n \\
&\phantom{=} {}- i g^2 \int_{-\infty}^0 dt \, e^{- \epsilon \abs{t}}
\int d^3 x \, \Delta_F (0 - t, \bf{0} - \bf{x})
\left[ \bar{\psi}_{\bf{p}_A, r}^A (t, \bf{x}) 
\psi_{\bf{p}'_A, r'}^A (t, \bf{x}) \right] \n \\
&\phantom{=} {}- i g^2 \int_{-\infty}^0 dt \, e^{- \epsilon \abs{t}}
\int d^3 x \left[ \bar{\psi}_{\bf{p}_A, r}^A (0, \bf{x}) 
\psi_{\bf{p}'_A, r'}^A (0, \bf{x}) \right] \Delta_F (0 - t, \bf{x} - \bf{0})
\:, \label{HBWfixed}
\eal
where the (renormalized) vacuum energy $E_V'$ differs from $E_V$
as determined from Eq.\ \eqref{vaccor} for the case of two dynamical fermions,
precisely because fermion $B$ has now been replaced by a fixed source
which is formally considered to be part of the vacuum.

Using the non-covariant expression \eqref{Fpropnc} for the propagators and
Eq.\ \eqref{wavef} for the fermionic wave functions, Eq.\ \eqref{HBWfixed} 
leads to the effective Schr\"odinger equation
\bmu
\sqrt{M_A^2 + \bf{p}^2} \, \phi (\bf{p})
- g^2 \int \frac{d^3 p'}{(2 \pi)^3} \sqrt{
\frac{E^A_{\bf{p}} + M_A}{2 E^A_{\bf{p}}} \,
\frac{E^A_{\bf{p}'} + M_A}{2 E^A_{\bf{p}'}}} \\
\times \frac{1}{2 \omega_{\bf{p} - \bf{p}'}} \left( \frac{1}{
\omega_{\bf{p} - \bf{p}'}} + \frac{1}{E_{\bf{p}}^A +
\omega_{\bf{p} - \bf{p}'} - E_{\bf{p}'}^A} \right)
\left[ 1 - \frac{\bf{p} \cdot \bm{\sigma}}{E^A_{\bf{p}} + M_A} \,
\frac{\bf{p}' \cdot \bm{\sigma}}{E^A_{\bf{p}'} + M_A} \right]
\phi (\bf{p}') \\[2mm]
= \left( E - E_V' \right) \phi (\bf{p}) \:, \label{effixed}
\emu
where the wave function is now defined as
\be
\phi (\bf{p}) = \sum_r \phi (\bf{p}, r) \chi (\bf{p}, r)
\equiv \sum_r \langle \bf{p}, r | \phi \rangle \chi (\bf{p}, r) \:.
\ee
Equation \eqref{effixed} is identical to Eq.\ \eqref{1Blimit} except
for an irrelevant shift in the vacuum energy and the fact that the
wave function in Eq.\ \eqref{1Blimit} includes the orientation of the
spin of fermion $B$ which, however, has no influence on the dynamics of
particle $A$. Self-consistency of the method in the one-body limit is
hence established.

We will now investigate how the effective Schr\"odinger equation changes
when we replace one of the constituents, say fermion $A$, by the
corresponding antiparticle. First of all, consider the 
one-$\bar{A}$-antifermion sector where the matrix elements of the
Bloch-Wilson Hamiltonian to lowest non-trivial order are given by
\bmu
\langle \bf{p}_{\bar{A}}, r | H_{\text{BW}} | \bf{p}'_{\bar{A}}, s \rangle 
= \left( E_V + E_{\bf{p}_{\bar{A}}}^A \right) (2 \pi)^3 
\delta (\bf{p}_{\bar{A}} - \bf{p}_{\bar{A}}') \, \delta_{rs} \\
+ i g^2 \int_{-\infty}^0 dt \, e^{- \epsilon \abs{t}} \int d^3 x \, d^3 x' 
\left[ \bar{\psi}_{\bf{p}'_{\bar{A}}, s}^{\bar{A}} (t, \bf{x}') 
S_F^A (t - 0, \bf{x}' - \bf{x}) \psi_{\bf{p}_{\bar{A}}, r}^{\bar{A}} 
(0, \bf{x}) \right] \Delta_F (0 - t, \bf{x} - \bf{x}') \\
+ i g^2 \int_{-\infty}^0 dt \, e^{- \epsilon \abs{t}} \int d^3 x \, d^3 x' 
\left[ \bar{\psi}_{\bf{p}'_{\bar{A}}, s}^{\bar{A}} (0, \bf{x}) 
S_F^A (0 - t, \bf{x} - \bf{x}') \psi_{\bf{p}_{\bar{A}}, r}^{\bar{A}} 
(t, \bf{x}') \right] \Delta_F (t - 0, \bf{x}' - \bf{x}) \label{oneapcor}
\emu
[compare with Eq.\ \eqref{onepcor} and note the change in sign]. Here we 
need the antifermion wave functions defined by
\be
\bar{\psi}_{\bf{p}, r}^{\bar{A}} (t, \bf{x}) = \frac{\bar{v}_A (\bf{p}, r)}
{\sqrt{2 E_{\bf{p}}^A}} \, e^{-i E_{\bf{p}}^A t + i \bf{p} \cdot \bf{x}}
\:.
\ee
We can convert Eq.\ \eqref{oneapcor} to the form corresponding to a particle
$A$ by introducing the charge conjugate wave functions
\be
\psi_{\bf{p}, r}^{\bar{A}, C} (t, \bf{x}) = 
C \left[ \bar{\psi}_{\bf{p}, r}^{\bar{A}} (t, \bf{x}) \right]^T
\label{awavef}
\ee
(the superindex $T$ stands for transposition), with the charge conjugation 
matrix $C = - i \gamma^0 \gamma^2$ in the Dirac-Pauli representation. 
Eq.\ \eqref{oneapcor} is then transformed into expression \eqref{onepcor} 
for the subspace of one fermion $A$ (including the sign) with the wave 
function $\psi_{\bf{p}, r}^A (t, \bf{x})$ being replaced by 
$\psi_{\bf{p}, r}^{\bar{A}, C} (t, \bf{x})$, which corresponds to the
replacement of the particle spinor $u_A (\bf{p}, r)$ with the
charge conjugate antiparticle spinor
\be
v_A^C (\bf{p}, r) = C \left[ \bar{v}_A (\bf{p}, r) \right]^T
\ee
for the description of the spin orientation of the antifermion. Since
$v_A^C (\bf{p}, r)$ is a positive-energy solution of the Dirac equation,
the mass renormalization from $m_A$ to $M_A$ for antifermions is identical 
to the one for fermions $A$.

We now turn to the two-particle sector with an antifermion $\bar{A}$ and a
fermion $B$. The matrix elements of the corresponding Bloch-Wilson 
Hamiltonian to lowest non-trivial order read
\bal
\lefteqn{\langle \bf{p}_{\bar{A}}, r; \bf{p}_B, s | H_{\text{BW}} 
| \bf{p}'_{\bar{A}}, r'; \bf{p}'_B, s' \rangle} \n \\
&= \left( E_V + \sqrt{M_A^2 + \bf{p}_A^2} + \sqrt{M_B^2 + \bf{p}_B^2}
\right) (2 \pi)^3 \delta (\bf{p}_{\bar{A}} - \bf{p}'_{\bar{A}}) 
\delta_{r r'} (2 \pi)^3 \delta (\bf{p}_B - \bf{p}'_B) \delta_{s s'} \n \\
&\phantom{=} {}+ i g^2 \int_{-\infty}^0 dt \, e^{- \epsilon \abs{t}}
\int d^3 x \, d^3 x' \left[ \bar{\psi}_{\bf{p}_B, s}^B (0, \bf{x}) 
\psi_{\bf{p}'_B, s'}^B (0, \bf{x}) \right]  \n \\
&\phantom{=} \hspace{6cm} {}\times \Delta_F (0 - t, \bf{x} - \bf{x}')
\left[ \bar{\psi}_{\bf{p}'_{\bar{A}}, r'}^{\bar{A}} (t, \bf{x}') 
\psi_{\bf{p}_{\bar{A}}, r}^{\bar{A}} (t, \bf{x}') \right] \n \\
&\phantom{=} {}+ i g^2 \int_{-\infty}^0 dt \, e^{- \epsilon \abs{t}}
\int d^3 x \, d^3 x' \left[ \bar{\psi}_{\bf{p}'_{\bar{A}}, r'}^{\bar{A}} 
(0, \bf{x}) \psi_{\bf{p}_{\bar{A}}, r}^{\bar{A}} (0, \bf{x}) \right] \n \\
&\phantom{=} \hspace{6cm} {}\times \Delta_F (0 - t, \bf{x} - \bf{x}')
\left[ \bar{\psi}_{\bf{p}_B, s}^B (t, \bf{x}') 
\psi_{\bf{p}'_B, s'}^B (t, \bf{x}') \right] \:, \label{twoapcor}
\eal
to be compared with Eq.\ \eqref{twopcor}. Again, the introduction of
the charge conjugate wave functions converts Eq.\ \eqref{twoapcor}
to the form of Eq.\ \eqref{twopcor}, with the wave 
function $\psi_{\bf{p}, r}^A (t, \bf{x})$ replaced by 
$\psi_{\bf{p}, r}^{\bar{A}, C} (t, \bf{x})$, or the spinor
$u_A (\bf{p}, r)$ by $v_A^C (\bf{p}, r)$. The effective Schr\"odinger 
equation for constituents $\bar{A} B$ can then be written in the spinorial 
form of Eq.\ \eqref{spinrep}, where in the definition \eqref{spinwavef} of 
the wave function the Pauli spinor $\chi_A (\bf{p}, r)$ has to be replaced 
by the charge conjugate spinor
\be
\xi_A^C (\bf{p}, r) = i \sigma^2 \xi_A^\ast (\bf{p}, r) \:,
\ee
$\xi_A (\bf{p}, r)$ being the Pauli spinor that describes the spin 
orientation of the corresponding negative-energy solution of the Dirac 
equation. The most important result is that, just as for the mass 
renormalization, the interaction via a scalar boson is ``charge conjugation 
blind'', i.e., does not distinguish between a fermion $A$ and its 
antifermion $\bar{A}$. This is, in fact, the expected behaviour.

The same arguments are used to describe an antifermion $\bar{A}$
interacting with a static source which leads to a Bloch-Wilson Hamiltonian
and an effective Schr\"odinger equation analogous to Eqs.\ \eqref{HBWfixed}
and \eqref{effixed}. Consequently, the one-body limit is self-consistent also
in the case of a (light) antifermion $\bar{A}$. More interesting is the
case of an antisource corresponding to the one-body limit $M_B \to \infty$
for an antifermion $\bar{B}$: a simple-minded argument replaces $\psi_B
(\bf{x})$ in the interaction Hamiltonian with a classical negative-energy
Dirac field, which leads to a probability density
\be
\rho (\bf{x}) = \psi^\dagger_B (\bf{x}) \psi_B (\bf{x}) \approx 
- \bar{\psi}_B (\bf{x}) \psi_B (\bf{x}) \:, \label{naivedneg}
\ee
if we suppose that the relevant momenta satisfy $\bf{p}^2 \ll M_B^2$.
Equation \eqref{naivedneg} with $\rho (\bf{x}) = \delta (\bf{x})$ would
be in conflict with the one-body limit $M_B \to \infty$ for constituents
$A \bar{B}$ or $\bar{A} \bar{B}$. The reason is, of course, that
$\rho (\bf{x})$ is to be interpreted physically as a charge density (when
multiplied with the charge of fermion $B$), and is \emph{negative} for
antifermions, hence it is $\bar{\psi}_B (\bf{x}) \psi_B (\bf{x})$ which
turns out to be positive and is to be replaced by $\delta (\bf{x})$.
The latter results are related to the use of anticommutators in the
quantization of the Dirac field.

In order to have a formally satisfactory description, we define an
approximately localized antifermion state with spin orientation $s$ as
\be
| \bf{x} = \bf{0}, \Delta x; s \rangle =
\int \frac{d^3 p_{\bar{B}}}{(2 \pi)^3} \left( \frac{8 \pi}{3} \Delta x^2
\right)^{3/4} e^{- \Delta x^2 \bf{p}_{\bar{B}}^2 /3} \, | \bf{p}_{\bar{B}},
s \rangle \:, \label{antiloc}
\ee
where $| \bf{p}_{\bar{B}}, s \rangle$ denotes a $\bar{B}$-antifermion
3-momentum eigenstate. If we choose $M_B$ large enough for
\be
\Delta x^2 M_B^2 \gg 1 \label{MBDelta}
\ee
to hold, we obtain
\be
\langle \bf{x} = \bf{0}, \Delta x; s | \bm{:} \bar{\psi}_B (\bf{x})
\psi_B (\bf{x}) \bm{:} | \bf{x} = \bf{0}, \Delta x; s \rangle 
= \left( \frac{3}{2 \pi \Delta x^2} \right)^{3/2} e^{-3 \bf{x}^2/
(2 \Delta x^2)} \:. \label{rightdpos}
\ee
In the limit $\Delta x \to 0$ the right-hand side of Eq.\ 
\eqref{rightdpos} tends towards $\delta (\bf{x})$. Equation \eqref{MBDelta}
implies that we need $M_B \to \infty$ (even faster) in this limit.
We take Eq.\ \eqref{rightdpos} as justification to replace
$\left[ \bm{:} \bar{\psi}_B (\bf{x}) \psi_B (\bf{x}) \bm{:} \right]$ with 
$\delta (\bf{x})$ for a fixed antisource, leading to the interaction 
Hamiltonian \eqref{defHfixed}. The one-body limit $M_B \to \infty$ is hence 
fully consistent also for the case of an antifermion $\bar{B}$.

We will now consider bound states of identical fermions, $A$-fermions
to be concrete. To this end, we calculate the Bloch-Wilson Hamiltonian
to lowest non-trivial order for the subspace
\be
\Omega_0 = \text{span} \left\{ | \bf{p}_A, r; \bf{p}_A', s \rangle |
\bf{p}_A, \bf{p}_A' \in \mathbb{R}^3, r, s = 1, 2 \right\}
\ee
of two-$A$-fermion states. As a consequence of the identity of the
constituents, ``crossed'' diagrams appear which carry a relative minus
sign due to the antisymmetry
\be
| \bf{p}_A, r; \bf{p}_A', s \rangle = - | \bf{p}_A', s; \bf{p}_A, r \rangle
\:.
\ee
However, mass renormalization works exactly as before, and also the
effective Schr\"odinger equation is the same as in the case of $AB$
bound states (with $M_B = M_A$) when we take into account the antisymmetry 
of the wave function
\be
\phi (\bf{p}_A, r; \bf{p}_A', s) = - \phi ( \bf{p}_A', s; \bf{p}_A, r) \:.
\ee
In the center-of-mass system, we have consequently
\be
\phi (\bf{p}) = - \phi (-\bf{p})^t \label{phiantis}
\ee
with the definitions \eqref{cmswavef} and \eqref{spinwavef}, where the 
transposition $t$ refers to the tensor product,
\be
\left[ \chi_A (-\bf{p}, r) \otimes \chi_A (\bf{p}, s) \right]^t
= \chi_A (\bf{p}, s) \otimes \chi_A (-\bf{p}, r) \:. \label{defttrans}
\ee
The consequences of antisymmetry for the solutions of the effective
Schr\"odinger equation will be discussed in the next section. In the
non-relativistic limit, however, the situation is particularly simple
because the spin degrees of freedom do not participate in the dynamics:
the solutions of Eq.\ \eqref{NRmom} with even orbital
angular momentum have symmetric spatial wave functions and hence 
antisymmetric spin states (total spin zero), while solutions with odd
orbital angular momentum have antisymmetric orbital wave functions
and hence necessarily symmetric spin states (total spin one).

In a theory which only contains $A$-fermions and scalar bosons, the
results for the mass renormalization and the $AA$ bound
states are the same as the ones presented above for a theory 
with $A$- and $B$-fermions, only the (irrelevant) corrections to the free 
vacuum energy, $E_V - E_0$, are different (the $B$-fermion vacuum loops 
are missing).

The last case we will consider in this section is the one of 
$A \bar{A}$ bound states, of one fermion and the corresponding antifermion.
In this case, there are additional contributions from the virtual 
annihilation diagrams to the effective Hamiltonian, so that the effective 
potential reads
\bal
\lefteqn{\langle \bf{p}_A, r; \bf{p}_{\bar{A}}, s | V 
| \bf{p}_A', r'; \bf{p}_{\bar{A}}', s' \rangle} \n \\
&= i g^2 \int_{-\infty}^0 dt \, e^{- \epsilon \abs{t}}
\int d^3 x \, d^3 x' \left[ \bar{\psi}_{\bf{p}'_{\bar{A}}, s'}^{\bar{A}} 
(0, \bf{x}) \psi_{\bf{p}_{\bar{A}}, s}^{\bar{A}} (0, \bf{x}) \right] \n \\
&\phantom{=} \hspace{6cm} {}\times \Delta_F (0 - t, \bf{x} - \bf{x}')
\left[ \bar{\psi}_{\bf{p}_A, r}^A (t, \bf{x}') 
\psi_{\bf{p}'_A, r'}^A (t, \bf{x}') \right] \n \\
&\phantom{=} {}+ i g^2 \int_{-\infty}^0 dt \, e^{- \epsilon \abs{t}}
\int d^3 x \, d^3 x' \left[ \bar{\psi}_{\bf{p}_A, r}^A (0, \bf{x}) 
\psi_{\bf{p}'_A, r'}^A (0, \bf{x}) \right]  \n \\
&\phantom{=} \hspace{6cm} {}\times \Delta_F (0 - t, \bf{x} - \bf{x}')
\left[ \bar{\psi}_{\bf{p}'_{\bar{A}}, s'}^{\bar{A}} (t, \bf{x}') 
\psi_{\bf{p}_{\bar{A}}, s}^{\bar{A}} (t, \bf{x}') \right] \n \\
&\phantom{=} {}- i g^2 \int_{-\infty}^0 dt \, e^{- \epsilon \abs{t}}
\int d^3 x \, d^3 x' \left[ \bar{\psi}_{\bf{p}_A, r}^A 
(0, \bf{x}) \psi_{\bf{p}_{\bar{A}}, s}^{\bar{A}} (0, \bf{x}) \right] \n \\
&\phantom{=} \hspace{6cm} {}\times \Delta_F (0 - t, \bf{x} - \bf{x}')
\left[ \bar{\psi}_{\bf{p}'_{\bar{A}}, s'}^{\bar{A}} (t, \bf{x}') 
\psi_{\bf{p}'_A, r'}^A (t, \bf{x}') \right] \n \\
&\phantom{=} {}- i g^2 \int_{-\infty}^0 dt \, e^{- \epsilon \abs{t}}
\int d^3 x \, d^3 x' \left[ \bar{\psi}_{\bf{p}'_{\bar{A}}, s'}^{\bar{A}} 
(0, \bf{x}) \psi_{\bf{p}'_A, r'}^A (0, \bf{x}) \right]  \n \\
&\phantom{=} \hspace{6cm} {}\times \Delta_F (0 - t, \bf{x} - \bf{x}')
\left[ \bar{\psi}_{\bf{p}_A, r}^A (t, \bf{x}') 
\psi_{\bf{p}_{\bar{A}}, s}^{\bar{A}} (t, \bf{x}') \right] \n \\
&= \frac{g^2}{\sqrt{2 E_{\bf{p}_A}^A \, 2 E_{\bf{p}_{\bar{A}}}^A \, 
2 E_{\bf{p}'_A}^A \, 2 E_{\bf{p}'_{\bar{A}}}^A}} \Bigg\{
\frac{1}{2 \omega_{\bf{p}_A - \bf{p}'_A}} \left( \frac{1}{E_{\bf{p}_A}^A +
\omega_{\bf{p}_A - \bf{p}'_A} - E_{\bf{p}'_A}^A} + 
\frac{1}{E_{\bf{p}_{\bar{A}}}^A +
\omega_{\bf{p}_{\bar{A}} - \bf{p}'_{\bar{A}}} - E_{\bf{p}'_{\bar{A}}}^A} 
\right) \n \\[2mm]
&\phantom{=} \hspace{5.5cm} {}\times \left[ \bar{u}_A (\bf{p}_A, r) \, 
u_A (\bf{p}'_A, r') \right] \left[ \bar{v}_A (\bf{p}'_{\bar{A}}, s') \, 
v_A (\bf{p}_{\bar{A}}, s) \right] \n \\[2mm]
&\phantom{=} {}- \frac{1}{2 \omega_{\bf{p}_A + \bf{p}_{\bar{A}}}} \left( 
\frac{1}{\omega_{\bf{p}_A + \bf{p}_{\bar{A}}} + E_{\bf{p}_A}^A
+ E_{\bf{p}_{\bar{A}}}^A} + \frac{1}{\omega_{\bf{p}'_A + \bf{p}'_{\bar{A}}} 
- E_{\bf{p}'_A}^A - E_{\bf{p}'_{\bar{A}}}^A} \right) \n \\
&\phantom{=} \hspace{1cm} {}\times \left[ \bar{u}_A (\bf{p}_A, r) \, 
v_A (\bf{p}_{\bar{A}}, s) \right] 
\left[ \bar{v}_A (\bf{p}'_{\bar{A}}, s') \, u_A (\bf{p}'_A, r') \right]
\Bigg\} (2 \pi)^3 \delta (\bf{p}_A + \bf{p}_{\bar{A}} - \bf{p}'_A 
- \bf{p}'_{\bar{A}}) \:.
\eal
We will leave the discussion of the effect of virtual annihilation on the
bound state energies in different theories and the related instability of 
these states for a future investigation.

To close this section, we will briefly comment on the corresponding
results in a purely bosonic theory, where the constituents
are chosen to be charged scalar bosons. The consistency of the 
non-relativistic and one-body limits in this case has been shown in detail 
in Ref.\ \cite{WL02}. If we substitute one or both of
the constituents by antibosons, there are, compared
to the fermionic case, no additional minus signs from anticommutation 
relations to take into account and, of course, no spinor structures, 
consequently the whole argument is much simpler than for fermionic 
constituents. The results are, however, finally the same: mass
renormalization is identical for antiparticles and for particles, and the
interaction due to scalar boson exchange is universally attractive and
does not distinguish particles from antiparticles. 

As for a static antisource, consider the charge density (properly to be 
multiplied by the charge of boson $B$)
\be
\rho (x) = \phi_B^\ast (x) i \frac{\d}{\d t} \phi_B (x) - \phi_B (x)
i \frac{\d}{\d t} \phi_B^\ast (x) \approx - 2 M_B \phi_B^\ast (x)
\phi_B (x)
\ee
for a classical negative-energy solution of the Klein-Gordon equation,
in case that the relevant momenta fulfill $\bf{p}^2 \ll M_B^2$. The
probability density is the negative of $\rho (x)$, hence we would
replace 
\be
\bm{:} \phi_B^\dagger (\bf{x}) \phi_B (\bf{x}) \bm{:} \: = \frac{1}{2 M_B} 
\, \delta (\bf{x}) \label{bosondpos}
\ee
in the interaction Hamiltonian
for a localized antisource. A more formal argument proceeds in analogy 
with Eqs.\ \eqref{antiloc}--\eqref{rightdpos} for the fermionic case, with 
the result \eqref{bosondpos}. The one-boson limit $M_B \to \infty$ is
then consistent also in the case of an antiboson $\bar{B}$.

Finally, in the case of identical bosonic constituents, the effective
Schr\"odinger equation is the same as for bosonic $AB$ bound states,
only that the wave function has to be symmetric under particle exchange
in this case. Since there are no spin degrees of freedom in the scalar
bosonic case, the spatial wave function has to be symmetric, hence only 
even angular momenta are allowed.

\section{Numerical solution}

In order to actually solve the effective Schr\"odinger equation in the
form \eqref{spinrep}, it is convenient first to separate off the angular and 
spin degrees of freedom. A direct numerical solution of Eq.\ 
\eqref{spinrep} would lead to numerical instabilities for equal and
opposite momenta, due to the presence of a singularity in the integrand.
The effective Hamiltonian is rotationally invariant, 
hence it is natural to consider total angular momentum eigenstates. To make
contact to the usual spectroscopy, we choose to couple first the
individual spins to a total spin $\bf{S}$ and then couple this spin
with the relative orbital angular momentum $\bf{L}$ to the total
angular momentum $\bf{J}$. The usual construction with Clebsch-Gordan
coefficients yields simultaneous eigenstates of $\bf{J}^2$, $J_z$,
$\bf{S}^2$, and $\bf{L}^2$ which we will denote as 
$\supfi{2S + 1}\cal{Y}_{lM}^J (\hat{\bf{p}})$, $\hat{\bf{p}} \equiv
\bf{p}/\abs{\bf{p}}$. Explicit expressions are given in Appendix
\ref{appang}.
 
The Hamiltonian \eqref{spinrep} contains the helicity operators 
$\hat{\bf{p}} \cdot \bm{\sigma}_A$ and $\hat{\bf{p}} \cdot \bm{\sigma}_B$. 
These operators are hermitian and unitary, and in particular
\be
(\hat{\bf{p}} \cdot \bm{\sigma}_A)^2 = (\hat{\bf{p}} \cdot \bm{\sigma}_B)^2
= 1 \:.
\ee
The helicity operators are invariant under spatial rotations, however, they
are odd under spatial parity transformations which maintain the spin 
directions unchanged. Since the Schr\"odinger equation \eqref{spinrep}
contains only even powers of helicity operators, the effective Hamiltonian
is parity even (the intrinsic parities of the constituent fermions have no
use in the present context, and we will not consider them in the following). 

We hence have the conservation of total angular momentum $J$ and spatial 
parity $(-1)^l$, but a priori not of relative orbital angular momentum $l$ 
or total spin $S$. For given $J$, $l$ can take the values $J$ (for $S=0$)
and $J, J \pm 1$ (for $S=1$), $l = J, J - 1$ being excluded for $J=0$,
$S=1$. Taking into account the conservation of $(-1)^l$, we can hence 
conclude without any explicit calculation that the effective Hamiltonian 
may mix states with $l=J$, $S=0$ and with $l=J$, $S=1$ on the one hand
(we will call this ``S-coupling''), and states $l=J-1$, $S=1$ and
$l=J+1$, $S=1$ (``L-coupling'') on the other. 
The effective Schr\"odinger equation
will then decay into pairs of coupled one-dimensional equations. In the
special case $J=0$, neither of the two mixings is possible. For future
use we remark that the use of helicity eigenstates is expected to
diagonalize the effective Hamiltonian in the S-coupled sector, thus slightly 
simplifying the calculations, although there is no reason why L-coupling 
should not occur in this case.

For the actual solution of the effective Schr\"odinger equation
\eqref{spinrep}, we need explicit expressions for the application of
the helicity operators. Again, it is clear from the fact that the
helicity operators preserve total angular momentum and change parity,
that the application of a helicity operator maps S-coupled states to
L-coupled states and vice versa. The explicit expressions, as well as a
rather pedestrian way to derive them, are presented in Appendix
\ref{appang}. Finally, the integration over the angles, i.e., over
$\hat{\bf{p}}'$, in Eq.\ \eqref{spinrep} can be performed with
the help of a partial wave decomposition combined with the spherical 
harmonics addition theorem,
\be
V (p, p', \cos \theta) = \sum_{l=0}^\infty \frac{4 \pi}{2l + 1} \, 
a_l (p, p') \sum_{m=-l}^l Y_{lm} (\hat{\bf{p}}) Y_{lm}^\ast (\hat{\bf{p}}')
\:, \label{addtheorem}
\ee
where $p \equiv \abs{\bf{p}}$, $\theta$ denotes the angle between $\bf{p}$ 
and $\bf{p}'$, and
\be
a_l (p, p') = \frac{2l + 1}{2} \int_{-1}^1 d \cos \theta \, P_l (\cos \theta)
V (p, p', \cos \theta) \:.
\ee
(Note that we have changed conventions relative to the ones employed in our 
earlier work \cite{WL02}.)

The whole procedure outlined above can be carried through independently
of the boson mass $\mu$. In what follows, I will focus on the case of 
massless bosons $\mu=0$, as discussed in the introduction. The explicit 
form of the effective Schr\"odinger equation in this case is, again, given 
in Appendix \ref{appang}. No additional difficulties are expected in the 
massive case in principle, even though the existence of a critical coupling 
constant changes the qualitative features of the spectrum.

The (pairs of) one-dimensional integral equations can now be
solved numerically. To this end, the equations were converted to
(continuous) matrix form and the corresponding two-dimensional integrals 
approximated by finite sums over a discrete two-dimensional grid. The 
distribution of abscissas took the logarithmic singularity of the 
integrand and the long range of the wave functions in configuration space 
into account. We have approximated the solutions by a finite linear 
combination of an appropriately chosen set of basis functions, the same
we had used before in the scalar case \cite{WL02}. Two parameters
that determine the shape of the basis functions were optimized variationally.
The orthogonality of the basis functions could be retained numerically to 11 
to 14 decimal places. Both energies and wave functions converged with 
increasing number of integration points and basis functions. However, in our 
experience \cite{WL02} convergence does not guarantee the correctness of 
the solutions if the choice of basis is inappropriate. For this reason, we
have also checked the residual $r_i (p)$ of the solutions $\phi_i (p)$ 
defined as $(E - H_{\text{BW}}) \phi_i (p) = r_i (p)$. The determination 
of the residual $r_i(p)$ and the ``point-wise" convergence of 
the wave functions were limited essentially by the redundancy 
of the grid points (up to 400) with respect to the number of basis functions 
(up to 40). The analogue of the Gibbs phenomenon in Fourier series was, in 
the worst case, of the order of two percent.

\begin{figure}
\begin{center}
\resizebox{11cm}{!}{\includegraphics*{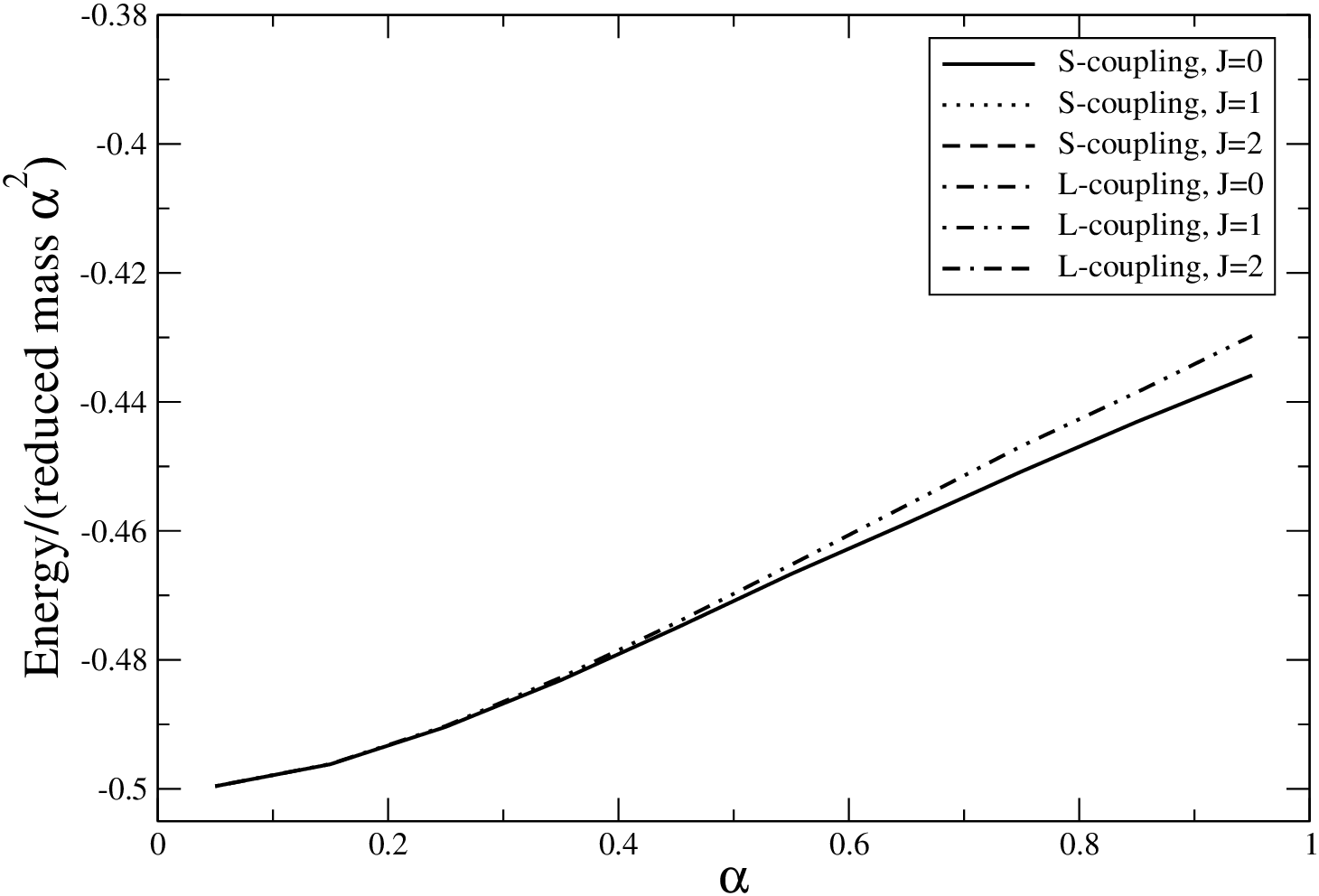}} \\[1cm]
\resizebox{11cm}{!}{\includegraphics*{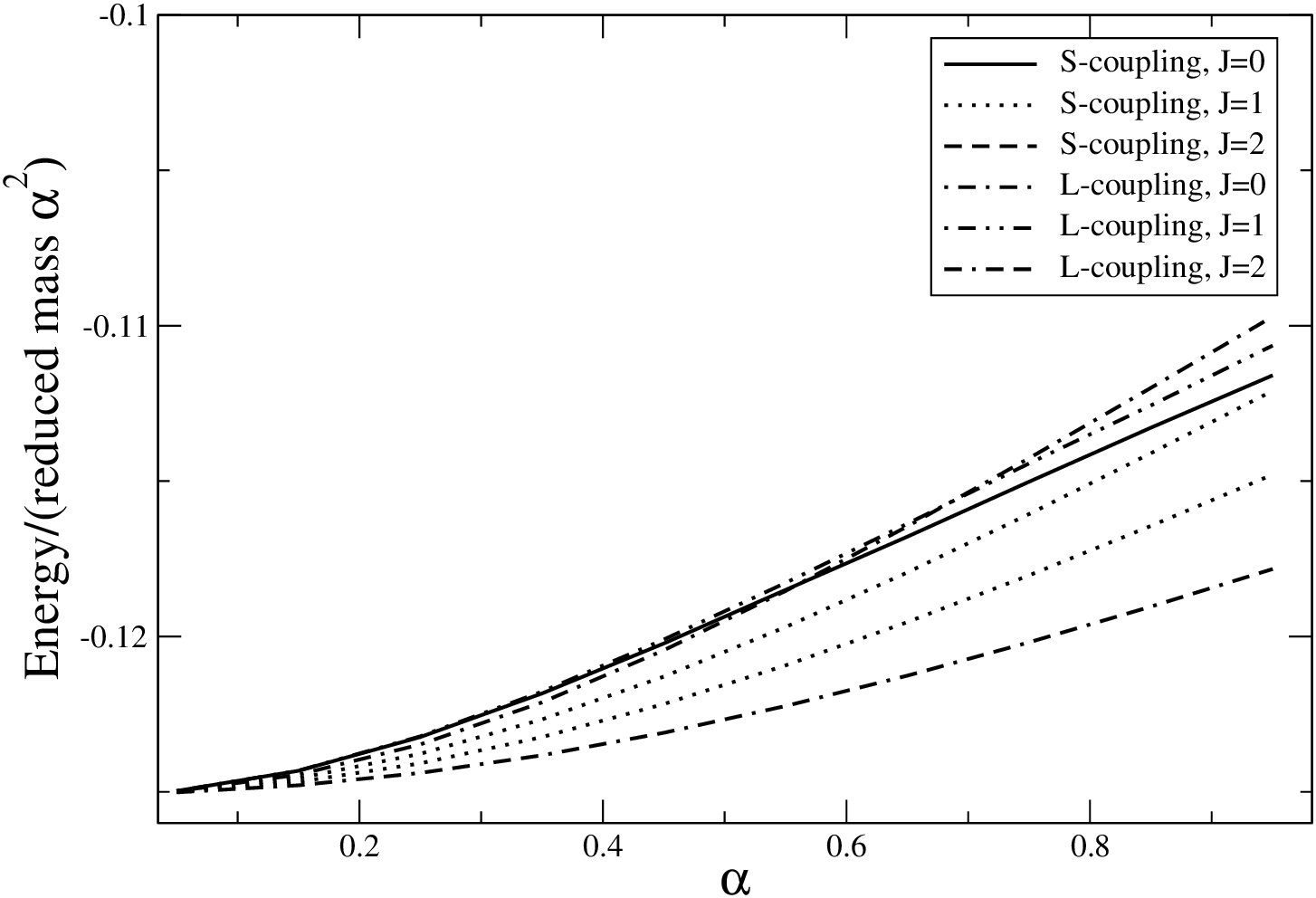}}
\end{center}
\caption{The lowest energy eigenvalues (principal quantum numbers $n=1$
and $n=2$) of the effective Hamiltonian as functions of the fine structure 
constant $\alpha = g^2/(4 \pi)$, for the case of equal fermion masses. 
\label{figq}}
\end{figure}

\begin{figure}
\begin{center}
\resizebox{11cm}{!}{\includegraphics*{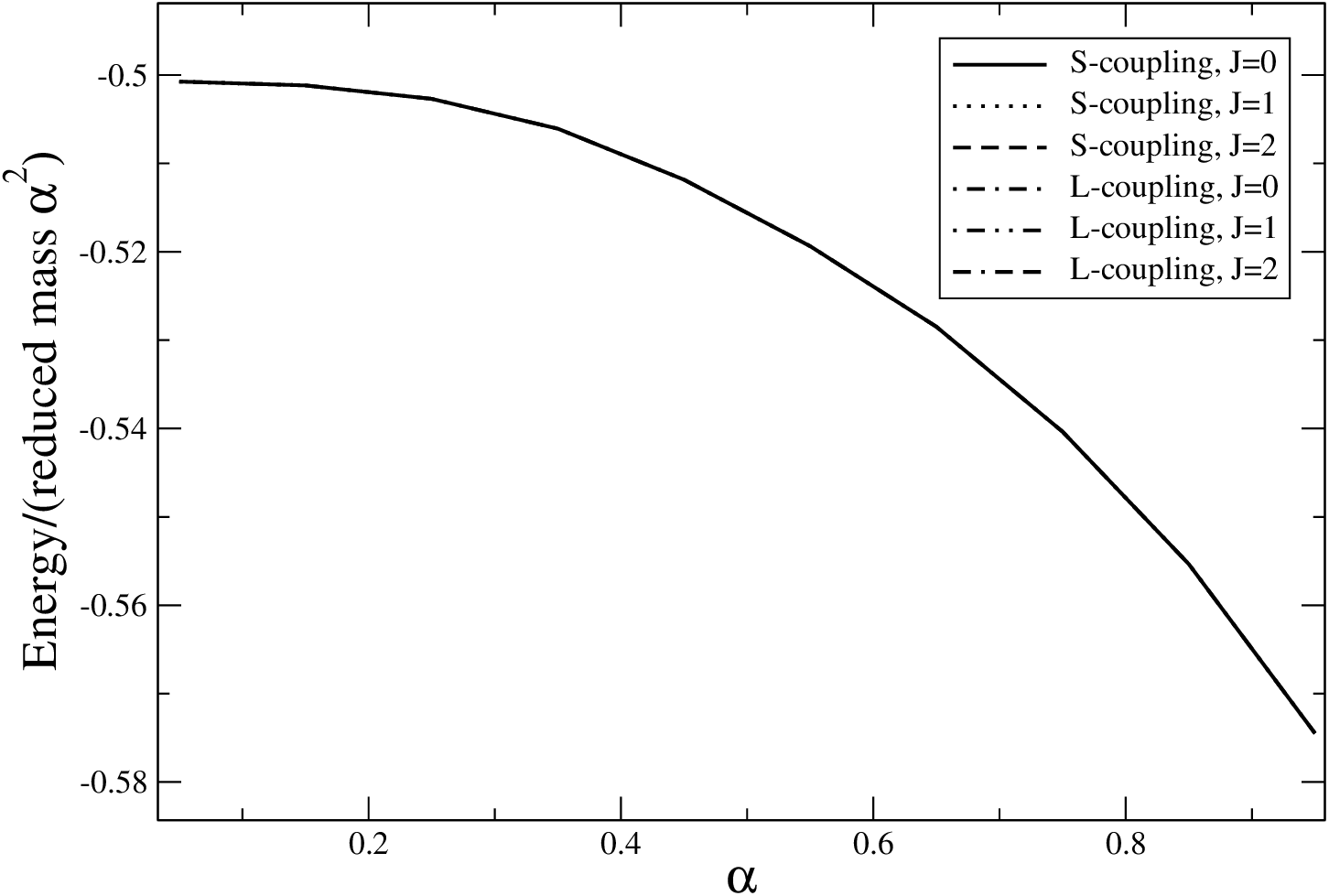}} \\[1cm]
\resizebox{11cm}{!}{\includegraphics*{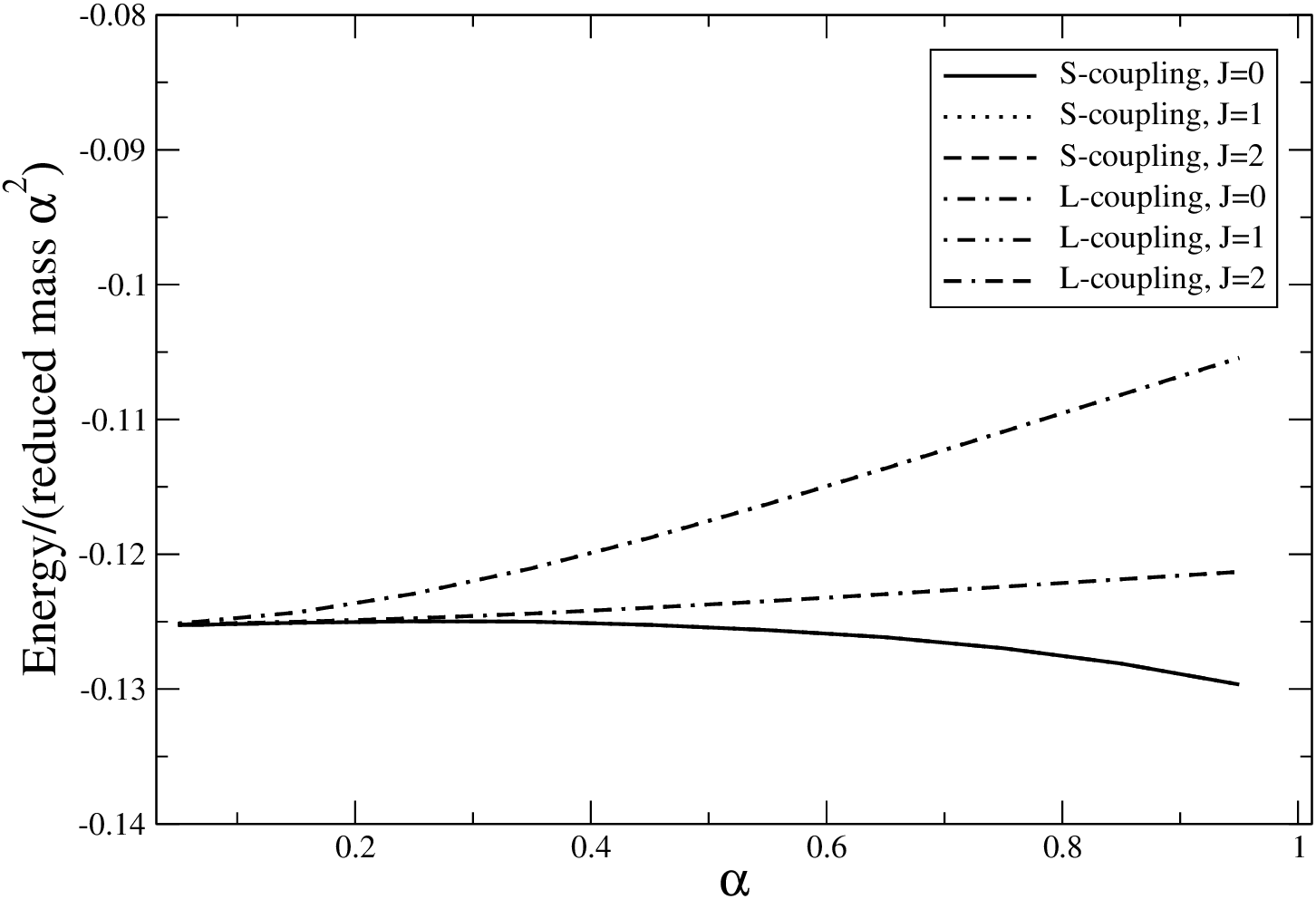}}
\end{center}
\caption{The lowest energy eigenvalues ($n=1$ and $n=2$) in the one-body 
limit $M_B \to \infty$. 
\label{figh}}
\end{figure}

\begin{figure}
\begin{center}
\resizebox{11cm}{!}{\includegraphics*{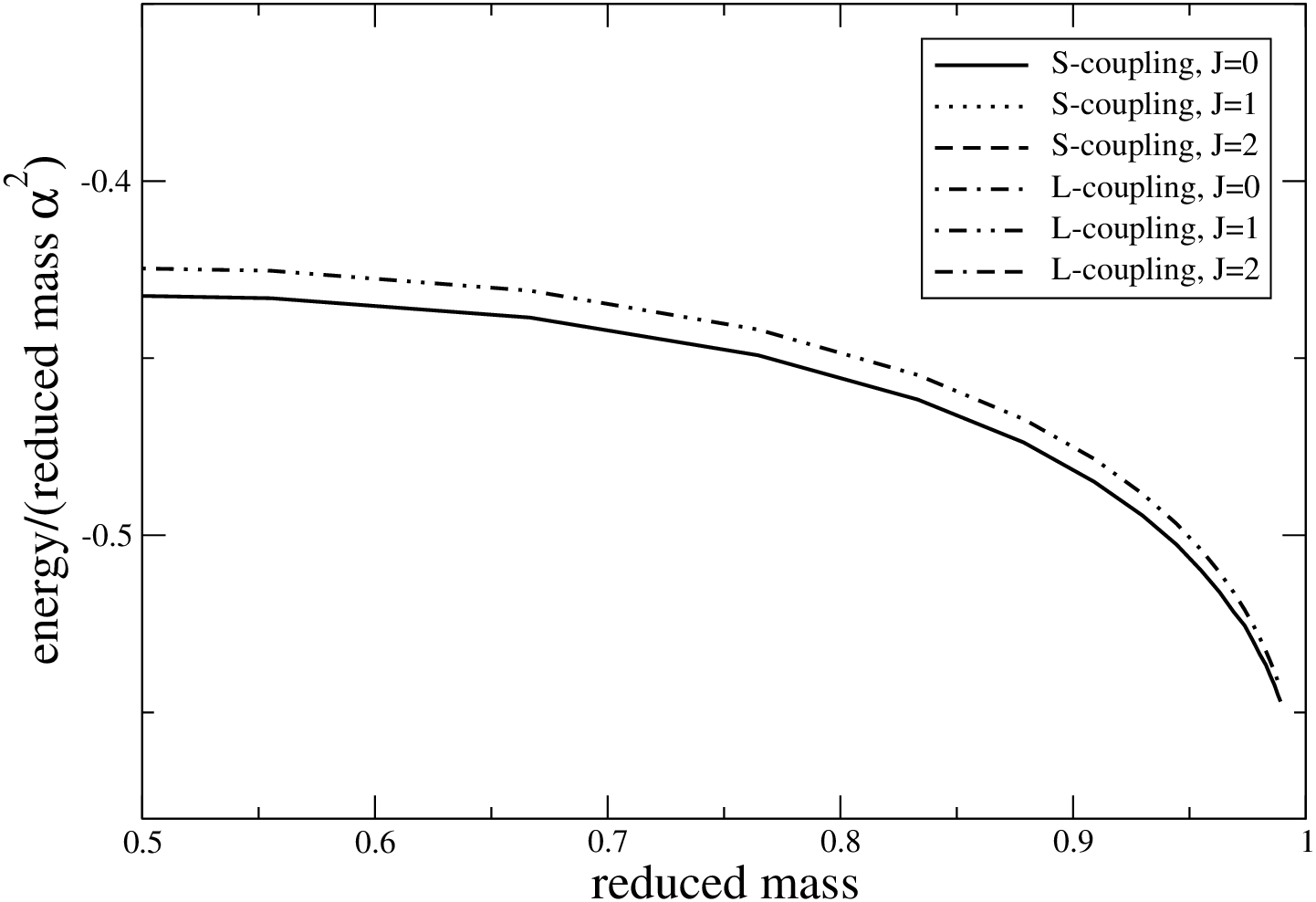}} \\[1cm]
\resizebox{11cm}{!}{\includegraphics*{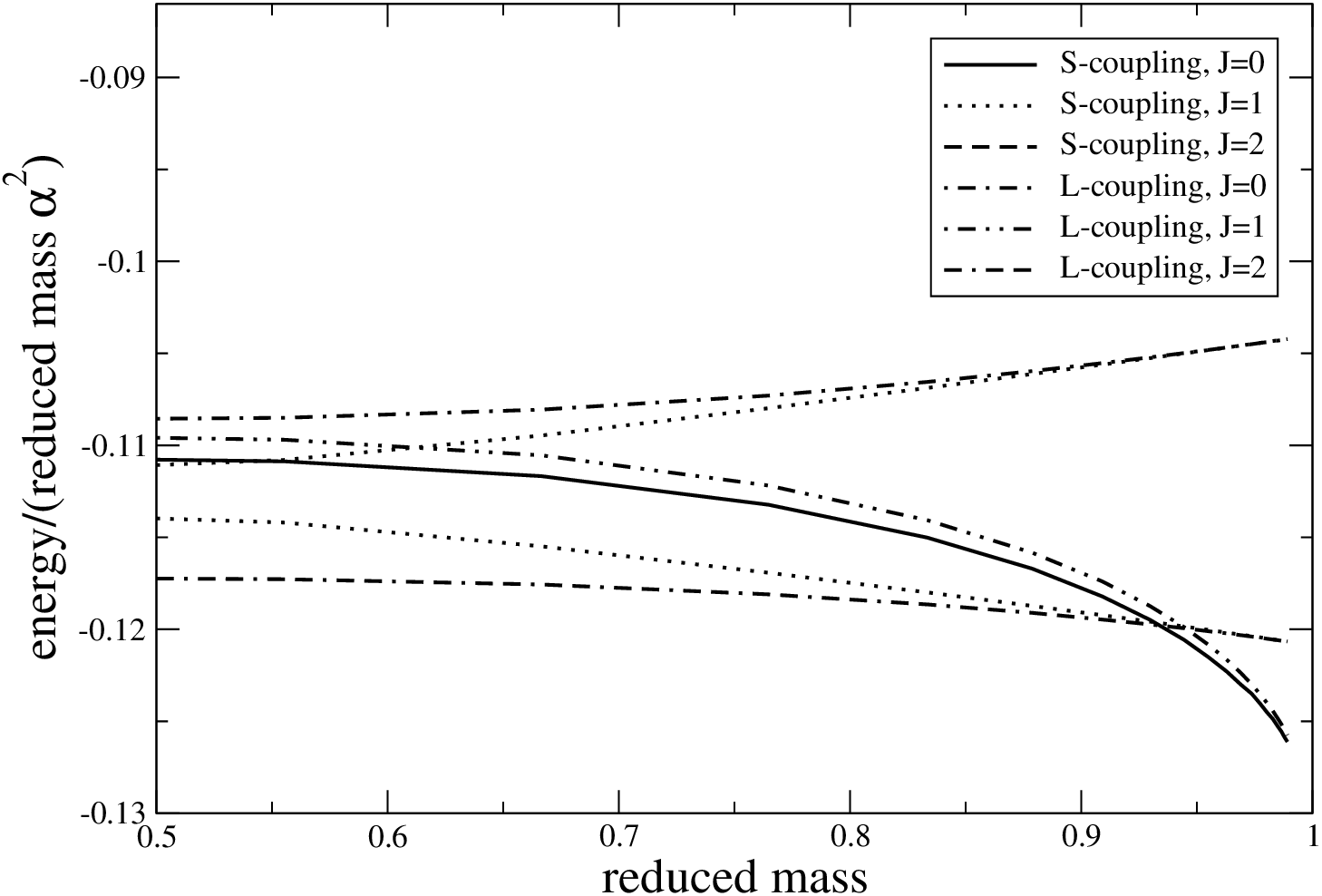}}
\end{center}
\caption{The variation of the lowest energy eigenvalues with the mass ratio 
$M_r/M_A$ between the extreme cases $M_r = M_A/2$ and $M_r = M_A$ depicted 
in Figs.\ 1 and 2, respectively, for fixed $\alpha = 1$.
\label{figr}}
\end{figure}

The numerical solutions (for $\mu = 0$) are shown in Figs.\ 
\ref{figq} and \ref{figh} for two extreme mass ratios, $M_A = M_B$ and
$M_B \to \infty$ (with $M_A$ fixed), as functions of the (Yukawa theory)
fine structure constant $\alpha = g^2/4 \pi$. 
Between these two extremes, the eigenvalues for fixed $\alpha$ can be seen to 
vary smoothly with the mass ratio in Fig.\ \ref{figr} for $\alpha = 1$. 
In all figures, the energy eigenvalues are represented normalized to twice 
the non-relativistic ionization energy in a Coulomb potential, $M_r \alpha^2$, 
where $M_r$ denotes the reduced mass. 

In Figs.\ \ref{figq} and \ref{figh}, 
it is seen that in the small-coupling limit $\alpha \to 0$, the energies 
tend to the non-relativistic Coulomb values
\be
- \frac{M_r \alpha^2}{2} \frac{1}{n^2} \:, \quad
n=1,2,3,\ldots \:, \label{nrel}
\ee
as expected for $\mu = 0$. In particular, we observe the characteristic
degeneracies in this limit. For instance, we expect and find that the
following states tend to the same energy eigenvalue $- M_r \alpha^2/8$
(principal quantum number $n=2$) for $\alpha \to 0$:
the ground states in the L-coupled $J=0$ and $J=2$ sectors (states
$2 \supfi{3} P_0$ and $2 \supfi{3} P_2$ in the usual spectroscopic notation
$n \supfi{2 S + 1} L_J$) and in the S-coupled $J=1$ sector ($2 \supfi{1} P_1$),
as well as the first excited states in the S-coupled $J=0$ and $J=1$
sectors ($2 \supfi{1} S_0$ and $2 \supfi{3} P_1$) and in the L-coupled
$J=1$ sector ($2 \supfi{3} S_1$). To be precise, the lowest-energy 
eigenstates in the S-coupled $J=1$ sector are linear combinations of the 
$2 \supfi{1} P_1$ and $2 \supfi{3} P_1$ states, the coefficients depending 
on the mass ratio. In Fig.\ \ref{figh}, only three of these degenerate
states are visible because there are always two curves lying on top of
each other (see the discussion of the one-body limit below).

Let us now discuss the equal-mass case of Fig.\ \ref{figq} in detail. 
The binding is \emph{weaker} in this case than predicted by the 
non-relativistic formula. Compared to an electromagnetic interaction 
(exchange of photons) as in positronium, where we eliminate the 
contribution of the virtual annihilation diagram, the sign of the 
relativistic corrections is just the opposite (with the exception of the 
$1 \supfi{3} S_1$ state). Also, the ordering of the levels is different.

In the case of equal masses, the effective 
Hamiltonian possesses an additional symmetry under the exchange of the 
fermions $A$ and $B$,
\be
\phi (\bf{p}) \to \phi (- \bf{p})^t
\ee
in terms of the spinorial wave function in the center-of-mass system
(see Eq.\ \eqref{defttrans}). This symmetry is implicit in our discussion 
of bound states of identical fermions in Section 3, where we saw that in 
the case of identical constituents the effective Schr\"odinger equation is 
the same as for $AB$ bound states, hence it must possess exchange symmetry.
The exchange parity for the angular momentum eigenstates is 
$(-1)^l (-1)^{S+1}$. Together with the symmetry under spatial parity $(-1)^l$
(remember that we do not take the intrinsic parities of the fermions 
into account), this new symmetry forbids the $S$-coupling, hence 
$S$ becomes a good quantum number in this case. This can also be seen
explicitly in the expressions for the matrix elements of the effective
potential in the S-coupled sector, Eq.\ \eqref{Scouplepot} in Appendix 
\ref{appang}, for $M_A = M_B$. In the case of identical fermionic
constituents, the wave function must be antisymmetric with respect to
particle exchange, hence, for $J$ even, in the S-coupled sector only
$S=0$ states are possible, while we have both $l = J - 1$ and $l = J + 1$
states in the L-coupled sector, in particular the coupling between these
states remains. On the other hand, for $J$ odd, we have only $S=1$ states
in the S-coupled sector, and \emph{no} possible state in the L-coupled
sector. Among the eight states shown in Fig.\ \ref{figq}, five are
antisymmetric under particle exchange, namely the states $1 \supfi{1} S_0$, 
$2 \supfi{1} S_0$, $2 \supfi{3} P_0$, $2 \supfi{3} P_1$, and 
$2 \supfi{3} P_2$. We can see explicitly in these examples that the absence 
of S-coupling is necessary for the antisymmetry of the states.

In the one-body limit $M_B \to \infty$ depicted in Fig.\ \ref{figh},
the sign of the relativistic corrections changes for several states 
with respect to the equal-mass case. Within numerical accuracy, there 
are always two exactly degenerate states. The reason for the degeneracy 
can be seen from the effective Schr\"odinger
equation in the one-body limit, Eq.\ \eqref{1Blimit}: the effective
Hamiltonian is invariant under rotations which involve the spatial
coordinates and the spin $\bf{s}_A$ of particle $A$ only, and
\emph{independently} under rotations of spin $\bf{s}_B$
(which does not affect the dynamics). The total angular momentum of 
fermion $A$, $\bf{j}_A = \bf{L} + \bf{s}_A$, is then a conserved
quantity, and it is natural to consider simultaneous eigenstates of
$\bf{j}^2_A$, $j_{A,z}$, $\bf{L}^2$, and $s_{B,z}$. Since $j_A = l
\pm 1/2$ and spatial parity $(-1)^l$ is conserved as before, it follows
that $l$ is a good quantum number in this limit.

As a further consequence, states that only differ in the value of
$j_{A, z}$ or $s_{B, z}$ are degenerate, and one may as well consider 
simultaneous eigenstates of $\bf{J}^2$, $J_z$, $\bf{j}_A^2$, and $\bf{L}^2$, 
where $\bf{J} = \bf{j}_A + \bf{s}_B$ with eigenvalues $J = j_A \pm 1/2$. 
Now states which only differ in the eigenvalue of $\bf{J}^2$ or $J_z$
are degenerate. We can express
the latter states in terms of the simultaneous eigenstates of $\bf{J}^2$, 
$J_z$, $\bf{L}^2$, and $\bf{S}^2$ that we are using in the numerical
calculations. From the beforegoing discussion, we expect the absence of
L-coupling and the degeneracy of one of the S-coupled states which is
eigenstate of $\bf{j}_A^2$ with eigenvalue $j_A = J + 1/2$ and $l=J$,
with an L-coupled state with $\bf{J}^2$-eigenvalue $J+1$ and $l=J$
(hence necessarily $j_A = J + 1/2$). The other S-coupled state with $l=J$
is an $\bf{j}_A^2$-eigenstate with $j_A = J - 1/2$, and is degenerate
with an L-coupled state with $\bf{J}^2$-eigenvalue $J- 1$, $l=J$, and 
$j_A = J - 1/2$. Explicit expressions for the eigenstates in the different 
coupling schemes and their relations are given in Appendix \ref{appang},
where we follow a different line of reasoning starting from the explicit
expressions for the matrix elements of the effective potential in the 
S- and L-coupled sectors, Eqs.\ \eqref{Scouplepot} and \eqref{Lcouplepot}.

These expectations are fully borne out in the results of the numerical
calculations. Among the eight states calculated in Fig.\ \ref{figh}, we
expect and find the following to be degenerate: $1 \supfi{1} S_0$ and
$1 \supfi{3} S_1$, $2 \supfi{1} S_0$ and $2 \supfi{3} S_1$, $2 \supfi{3} P_0$
with one linear combination of $2 \supfi{1} P_1$ and $2 \supfi{3} P_1$,
and $2 \supfi{3} P_2$ with the orthogonal linear combination of
$2 \supfi{1} P_1$ and $2 \supfi{3} P_1$. The coefficients of these linear
combinations are predicted and found to be $-\sqrt{1/3}$ and $\sqrt{2/3}$,
and $\sqrt{2/3}$ and $\sqrt{1/3}$, respectively (see Eq.\ \eqref{jAcouple}
in Appendix \ref{appang}), and correspond to eigenstates of
$\bf{j}_A^2$ with eigenvalues $j_A = 1/2$ and $3/2$.
Unlike for the Dirac equation with an electromagnetic Coulomb potential, 
states with the same $j_A$ but different $l$, here $j_A = 1/2$ and $l=0$
(states $2 \supfi{1} S_0$ and $2 \supfi{3} S_1$) or $l=1$ (states 
$2 \supfi{3} P_0$ and the first linear combination of $2 \supfi{1} P_1$ and 
$2 \supfi{3} P_1$), are not degenerate. In addition, the ordering of the 
$l=1$ states is opposite to the electromagnetic case.

Figure \ref{figr} shows the smooth transition between the two extreme mass
ratios $M_A = M_B$ and $M_A/M_B = 0$ for fixed $\alpha = 1$, and the
appearance of the characteristic degeneracies in the limit $M_A/M_B \to 0$.
One would like to compare the coefficients of the mixing of the S-coupled
states $2 \supfi{1} P_1$ and $2 \supfi{3} P_1$ against theoretical
predictions. However, the diagonalization of the effective potential matrix
\eqref{Scouplepot} is generally not possible without solving the entire 
equation, i.e., it cannot be isolated from the $p$-dependence of the wave 
function. To the order $\alpha^4$ of the first relativistic corrections,
the diagonalization can still be performed analytically and turns out to
factorize from the ``radial'' $p$-dependence. The results are presented in
Appendix \ref{appert} and provide very satisfactory approximations
to the corresponding results of the numerical calculations. In particular, 
the analytical results indicate that there is no hyperfine splitting to 
the order $\alpha^4$ and $M_A/M_B$, for the energy levels that are 
degenerate in the one-body limit. In fact, even for $\alpha = 1$ no 
hyperfine splitting of the order $M_A/M_B$ is visible in Fig.\ \ref{figr}.

\section{Conclusions}

We have presented what appears to be the first consistent treatment of
bound states in Yukawa theory. It is the result of a straightforward
application of the generalized Gell-Mann--Low theorem. The
consistency of the method has been checked thoroughly. In particular,
we have shown that mass renormalization can be performed exactly as in
a manifestly covariant formulation, even though the renormalization 
conditions were imposed entirely within our Hamiltonian framework.
We have checked the non-relativistic and one-body limits, replaced the
fermionic constituents by antifermions, and considered the case of identical
constituents. In all these cases, the formalism generates the correct
results in a very natural way. In the numerical calculations, no abnormal
solutions have been found (nor were there expected to be any, due to the
absence of relative time or energy as a dynamical variable), and all the
characteristic degeneracies in the non-relativistic and one-body limits
show up in the numerical results with very good accuracy.

In general terms, the framework presented here has several advantages over
other formulations of quantum field theoretic bound state equations. As
we have shown, the derivation of the effective Schr\"odinger equation is
straightforward and presents no essential complications in the case of 
fermionic constituents as compared to scalar bosons. In principle, the
complete bound state spectrum can be obtained as we have demonstrated
by numerically determining the eight lowest-lying states (corresponding to
the non-relativistic principal quantum numbers $n=1$ and $n=2$). The
wave functions for these states are also obtained in the course of the
(approximate) diagonalization of the effective Hamiltonian.

Several rather technical issues, which are nonetheless expected to be important
for related work in the near future, have been treated in detail in the
appendices. In particular, the relation between manifestly covariant and
non-covariant representations of the relevant loop integrals has been
established (Appendix \ref{appint}). We have discussed dimensional, 
Pauli-Villars, Schwinger proper time and covariant and non-covariant
cutoff regularizations for the appearing one-loop integrals from a
non-covariant Hamiltonian perspective (Appendix \ref{appreg}). Finally,
we have presented explicit expressions for the angular momentum eigenstates 
in different coupling schemes, for the application of helicity operators and
the coefficient functions in a partial wave expansion, all necessary
ingredients for the separation of angular and spin variables in the
effective Schr\"odinger equation (Appendix \ref{appang}).

The results of this work, if only as a point of departure for the application
to more realistic physical situations in the future, bear on fundamental 
issues in nuclear and high energy physics, as for example the nucleon-nucleon
interaction. In this respect, one interesting particular result of the
numerical computations is the qualitative difference between the relativistic
bound state spectra for scalar (boson exchange) and electromagnetic
interactions.

\subsection*{Acknowledgments}
One of us (A.W.) gratefully acknowledges support by CIC-UMSNH and Conacyt 
grant 32729-E. Part of the research (by N.L.) was done at the Department
of Physics and Astronomy of the University of Pittsburgh in the group of
Eric Swanson and Steve Dytman.

\begin{appendix}

\section{Covariant and non-covariant representations of loop integrals
\label{appint}}

In this appendix, we will relate different expressions for loop integrals in
momentum space, where the manifestly covariant representations arise directly
from the momentum space Feynman rules, while the non-covariant representations
result naturally from the application of the Gell-Mann--Low theorem.

Let us begin with the lowest-order correction to the vacuum energy density
Eq.\ \eqref{vacmom}. The equivalence will be established by performing
the integrations over $p_0$ and $p'_0$ through the use of complex
integration theory. For greater clarity, we will first discuss the
analogous problem in a purely scalar theory \cite{WL02}. The corresponding
formula differs from Eq.\ \eqref{vacmom} by the global sign (for both 
expressions) and the numerators of the integrands which are simply equal
to one. Considering the first term (for particle $A$) for concreteness,
we begin with the integral over $p'_0$,
\be
\int_{- \infty}^\infty \frac{d p'_0}{2 \pi} \,
\frac{1}{[ (p - p')^2 - \mu^2 + i \epsilon ] [ {p'}^2 - m_A^2 + i \epsilon ]} 
\:. \label{vacint}
\ee
Standard application of the residue theorem, closing the integration
contour through the usual large semicircle in the upper half plane, gives
\be
-i \left\{ \frac{1}{2 E_{\bf{p}'}^A \left[ (p_0 + E_{\bf{p}'}^A)^2
- \omega_{\bf{p} - \bf{p}'}^2 + i \epsilon \right]} +
\frac{1}{2 \omega_{\bf{p} - \bf{p}'} \left[ (p_0 - \omega_{\bf{p} - \bf{p}'})^2
- (E_{\bf{p}'}^A)^2 + i \epsilon \right]} \right\} \:.
\ee
This expression becomes much more transparent after a decomposition in
partial fractions with respect to $p_0$,
\bmu
\frac{i}{2 \omega_{\bf{p} - \bf{p}'} \, 2 E_{\bf{p}'}^A} \bigg[
\frac{1}{p_0 + \omega_{\bf{p} - \bf{p}'} + E_{\bf{p}'}^A - i \eta}
- \frac{1}{p_0 - \omega_{\bf{p} - \bf{p}'} - E_{\bf{p}'}^A + i \eta} \\
+ 2 \pi i \, \delta \left( p_0 - \omega_{\bf{p} - \bf{p}'} + E_{\bf{p}'}^A
\right) \bigg] \:, \label{parfrac}
\emu
where we have used the formula
\be
\frac{1}{\omega - i \eta} - \frac{1}{\omega + i \eta} = 2 \pi i \delta
(\omega)
\ee
(for $\eta \to 0$). Equation \eqref{parfrac} \emph{cannot be
correct} as it stands: the original integral \eqref{vacint} is even under
$p_0 \to - p_0$ (by the substitution $p_0' \to - p_0'$ of the integration
variable), and this symmetry is manifestly broken by the delta function
in Eq.\ \eqref{parfrac}.

However, the term with the delta function only contributes for
$p_0 = \omega_{\bf{p} - \bf{p}'} - E_{\bf{p}'}^A$, and this is precisely
the value of $p_0$ where the two poles in the upper half plane coincide.
The correct evaluation of the residue at the double pole gives for the
integral in this case
\be
\frac{i}{2 \omega_{\bf{p} - \bf{p}'} \, 2 E_{\bf{p}'}^A} \bigg[
\frac{1}{p_0 + \omega_{\bf{p} - \bf{p}'} + E_{\bf{p}'}^A - i \eta}
- \frac{1}{p_0 - \omega_{\bf{p} - \bf{p}'} - E_{\bf{p}'}^A + i \eta}
\bigg]_{p_0 = \omega_{\bf{p} - \bf{p}'} - E_{\bf{p}'}^A} \:,
\ee
so the result \eqref{parfrac} is wrong for this value of $p_0$, and the
term with the delta function has to be omitted in \eqref{parfrac}. The
integration over $p_0$ can then be performed easily, using the residue
theorem again. The result, after substituting $\bf{p}' \to - \bf{p}'$,
 is the one expected from Eq.\	\eqref{vacmom}
or rather its scalar analogue. Note that the use of Eq.\ \eqref{parfrac}
including the delta function would lead to additional (incorrect) terms
in Eq.\ \eqref{vacmom}.

In the Yukawa case, Eq.\ \eqref{vacmom} proper, the argument is nearly
identical, only the expressions are slightly more complicated. The
result for the naive $p_0'$-integration, after the decomposition in
partial fractions, reads
\bal
&\phantom{=} \int_{- \infty}^\infty \frac{d p'_0}{2 \pi} \, 
\frac{4 (p \cdot p' + m_A^2)}
{[ (p - p')^2 - \mu^2 + i \epsilon ] [ {p'}^2 - m_A^2 + i \epsilon ]} \n \\
&= \frac{4i}{2 \omega_{\bf{p} - \bf{p}'} \, 2 E_{\bf{p}'}^A} \bigg\{
-2 E_{\bf{p}'}^A + \Big[ E_{\bf{p}'}^A ( E_{\bf{p}'}^A + 
\omega_{\bf{p} - \bf{p}'} ) - \bf{p} \cdot \bf{p}' + m_A^2 \Big]
\bigg[ \frac{1}{p_0 + \omega_{\bf{p} - \bf{p}'} + E_{\bf{p}'}^A - i \eta} \n \\
&\phantom{=} {}- \frac{1}{p_0 - \omega_{\bf{p} - \bf{p}'} - E_{\bf{p}'}^A 
+ i \eta} \bigg] + \Big[ E_{\bf{p}'}^A ( E_{\bf{p}'}^A - 
\omega_{\bf{p} - \bf{p}'} ) - \bf{p} \cdot \bf{p}' + m_A^2 \Big]
2 \pi i \, \delta \left( p_0 - \omega_{\bf{p} - \bf{p}'} + E_{\bf{p}'}^A 
\right) \bigg\} \:. \label{parfrac2}
\eal
By a calculation of the residue at the double pole for the special case
$p_0 = \omega_{\bf{p} - \bf{p}'} - E_{\bf{p}'}^A$, one can again show
that the term with the delta function in Eq.\ \eqref{parfrac2} is spurious.
Integration over $p_0$ of the rest gives the desired result.

We now turn to the lowest-order corrections to the mass, Eq.\
\eqref{onepmom}. As before, we begin with the scalar case where all the
numerators in Eq.\ \eqref{onepmom} are replaced by one \cite{WL02}. Then 
the result of the $p_0'$-integration in the manifestly covariant expression 
is given precisely by Eq.\ \eqref{parfrac} above, with $p_0$ to be replaced 
by $E_{\bf{p}}^A$. The delta function in Eq.\ \eqref{parfrac} is, of course,
again spurious, although it cannot give any contribution anyway as long as
$\mu^2 < 4 m_A^2$.

In Yukawa theory, where Eq.\ \eqref{onepmom} properly applies, the
$p_0'$-integration gives the following result \cite{Est03}, after a 
decomposition in partial fractions (with respect to $p_0$),
\bmu
\int_{- \infty}^\infty \frac{d p'_0}{2 \pi} \,
\frac{p' \cdot \gamma + m_A}{[ (p - p')^2 - \mu^2 + i \epsilon ] 
[ {p'}^2 - m_A^2 + i \epsilon ]} \\
= \frac{i}{2 \omega_{\bf{p} - \bf{p}'} \, 2 E_{\bf{p}'}^A} \bigg[
\frac{- E_{\bf{p}'}^A \gamma_0 - \bf{p}' \cdot \bm{\gamma} + m_A}
{p_0 + \omega_{\bf{p} - \bf{p}'} + E_{\bf{p}'}^A - i \eta}
- \frac{E_{\bf{p}'}^A \gamma_0 - \bf{p}' \cdot \bm{\gamma} + m_A}
{p_0 - \omega_{\bf{p} - \bf{p}'} - E_{\bf{p}'}^A + i \eta} \\
+ \left( - E_{\bf{p}'}^A \gamma_0 - \bf{p}' \cdot \bm{\gamma} + m_A \right)
2 \pi i \, \delta \left( p_0 - \omega_{\bf{p} - \bf{p}'} + E_{\bf{p}'}^A
\right) \bigg] \:, \label{parfrac3}
\emu
putting $p_0 = E_{\bf{p}}^A$. Again, the delta function turns out to be
spurious (by explicitly considering the case of a double pole in the upper
half plane), which leads to the non-covariant expression in Eq.\ 
\eqref{onepmom}.

\section{Regularization of one-loop integrals \label{appreg}}

The aim of this appendix is to derive Eq.\ \eqref{Gform} starting from
the non-covariant expression for $G (\bf{p})$  in Eq.\ \eqref{onepmom} in
a suitably regularized form, so that all integrals appearing in the
derivation are well-defined. By far the simplest way is to go through the
manifestly covariant form also presented in Eq.\ \eqref{onepmom},
regularized correspondingly for the present purpose. For the complications
that arise in a direct derivation in the non-covariant formulation for the
simpler case of a purely scalar theory where it has to be shown that
$G (\bf{p})$ is actually independent of $\bf{p}$ (and only depends on
$\left. p^2 \right|_{p_0 = E_{\bf{p}}^A} = m_A^2$), see Ref.\ 
\cite{KG00}.

\subsection{Dimensional regularization}

The technically simplest regularization scheme (although not the most
natural one in the present context, see below) is dimensional
regularization. The idea is to continuously change the dimension of space
to smaller values where all the integrals are well-defined, to be save
to spatial dimensions smaller than two, i.e., space-time dimensions
$D < 3$. We can then establish the relation between the non-covariant
and covariant expression in Eq.\ \eqref{onepmom} for these dimensions and 
show that the dependence on $D$ is analytical, with a simple pole appearing 
at $D = 4$. As a consequence, Eq.\ \eqref{Gform} can be
shown to hold true for space-time dimensions arbitrarily close to (but 
smaller than) four by analytical continuation. However, the
definition of the integrals in arbitrary (non-integer) dimensions is
made precise only in the (Euclidean) covariant formulation, which makes 
this form of regularization somewhat unnatural for the non-covariant 
expressions.

In detail, we begin by establishing the relation between the non--covariant
and covariant expressions for $G(\bf{p})$ in Eq.\ \eqref{onepmom}, but
for $(D - 1)$ spatial dimensions instead of three where, to begin with,
$D < 3$ in order that all integrals are well-defined. The analogue of
Eq.\ \eqref{onepmom} in $D$ dimensions can then be shown by
integrating over $p_0'$ in the covariant form exactly as detailed in
Appendix \ref{appint} [see Eq.\ \eqref{parfrac3} and the following remarks].
The $d^{D-1} p'$-integration is not touched in this process. With the
covariant form at hand, we can introduce Feynman parameters in the usual way:
\bal
G^{\text{DR}}_\varepsilon (\bf{p})
&= \left. i g^2 \int \frac{d^D p'}{(2 \pi)^D} \int_0^1 dx
\frac{p' \cdot \gamma + m_A}{ \left[ (p - p')^2 x - \mu^2 x + {p'}^2 (1 - x) 
- m_A^2 (1 - x) + i \epsilon \right]^2} \right|_{p_0 = E_{\bf{p}}^A} \n \\
&= \left. i g^2 \int_0^1 dx \int \frac{d^D p'}{(2 \pi)^D} 
\frac{p' \cdot \gamma + m_A}{ \left[ (p' - x p)^2 + x(1 - x) p^2 - x \mu^2 
- (1 - x) m_A^2 + i \epsilon \right]^2} \right|_{p_0 = E_{\bf{p}}^A} \n \\
&= \left. i g^2 \int_0^1 dx \int \frac{d^D q}{(2 \pi)^D} 
\frac{x p \cdot \gamma + m_A}{ \left[ q^2 + x(1 - x) p^2 - x \mu^2 
- (1 - x) m_A^2 + i \epsilon \right]^2} \right|_{p_0 = E_{\bf{p}}^A} \:,
\label{Gform2}
\eal
denoting the dimensionally regularized form of $G (\bf{p})$ as
$G^{\text{DR}}_\varepsilon (\bf{p})$, where $\varepsilon = 4 - D$. In the last 
step, we have shifted the integration variable to $q = p' - x p$ and used 
that for the term $q \cdot \gamma$ emerging in the numerator, the integrand 
is odd. These manipulations are unproblematic as long as we stick to 
dimensions $D < 3$ where the integrals are well-defined.

It is now convenient to decompose $G^{\text{DR}}_\varepsilon (\bf{p})$
in two parts in analogy with Eq.\ \eqref{Gform},
\be
G^{\text{DR}}_\varepsilon (\bf{p}) = \left[ G_{1, \varepsilon}^{\text{DR}} 
(p^2) p \cdot \gamma + G_{0, \varepsilon}^{\text{DR}} (p^2) m_A 
\right]_{p_0 = E_{\bf{p}}^A} \:. \label{GformD}
\ee
However, all space-time vectors in \textit{this} equation are still
$D$-dimensional. Replacing $\left. p^2 \right|_{p_0 = E_{\bf{p}}^A}$
by $m_A^2$ and Wick rotating to Euclidean space gives
\be
G^{\text{DR}}_{0, \varepsilon} (m_A^2)
= - g^2 \int_0^1 dx \int \frac{d^D q_E}{(2 \pi)^D} 
\frac{1}{\left[ q_E^2 + (1 - x)^2 m_A^2 + x \mu^2 \right]^2}
\label{Eucl}
\ee
The corresponding expression for $G^{\text{DR}}_{1, \varepsilon} (m_A^2)$ is 
equal to Eq.\ 
\eqref{Eucl} except for an additional factor of $x$ in the integrand.
It is these latter integrals which are rigorously defined for arbitrary, 
continuous values of $D$. They are actually defined in such a way as to make 
them analytical functions of $D$, with a simple pole at $D = 4$. We can
use analytic continuation to define all the beforegoing integrals for
$3 \le D < 4$, leaving the established relations intact, in particular
Eq.\ \eqref{GformD}. In the limit $\varepsilon \to 0$ or $D \to 4$, via the 
standard formulae of dimensional regularization,
\be
G^{\text{DR}}_{0, \varepsilon} (m_A^2)
= - \frac{g^2}{(4 \pi)^2} \int_0^1 dx \left[ 
\frac{2}{\varepsilon} - \gamma_E + \ln (4 \pi) -
\ln \frac{(1 - x)^2 m_A^2 + x \mu^2}{\kappa^2} \right] \:, \label{dimreG}
\ee
plus terms which tend to zero in this limit. In Eq.\ \eqref{dimreG}, 
$\kappa$ is the renormalization scale, and $g$ is 
left dimensionless also for $D \neq 4$. The corresponding expression for 
$G^{\text{DR}}_{1, \varepsilon} (m_A^2)$ is, again, equal to Eq.\ 
\eqref{dimreG} except for an additional factor of $x$ in the integrand.

After an integration by parts, the evaluation of the integrals over the
Feynman parameter $x$ is straightforward and yields \cite{Est03}
\bal
\lefteqn{G^{\text{DR}}_{0, \varepsilon} (m_A^2) = - \frac{g^2}{(4 \pi)^2} 
\Bigg\{ \frac{2}{\varepsilon} - \gamma_E + \ln (4 \pi) - 
\ln \frac{m_A^2}{\kappa^2} + 2 - \frac{1}{2} \frac{\mu^2}{m_A^2}
\ln \frac{\mu^2}{m_A^2}} \n \\
&\phantom{=} \hspace{6cm} {}- 2
\sqrt{\frac{\mu^2}{m_A^2} \left( 1 - \frac{1}{4} \frac{\mu^2}{m_A^2} \right)} 
\, \text{arcctg} \sqrt{\frac{\mu^2/(4 m_A^2)}{1 - \mu^2/(4 m_A^2)}} \Bigg\} 
\:, \label{dimreG2a} \\
\lefteqn{G^{\text{DR}}_{1, \varepsilon} (m_A^2) = - \frac{g^2}{(4 \pi)^2} 
\Bigg\{ \frac{1}{2} \left[ \frac{2}{\varepsilon} - \gamma_E + \ln (4 \pi) 
\right] - \frac{1}{2} \ln \frac{m_A^2}{\kappa^2} + \frac{3}{2} - \frac{1}{2} 
\frac{\mu^2}{m_A^2}} \n \\
&\phantom{=} {}- \left( \frac{\mu^2}{m_A^2} - \frac{1}{4} \frac{\mu^4}{m_A^4} 
\right) \ln \frac{\mu^2}{m_A^2} - \left( 2 - \frac{\mu^2}{m_A^2} \right) 
\sqrt{\frac{\mu^2}{m_A^2} \left( 1 - \frac{1}{4} \frac{\mu^2}{m_A^2} \right)} 
\, \text{arcctg} \sqrt{\frac{\mu^2/(4 m_A^2)}{1 - \mu^2/(4 m_A^2)}} \Bigg\} 
 \label{dimreG2b}
\eal
(for $\varepsilon \to 0$). Observe the simple pole at $\varepsilon = 0$ or 
$D=4$ and the absence of IR divergences for $\mu \to 0$.

The result for $G^{\text{DR}}_{0, \varepsilon} (m_A^2)$ gives immediately the 
explicit dimensionally regularized expression for $\Delta m_A^2$ in the 
purely scalar case considered in Ref.\ \cite{WL02} (where the coupling 
constant $g$ has the dimension of mass). For Yukawa theory, we define 
$\Delta m_A^2$ by
\be
\Delta m_A^2 = 2 m_A^2 \left[ G^{\text{DR}}_{0, \varepsilon} (m_A^2) + 
G^{\text{DR}}_{1, \varepsilon} (m_A^2) \right] \:,
\ee
so that
\be
\bar{u}_A^\varepsilon (\bf{p}, r) G^{\text{DR}}_\varepsilon (\bf{p}) 
u_A^\varepsilon (\bf{p}, s) = \Delta m_A^2 \, \delta_{rs}
\ee
from the analogue of Eq.\ \eqref{dirac}, where the Dirac equation in $D$
dimensions is used. Explicitly, from Eqs.\ \eqref{dimreG2a} and
\eqref{dimreG2b},
\bal
\frac{\Delta m_A^2}{2 m_A^2} &= 
- \frac{g^2}{(4 \pi)^2} \Bigg\{ \frac{3}{2} \left[ 
\frac{2}{\varepsilon} - \gamma_E + \ln (4 \pi) \right] - \frac{3}{2}
\ln \frac{m_A^2}{\kappa^2} + \frac{7}{2} - \frac{1}{2} \frac{\mu^2}{m_A^2} 
- \left( \frac{3}{2} \frac{\mu^2}{m_A^2} -
\frac{1}{4} \frac{\mu^4}{m_A^4} \right) \ln \frac{\mu^2}{m_A^2} \n \\
&\phantom{=} {}- \left( 4 - \frac{\mu^2}{m_A^2} \right) 
\sqrt{\frac{\mu^2}{m_A^2} \left( 1 - \frac{1}{4} \frac{\mu^2}{m_A^2} \right)} 
\, \text{arcctg} \sqrt{\frac{\mu^2/(4 m_A^2)}{1 - \mu^2/(4 m_A^2)}} \Bigg\}
\:,
\eal
which can now be used to redefine $m_A$ in terms of $M_A$ as in Eq.\ 
\eqref{rencond} and the limit $\varepsilon \to 0$ be taken with the (finite)
physical mass $M_A$ held fixed.

\subsection{Pauli-Villars regularization}

We now consider Pauli-Villars regularization which turns out to be the
most natural regularization scheme in the context of Hamiltonian
(non-covariant) perturbation theory. It is effected by subtracting from the 
expression in Eq.\ \eqref{onepmom}, denoted for the time being as 
$G (\bf{p}, \mu)$ to make its dependence on the boson mass $\mu$ explicit, a 
similar contribution for a ficticious ``heavy'' boson of mass $\Lambda$ to 
define
\be
G^{\text{PV}}_\Lambda (\bf{p}) = G (\bf{p}, \mu) - G (\bf{p}, \Lambda) \:.
\ee
It is then easy to verify by power counting that 
$G^{\text{PV}}_\Lambda (\bf{p})$ is UV finite, both from the non-covariant 
and the manifestly covariant expression.

The $p_0'$-integration in the covariant expression can then be performed
again as in Appendix \ref{appint}, only that the following
integration over $d^3 p'$ is now well-defined. This establishes the
equivalence of the non-covariant and manifestly covariant expressions for
$G^{\text{PV}}_\Lambda (\bf{p})$ [i.e., the regularized form of Eq.\ 
\eqref{onepmom}] rigorously. Starting from the covariant form,  we now 
introduce Feynman parameters and shift the integration variable as in Eq.\ 
\eqref{Gform2} to obtain
\be
G (\bf{p}, \mu)
= \left. i g^2 \int_0^1 dx \int \frac{d^4 q}{(2 \pi)^4} 
\frac{x p \cdot \gamma + m_A}{ \left[ q^2 + x(1 - x) p^2 - x \mu^2 
- (1 - x) m_A^2 + i \epsilon \right]^2} \right|_{p_0 = E_{\bf{p}}^A}
\label{Gform2PV}
\ee
and the analogous expression for $G (\bf{p}, \Lambda)$ [remembering that
its difference $G^{\text{PV}}_\Lambda (\bf{p})$ is well-defined in 
4 dimensions]. From Eq.\ \eqref{Gform2PV} and the corresponding expression 
for $G (\bf{p}, \Lambda)$, it is clear that $G^{\text{PV}}_\Lambda (\bf{p})$ 
is of the form described in Eq.\ \eqref{Gform}. The mass renormalization can 
then already be effected at this stage as detailed following Eq.\ 
\eqref{Gform}, reading $G^{\text{PV}}_\Lambda (\bf{p})$ instead of 
$G (\bf{p})$.

To evaluate $G^{\text{PV}}_\Lambda (\bf{p})$ analytically, it is easiest to 
apply dimensional regularization to $G (\bf{p}, \mu)$ and 
$G (\bf{p}, \Lambda)$ separately and analytically continue the result for the 
difference $G^{\text{PV}}_\Lambda (\bf{p})$ back to four space-time 
dimensions. We start with the form obtained in Eq.\ \eqref{Gform2PV} above, 
replace $\left. p^2 \right|_{p_0 = E_{\bf{p}}^A}$ by $m_A^2$, and separate 
$G^{\text{PV}}_\Lambda (\bf{p})$ in two parts
$G^{\text{PV}}_{1, \Lambda} (m_A^2)$ and 
$G^{\text{PV}}_{0, \Lambda} (m_A^2)$, analogous to Eq.\
\eqref{Gform}. The continuation to $D$ space-time dimensions and a Wick 
rotation to Euclidean space leads to the expression \eqref{Eucl}
for the contribution $G_0 (m_A^2, \mu)$. The integrations over
$d^D q_E$ and the Feynman parameter then yield the results given in Eqs.\
\eqref{dimreG2a} and \eqref{dimreG2b} for the contributions $G_0 (m_A^2, \mu)$ 
and $G_1 (m_A^2, \mu)$, respectively. 
The expressions for $G_0 (m_A^2, \Lambda)$ and $G_1 (m_A^2, \Lambda)$ are
obtained by replacing $\mu \to \Lambda$. In the (relevant) limit of large 
$\Lambda$, one analytically continues the last terms in Eqs.\ \eqref{dimreG2a} 
and \eqref{dimreG2b} to $\Lambda^2 > 4 m_A^2$ and expands the complete 
expression in powers of $m_A^2/\Lambda^2$ (up to second order). This gives,
after several cancellations,
\bal
G_0 (m_A^2, \Lambda) &\to - \frac{g^2}{(4 \pi)^2} \Bigg\{ 
\frac{2}{\varepsilon} - \gamma_E + \ln (4 \pi) - 
\ln \frac{\Lambda^2}{\kappa^2} + 1 \Bigg\} \:, \n \\
G_1 (m_A^2, \Lambda) &\to - \frac{g^2}{(4 \pi)^2} \Bigg\{ \frac{1}{2} 
\left[ \frac{2}{\varepsilon} - \gamma_E + \ln (4 \pi) \right] - \frac{1}{2}
\ln \frac{\Lambda^2}{\kappa^2} + \frac{1}{4} \Bigg\}
\eal
for $\Lambda^2 \gg m_A^2$. The same results can be obtained somewhat easier 
by replacing $(1 - x)^2 m_A^2 + x \Lambda^2 \to x \Lambda^2$ in the
argument of the logarithm in Eq.\ \eqref{dimreG} [and analogously for
$G_1 (m_A^2, \Lambda)$] from the start.

We hence obtain, finally, for $G_0 (m_A^2)$ and $G_1 (m_A^2)$ in 
Pauli-Villars regularization
\bal
G^{\text{PV}}_{0, \Lambda} (m_A^2) &= - \frac{g^2}{(4 \pi)^2} \Bigg\{ 
\ln \frac{\Lambda^2}{m_A^2} + 1 - \frac{1}{2} 
\frac{\mu^2}{m_A^2} \ln \frac{\mu^2}{m_A^2} \n \\
&\phantom{=} {}- 2
\sqrt{\frac{\mu^2}{m_A^2} \left( 1 - \frac{1}{4} \frac{\mu^2}{m_A^2} \right)} 
\, \text{arcctg} \sqrt{\frac{\mu^2/(4 m_A^2)}{1 - \mu^2/(4 m_A^2)}} \Bigg\}
\:, \label{pauvill1a} \\
G^{\text{PV}}_{1, \Lambda} (m_A^2) &= - \frac{g^2}{(4 \pi)^2} \Bigg\{ 
\frac{1}{2} \ln \frac{\Lambda^2}{m_A^2} + \frac{5}{4} - \frac{1}{2} 
\frac{\mu^2}{m_A^2} - \left( \frac{\mu^2}{m_A^2} -
\frac{1}{4} \frac{\mu^4}{m_A^4} \right) \ln \frac{\mu^2}{m_A^2} \n \\
&\phantom{=} {}- \left( 2 - \frac{\mu^2}{m_A^2} \right) 
\sqrt{\frac{\mu^2}{m_A^2} \left( 1 - \frac{1}{4} \frac{\mu^2}{m_A^2} \right)} 
\, \text{arcctg} \sqrt{\frac{\mu^2/(4 m_A^2)}{1 - \mu^2/(4 m_A^2)}} \Bigg\}
\label{pauvill1b}
\eal
($\Lambda^2 \gg m_A^2$). We obtain a logarithmic UV divergence with 
$\Lambda \to \infty$ and, again, the absence of IR divergences for $\mu \to 0$.
We emphasize that we have used dimensional regularization only as a
convenient calculational tool here, and that 
$G^{\text{PV}}_{0, \Lambda} (m_A^2)$ and $G^{\text{PV}}_{1, \Lambda} (m_A^2)$ 
are well-defined in four space-time dimensions from the start.

The result for $G^{\text{PV}}_{0, \Lambda} (m_A^2)$ gives directly the 
expression for $\Delta m_A^2$ in Pauli-Villars regularization for the purely 
scalar case of Ref.\ \cite{WL02}. For Yukawa theory, we have from Eq.\ 
\eqref{DeltamAdef}
\bal
\frac{\Delta m_A^2}{2 m_A^2} &= - \frac{g^2}{(4 \pi)^2} \Bigg\{ \frac{3}{2} 
\ln \frac{\Lambda^2}{m_A^2} + \frac{9}{4} - \frac{1}{2} 
\frac{\mu^2}{m_A^2} - \left( \frac{3}{2} \frac{\mu^2}{m_A^2} -
\frac{1}{4} \frac{\mu^4}{m_A^4} \right) \ln \frac{\mu^2}{m_A^2} \\
&\phantom{=} {}- \left( 4 - \frac{\mu^2}{m_A^2} \right) 
\sqrt{\frac{\mu^2}{m_A^2} \left( 1 - \frac{1}{4} \frac{\mu^2}{m_A^2} \right)} 
\, \text{arcctg} \sqrt{\frac{\mu^2/(4 m_A^2)}{1 - \mu^2/(4 m_A^2)}} \Bigg\} 
\:. \label{pauvill2}
\eal

\subsection{Schwinger proper time regularization}

The other two regularization schemes that we will discuss, Schwinger
proper time and momentum cutoff regularization, are only effective in
Euclidean space. Hence we have to write the non-covariant expression
in Eq.\ \eqref{onepmom} in Euclidean space in order to regularize. After
evaluation, it can then be analytically continued to physical values of
the external variables.

Let us begin with the diagrammatic expressions in Eq.\ \eqref{onepcor}.
We rewrite the first contribution to the mass renormalization by use
of the covariant representations \eqref{Fpropcov} of the propagators and 
Eq.\ \eqref{wavef} for the fermionic wave functions as
\bmu
- i g^2 \int_{-\infty}^0 dt \, e^{- \epsilon \abs{t}}
\int d^3 x \, d^3 x' \left[ \bar{\psi}_{\bf{p}_A, r}^A (0, \bf{x}) 
S_F^A (0 - t, \bf{x} - \bf{x}') \psi_{\bf{p}_A', s}^A (t, \bf{x}') \right] 
\Delta_F (0 - t, \bf{x} - \bf{x}') \\
= i g^2 \int \frac{d^3 p'}{(2 \pi)^3} \frac{d^3 k}{(2 \pi)^3} 
\int d^3 x \, d^3 x' \, \frac{e^{-i \bf{p}_A \cdot \bf{x}}}
{\sqrt{2 E_{\bf{p}_A}^A}} 
\int \frac{d p_0'}{2 \pi} \, \frac{\left[ \bar{u}_A (\bf{p}_A, r)
\left( p' \cdot \gamma + m_A \right) u_A (\bf{p}_A', s) \right] 
e^{i \bf{p}' \cdot (\bf{x} - \bf{x}')}}
{p'_0{}^2 - \bf{p}'{}^2 - m_A^2 + i \epsilon} \\
\times \int \frac{d k_0}{2 \pi} \, \frac{e^{i \bf{k} \cdot (\bf{x} - \bf{x}')}}
{k_0^2 - \bf{k}^2 - \mu^2 + i \epsilon}
\, \frac{e^{i \bf{p}_A' \cdot \bf{x}'}}{\sqrt{2 E_{\bf{p}_A'}^A}}
\left. \int_{-\infty}^0 d t \, e^{\epsilon t} e^{i (p'_0 + k_0 - p_0) t}
\right|_{p_0 = E_{\bf{p}_A'}^A} \label{ncovW}
\emu
In the latter expression we Wick rotate in the mathematically positive sense
\be
p'_0 \to p'_0 = i p_0^{\prime E} \:, \quad k_0 \to k_0 = i k_0^E \:,
\ee
where $p'_0{}^E$ and $k_0^E$ are real after the rotation. The sense of
the rotation is determined by the position of the poles in the integrand
(by the $i \epsilon$-prescription). Since $p'_0{}^E$ and $k_0^E$ take both
positive and negative values, it is imperative to Wick rotate $t$, too,
keeping the exponents imaginary in order that the integrand do not blow up.
This implies the rotation
\be
t \to t = - i t^E
\ee
in the negative sense, where $-\infty < t^E \le 0$. Keeping $p_0$ fixed at
$p_0 = E_{\bf{p}_A'}^A$ would then lead to a divergence in the 
$t^E$-integration, so that we have to rotate, in addition,
\be
p_0 \to p_0 = i p_0^E \:.
\ee
After performing the integration, we then analytically continue the
result to $p_0^E = -i E_{\bf{p}_A'}^A$.

As a result of this Wick rotation, the expression \eqref{ncovW} is written as
\bmu
- \frac{g^2}{\sqrt{2 E_{\bf{p}_A}^A \, 2 E_{\bf{p}_A'}^A}} 
\int \frac{d^3 p'}{(2 \pi)^3} \frac{d^3 k}{(2 \pi)^3} 
\int d^3 x \, d^3 x' \, e^{-i \bf{p}_A \cdot \bf{x}} \\
\times \int \frac{d p_0^{\prime E}}{2 \pi} \, 
\frac{\left[ \bar{u}_A (\bf{p}_A, r)
\left( p_0^{\prime E} \gamma_0^E + \bf{p}' \cdot \bm{\gamma}^E + m_A \right) 
u_A (\bf{p}_A', s) \right] 
e^{i \bf{p}' \cdot (\bf{x} - \bf{x}')}}
{(p_0^{\prime E})^2 + \bf{p}'{}^2 + m_A^2} \\
\times \int \frac{d k_0^E}{2 \pi} \, 
\frac{e^{i \bf{k} \cdot (\bf{x} - \bf{x}')}}
{(k_0^E)^2 + \bf{k}^2 + \mu^2}
\, e^{i \bf{p}_A' \cdot \bf{x}'} \left. \int_{-\infty}^0 d t^E \, 
e^{\epsilon^E t^E} e^{i (p_0^{\prime E} + k_0^E - p_0^E) t^E}
\right|_{p_0^E \to -i E_{\bf{p}_A'}^A} \:,
\emu
where we have defined $\epsilon^E = -i \epsilon$ to assure the convergence
of the $t^E$-integration. Alternatively, we can assume $\epsilon$ to
have a positive imaginary part from the beginning. As for the $\gamma$
matrices, we have chosen the common convention where
\be
p_0^{\prime E} \gamma_0^E + \bf{p}' \cdot \bm{\gamma}^E = p' \cdot \gamma \:,
\ee
i.e., $\gamma_0^E = i \gamma_0$ and $\gamma_i^E = \gamma^{i E} = - \gamma^i$.
Finally, for the momentum-space representation, we integrate over $d^3 x$, 
$d^3 x'$ and $d^3 k$ (taking advantage of the three-dimensional 
$\delta$-function resulting from the $x$- and $x'$-integrations). The 
resulting contribution to $G (\bf{p})$ [compare with Eq.\ \eqref{onepform}] 
is
\bmu
- g^2 \int \frac{d^3 p'}{(2 \pi)^3} \int \frac{d p_0^{\prime E}}{2 \pi} \, 
\frac{p_0^{\prime E} \gamma_0^E + \bf{p}' \cdot \bm{\gamma}^E + m_A}
{(p_0^{\prime E})^2 + \bf{p}'{}^2 + m_A^2} \\
\times \int \frac{d k_0^E}{2 \pi} \, 
\frac{1}{(k_0^E)^2 + (\bf{p} - \bf{p}')^2 + \mu^2}
\left. \int_{-\infty}^0 d t^E \, 
e^{\epsilon^E t^E} e^{i (p_0^{\prime E} + k_0^E - p_0^E) t^E}
\right|_{p_0^E \to -i E_{\bf{p}}^A} \:. \label{onepWa}
\emu

Turning now to the second contribution to the mass renormalization in 
Eq.\ \eqref{onepcor},
\bmu
- i g^2 \int_{-\infty}^0 dt \, e^{- \epsilon \abs{t}}
\int d^3 x \, d^3 x' \left[ \bar{\psi}_{\bf{p}_A, r}^A (t, \bf{x}') 
S_F^A (t - 0, \bf{x}' - \bf{x}) \psi_{\bf{p}'_A, s}^A (0, \bf{x}) \right] 
\Delta_F (t - 0, \bf{x}' - \bf{x}) \\
= i g^2 \int \frac{d^3 p'}{(2 \pi)^3} \frac{d^3 k}{(2 \pi)^3} 
\int d^3 x \, d^3 x' \, \frac{e^{-i \bf{p}_A \cdot \bf{x}'}}
{\sqrt{2 E_{\bf{p}_A}^A}} 
\int \frac{d p_0'}{2 \pi} \, \frac{\left[ \bar{u}_A (\bf{p}_A, r)
\left( p' \cdot \gamma + m_A \right) u_A (\bf{p}_A', s) \right] 
e^{-i \bf{p}' \cdot (\bf{x} - \bf{x}')}}
{p'_0{}^2 - \bf{p}'{}^2 - m_A^2 + i \epsilon} \\
\times \int \frac{d k_0}{2 \pi} \, 
\frac{e^{-i \bf{k} \cdot (\bf{x} - \bf{x}')}}
{k_0^2 - \bf{k}^2 - \mu^2 + i \epsilon}
\, \frac{e^{i \bf{p}_A' \cdot \bf{x}}}{\sqrt{2 E_{\bf{p}_A'}^A}}
\left. \int_{-\infty}^0 d t \, e^{\epsilon t} e^{-i (p'_0 + k_0 - p_0) t}
\right|_{p_0 = E_{\bf{p}_A}^A} \:,
\emu
and going through the steps following Eq.\ \eqref{ncovW}, we arrive at the
following contribution to $G (\bf{p})$:
\bmu
- g^2 \int \frac{d^3 p'}{(2 \pi)^3} \int \frac{d p_0^{\prime E}}{2 \pi} \, 
\frac{p_0^{\prime E} \gamma_0^E + \bf{p}' \cdot \bm{\gamma}^E + m_A}
{(p_0^{\prime E})^2 + \bf{p}'{}^2 + m_A^2} \\
\times \int \frac{d k_0^E}{2 \pi} \, 
\frac{1}{(k_0^E)^2 + (\bf{p} - \bf{p}')^2 + \mu^2}
\left. \int_{-\infty}^0 d t^E \, 
e^{\epsilon^E t^E} e^{-i (p_0^{\prime E} + k_0^E - p_0^E) t^E}
\right|_{p_0^E \to -i E_{\bf{p}}^A} \:. \label{onepWb}
\emu
It is easy to check the correctness of the results \eqref{onepWa} and
\eqref{onepWb} by performing the integrations over $p_0^{\prime E}$
and $k_0^E$ (using the residue theorem) and finally over $t^E$, leading
to the non-covariant expression in Eq.\ \eqref{onepmom}.

After this preparation, we are now in a position to introduce the
Schwinger proper time regularization by replacing
\be
\frac{1}{(p_0^{\prime E})^2 + (E_{\bf{p}'}^A)^2} = \int_0^\infty d \alpha \,
e^{- \alpha [(p_0^{\prime E})^2 + (E_{\bf{p}'}^A)^2]} \to
\int_{1/\Lambda^2}^\infty d \alpha \,
e^{- \alpha [(p_0^{\prime E})^2 + (E_{\bf{p}'}^A)^2]} \:, \label{Sreg}
\ee
and analogously for the other (scalar) propagator. We emphasize that only
in Euclidean space the modification of the lower limit for the parameter
integration corresponds to the exponential suppression of the propagator
for large momenta. Consequently, the proper time regularization of the
first contribution Eq.\ \eqref{onepWa} reads
\bmu
- g^2 \int \frac{d^3 p'}{(2 \pi)^3} \int_{1/\Lambda^2}^\infty d \alpha \,
d \beta \, e^{-\alpha (E_{\bf{p}'}^A)^2 - \beta \omega_{\bf{p} - \bf{p}'}^2}
\int_{-\infty}^0 d t^E \, e^{\epsilon^E t^E} e^{-i p_0^E t^E} \\
\times \int \frac{d p_0^{\prime E}}{2 \pi} \, 
\left[ p_0^{\prime E} \gamma_0^E + \bf{p}' \cdot \bm{\gamma}^E + m_A \right]
e^{-\alpha (p_0^{\prime E})^2 + i p_0^{\prime E} t^E}
\left. \int \frac{d k_0^E}{2 \pi} \, 
e^{-\beta (k_0^E)^2 + i k_0^E t^E} \right|_{p_0^E \to -i E_{\bf{p}}^A} \:.
\label{proptime1}
\emu

The integrations over $p_0^{\prime E}$ and $k_0^E$ in Eq.\ \eqref{proptime1}
are readily performed. After shifting the Euclidean time variable $t^E \to
t^{\prime E} = t^E + 2 i \alpha \beta p_0^E/(\alpha + \beta)$, one may
analytically continue $p_0^E$ to $-i E_{\bf{p}_A}^A$ directly in the
integrand to obtain
\bmu
- g^2 \int \frac{d^3 p'}{(2 \pi)^3} \int_{1/\Lambda^2}^\infty 
\frac{d \alpha \, d \beta}{4 \pi \sqrt{\alpha \beta}}
\, e^{-\alpha (E_{\bf{p}'}^A)^2 - \beta \omega_{\bf{p} - \bf{p}'}^2
+ \alpha \beta (E_{\bf{p}}^A)^2/(\alpha + \beta)} \\
\times \int_{-\infty}^{2 \alpha \beta E_{\bf{p}}^A/(\alpha + \beta)} 
d t^{\prime E} \left[ \left( \frac{\beta E_{\bf{p}}^A}{\alpha + \beta} 
- \frac{t^{\prime E}}{2 \alpha} \right) \gamma_0 - 
\bf{p}' \cdot \bm{\gamma} + m_A \right] 
e^{- (\alpha + \beta) (t^{\prime E})^2/(4 \alpha \beta)} \:.
\emu
A little algebra shows the first exponent in this expression to be negative.
Integration over $t^{\prime E}$ gives \cite{GR00}
\bmu
- g^2 \int \frac{d^3 p'}{(2 \pi)^3} \int_{1/\Lambda^2}^\infty 
\frac{d \alpha \, d \beta}{4 \pi \sqrt{\alpha \beta}}
\, e^{-\alpha (E_{\bf{p}'}^A)^2 - \beta \omega_{\bf{p} - \bf{p}'}^2}
\Bigg\{ \frac{\beta}{\alpha + \beta} \, \gamma_0 \\
+ \sqrt{\frac{\pi \alpha \beta}{\alpha + \beta}}
\left[ \frac{\beta E_{\bf{p}}^A}{\alpha + \beta} \, \gamma_0 - 
\bf{p}' \cdot \bm{\gamma} + m_A \right]
e^{\alpha \beta (E_{\bf{p}}^A)^2/(\alpha + \beta)} \, 
\text{erfc} \left( - \sqrt{\frac{\alpha \beta}{\alpha + \beta}} \,
E_{\bf{p}}^A \right) \Bigg\} \:, \label{onepSa}
\emu
where the complementary error function is defined as
\be
\text{erfc} (x) = \frac{2}{\sqrt{\pi}} \int_x^\infty d t \,
e^{- t^2} = \frac{2}{\sqrt{\pi}} \int_{- \infty}^{-x} dt \,
e^{- t^2} \:. \label{erfc}
\ee

For the second contribution to the mass renormalization we start from
Eq.\ \eqref{onepWb}, implement the Schwinger proper time regularization as in
Eq.\ \eqref{Sreg} and integrate over $p_0^{\prime E}$, $k_0^E$, and $t^E$
(after a shift and the continuation $p_0^E \to - i E_{\bf{p}}^A$) to
arrive at
\bmu
- g^2 \int \frac{d^3 p'}{(2 \pi)^3} \int_{1/\Lambda^2}^\infty 
\frac{d \alpha \, d \beta}{4 \pi \sqrt{\alpha \beta}}
\, e^{-\alpha (E_{\bf{p}'}^A)^2 - \beta \omega_{\bf{p} - \bf{p}'}^2}
\Bigg\{ -\frac{\beta}{\alpha + \beta} \, \gamma_0 \\
+ \sqrt{\frac{\pi \alpha \beta}{\alpha + \beta}}
\left[ \frac{\beta E_{\bf{p}}^A}{\alpha + \beta} \, \gamma_0 - 
\bf{p}' \cdot \bm{\gamma} + m_A \right]
e^{\alpha \beta (E_{\bf{p}}^A)^2/(\alpha + \beta)} \, 
\text{erfc} \left( \sqrt{\frac{\alpha \beta}{\alpha + \beta}} \,
E_{\bf{p}}^A \right) \Bigg\} \:. \label{onepSb}
\emu

The results \eqref{onepSa} and \eqref{onepSb} represent the proper time
regularization of the non-covariant expressions in Eq.\ \eqref{onepmom}.
When we add them up, we arrive at the proper time regularized equivalent
to Eq.\ \eqref{onepmom},
\bmu
G^{\text{PT}}_\Lambda (\bf{p}) = - g^2 \int \frac{d^3 p'}{(2 \pi)^3} 
\int_{1/\Lambda^2}^\infty 
\frac{d \alpha \, d \beta}{2 \pi} \sqrt{\frac{\pi}{\alpha + \beta}}
\left[ \frac{\beta E_{\bf{p}}^A}{\alpha + \beta} \, \gamma_0 - 
\bf{p}' \cdot \bm{\gamma} + m_A \right] \\
\times e^{-\alpha (E_{\bf{p}'}^A)^2 - \beta \omega_{\bf{p} - \bf{p}'}^2
+ \alpha \beta (E_{\bf{p}}^A)^2/(\alpha + \beta)} \:, \label{onepSc}
\emu
where we have used that [see Eq.\ \eqref{erfc}]
\be
\text{erfc} (x) + \text{erfc} (-x) = \frac{2}{\sqrt{\pi}}
\int_{-\infty}^\infty d t \, e^{-t^2} = 2 \:.
\ee

Next, we show that the sum \eqref{onepSc} of the non-covariant expressions
coincides with the Schwinger proper time regularized equivalent of the
covariant form in Eq.\ \eqref{onepmom}. To this end, first perform a Wick
rotation to rewrite the covariant expression in Eq.\ \eqref{onepmom} in 
Euclidean space,
\be
G (\bf{p}) = - g^2 \int \frac{d^4 p^{\prime E}}{(2 \pi)^4} \left.
\frac{p^{\prime E} \cdot \gamma^E + m_A}{\left[ (p^E - p^{\prime E})^2
+ \mu^2 \right] \left[ (p^{\prime E})^2 + m_A^2 \right]}
\right|_{p_0^E \to -i E_{\bf{p}}^A} \:, \label{onepEucov}
\ee
where all the products of Euclidean 4-vectors (including squares) are
understood to be Euclidean scalar products, i.e., taken with the positive
Euclidean metric. The proper time regularization is introduced as in Eq.\
\eqref{Sreg} and yields
\bmu
G^{\text{PT}}_\Lambda (\bf{p}) = - g^2 \int_{1/\Lambda^2}^\infty d \alpha \, 
d \beta \int \frac{d^4 p^{\prime E}}{(2 \pi)^4} \left[ p_0^{\prime E} 
\gamma_0^E + \bf{p}' \cdot \bm{\gamma}^E + m_A \right] \\
\times e^{- \alpha [ (p_0^{\prime E})^2 + \bf{p}^{\prime 2} + m_A^2 ] 
- \beta [ (p_0^E - p_0^{\prime E})^2 + (\bf{p} - \bf{p}')^2 + \mu^2 ]}
\bigg|_{p_0^E \to -i E_{\bf{p}}^A} \:. \label{onepcovS}
\emu
Integrating over $p_0^{\prime E}$ and continuing $p_0^E$ to $-i E_{\bf{p}}^A$
results in Eq.\ \eqref{onepSc}, thus establishing the equivalence of the
non-covariant and covariant proper time regularized expressions.

Note that there is a much faster way to establish this equivalence by
adding up Eq.\ \eqref{proptime1} and its counterpart from Eq.\ 
\eqref{onepWb}, substituting $t^E \to - t^E$ in the latter expression, and 
using the formula
\be
\int_{-\infty}^\infty d t^E \, e^{- \epsilon^E \abs{t^E}} e^{i \omega
t^E} = 2 \pi \delta (\omega) \:. \label{fordelta}
\ee
This procedure directly leads to the form \eqref{onepcovS}, so it indeed
represents a considerable shortcut. However, we were interested in 
writing down the proper time regularized equivalent of Eq.\ 
\eqref{onepmom}, which is why we followed the calculation through to
expressions \eqref{onepSa} and \eqref{onepSb}.

To establish the form \eqref{Gform}, we start again from the covariant 
expression \eqref{onepcovS}, but integrate over all four
components of $p^{\prime E}$ this time with the result
\be
G^{\text{PT}}_\Lambda (\bf{p}) = - \frac{g^2}{(4 \pi)^2}
\int_{1/\Lambda^2}^\infty \frac{d \alpha \, d \beta}{(\alpha + \beta)^2}
\left[ \frac{\beta}{\alpha + \beta} \, p \cdot \gamma + m_A \right]
e^{- \alpha m_A^2 - \beta \mu^2 + \alpha \beta p^2/(\alpha + \beta)}
\bigg|_{p_0 = E_{\bf{p}}^A} \:, \label{GformS}
\ee
from where relation \eqref{Gform} can be established. Note that the
exponent in Eq.\ \eqref{GformS} is manifestly negative for 
$p^2 |_{p_0 = E_{\bf{p}}^A} = m_A^2$.

In order to evaluate the functions $G^{\text{PT}}_{0, \Lambda} (m_A^2)$ and
$G^{\text{PT}}_{1, \Lambda} (m_A^2)$ in proper time regularization explicitly, 
we change variables $(\alpha, \beta)$ to $(\rho, \sigma)$, where
\bga
\rho = \alpha + \beta \:, \quad \sigma = \frac{\beta}{\alpha + \beta} 
\:, \n \\
0 < \sigma < 1 \:, \quad \rho \ge P (\sigma, \Lambda^2) \equiv
\max \left( \frac{1}{\sigma \Lambda^2}, \frac{1}{(1 - \sigma) \Lambda^2}
\right) \:. \label{vartrans}
\ega
The integration over $\rho$ can now be performed and leads to
\be
G^{\text{PT}}_\Lambda (\bf{p}) = -\frac{g^2}{(4 \pi)^2} \int_0^1 d \sigma 
\left[ \sigma \, p \cdot \gamma + m_A \right] \,
E_1 \left( [(1 - \sigma)^2 m_A^2 + \sigma \mu^2] P (\sigma, \Lambda^2)
\right) \:, \label{GformS2}
\ee
where we have introduced the exponential integral function \cite{AS65}
\be
E_1 (x) = \int_x^\infty d t \, \frac{e^{-t}}{t} 
= \int_1^\infty d t \, \frac{e^{- x t}}{t} \:, \quad x > 0 \:.
\ee
Using the expansion of $E_1 (x)$ for small values of the argument
\cite{AS65}, we can approximate $G^{\text{PT}}_\Lambda (\bf{p})$ by
\be
\frac{g^2}{(4 \pi)^2} \int_0^1 d \sigma 
\left[ \sigma \, p \cdot \gamma + m_A \right] \left[ \gamma_E +
\ln \left( [(1 - \sigma)^2 m_A^2 + \sigma \mu^2] P (\sigma, \Lambda^2)
\right) \right] \:, \label{Sapprox}
\ee
all higher terms in the expansion being suppressed by powers of
$1/\Lambda$. For the sake of readability, the demonstration of this latter
assertion is relegated to the end of this section. The results 
\eqref{dimreG2a} and \eqref{dimreG2b} and the integral of 
$\ln ( \Lambda^2 P (\sigma, \Lambda^2))$ which is elementary, can then be 
used to establish that
\bal
G^{\text{PT}}_{0, \Lambda} (m_A^2) &= - \frac{g^2}{(4 \pi)^2} \Bigg\{ 
\ln \frac{\Lambda^2}{m_A^2} - \gamma_E - \ln 2 + 1
- \frac{1}{2} \frac{\mu^2}{m_A^2} \ln \frac{\mu^2}{m_A^2} \n \\
&\phantom{=} {}- 2 \sqrt{\frac{\mu^2}{m_A^2} \left( 1 - \frac{1}{4} 
\frac{\mu^2}{m_A^2} \right)} \, \text{arcctg} \sqrt{\frac{\mu^2/(4 m_A^2)}
{1 - \mu^2/(4 m_A^2)}} \Bigg\} \:, \label{SPTreG2a} \\
G^{\text{PT}}_{1, \Lambda} (m_A^2) &= - \frac{g^2}{(4 \pi)^2} \Bigg\{ 
\frac{1}{2} \left[ \ln \frac{\Lambda^2}{m_A^2} - \gamma_E - \ln 2 \right] + 1
- \frac{1}{2} \frac{\mu^2}{m_A^2} - \left( \frac{\mu^2}{m_A^2} -
\frac{1}{4} \frac{\mu^4}{m_A^4} \right) \ln \frac{\mu^2}{m_A^2} \n \\
&\phantom{=} {}- \left( 2 - \frac{\mu^2}{m_A^2} \right) 
\sqrt{\frac{\mu^2}{m_A^2} \left( 1 - \frac{1}{4} \frac{\mu^2}{m_A^2} 
\right)} \, \text{arcctg} \sqrt{\frac{\mu^2/(4 m_A^2)}{1 - \mu^2/(4 m_A^2)}} 
\Bigg\} \:. \label{SPTreG2b}
\eal
As before, $G^{\text{PT}}_{0, \Lambda} (m_A^2)$ directly gives $\Delta m_A^2$ 
in the purely scalar theory, while in the Yukawa case we have
\bal
\frac{\Delta m_A^2}{2 m_A^2} &= 
- \frac{g^2}{(4 \pi)^2} \Bigg\{ \frac{3}{2} \left[ 
\ln \frac{\Lambda^2}{m_A^2} - \gamma_E - \ln 2 \right]
+ 2 - \frac{1}{2} \frac{\mu^2}{m_A^2} 
- \left( \frac{3}{2} \frac{\mu^2}{m_A^2} -
\frac{1}{4} \frac{\mu^4}{m_A^4} \right) \ln \frac{\mu^2}{m_A^2} \n \\
&\phantom{=} {}- \left( 4 - \frac{\mu^2}{m_A^2} \right) 
\sqrt{\frac{\mu^2}{m_A^2} \left( 1 - \frac{1}{4} \frac{\mu^2}{m_A^2} \right)} 
\, \text{arcctg} \sqrt{\frac{\mu^2/(4 m_A^2)}{1 - \mu^2/(4 m_A^2)}} \Bigg\}
\:. \label{SPTreGmA}
\eal

Finally, we will give the demonstration of Eq.\ \eqref{Sapprox} which was 
relegated to this point. To begin with, let us note that the approximation
\eqref{Sapprox} is not trivial because the function $P (\sigma, \Lambda^2)$
takes large values for $\sigma$ close to the limits of integration, 0 and 1.
We further remark that the extension of the integration domain to 
$0 < \sigma < 1$, $\rho \ge 2/\Lambda^2$ [compare with Eq.\ \eqref{vartrans}],
in which case the approximation corresponding to Eq.\ \eqref{Sapprox} is
straightforward, leads to a \textit{different} result.

This being said, we start from Eq.\ \eqref{GformS2}
and integrate by parts in order to get rid of the exponential integral
function. The result can be written as
\bmu
G^{\text{PT}}_\Lambda (\bf{p}) = - \frac{g^2}{(4 \pi)^2} \int_0^1 d \sigma 
\left[ \frac{\sigma^2}{2} \, p \cdot \gamma + \sigma m_A \right] \,
e^{- [(1 - \sigma)^2 m_A^2 + \sigma \mu^2] P (\sigma, \Lambda^2)} \\
\times \frac{d}{d \sigma} \ln \left( [(1 - \sigma)^2 m_A^2 + \sigma \mu^2] 
P (\sigma, \Lambda^2) \right) \:. \label{GformSibp}
\emu
In the exponent in this expression, we keep
the divergent terms (for $\sigma \to 0$ and $\sigma \to 1$) and expand the
rest in powers of $\sigma$ and $(1 - \sigma)$, respectively, leading to
\bal
&- \frac{g^2}{(4 \pi)^2} \Bigg\{ \int_0^{1/2} d \sigma 
\left[ \frac{\sigma^2}{2} \, p \cdot \gamma + \sigma m_A \right] \,
e^{- m_A^2/(\Lambda^2 \sigma)} \frac{d}{d \sigma} \bigg[ \ln 
\left( \frac{(1 - \sigma)^2 m_A^2 + \sigma \mu^2}{\Lambda^2} \right) - 
\ln \sigma \bigg] \n \\
&{}+ \int_{1/2}^1 d \sigma 
\left[ \frac{\sigma^2}{2} \, p \cdot \gamma + \sigma m_A \right] \,
e^{- \mu^2/(\Lambda^2 (1 - \sigma))} \frac{d}{d \sigma} \bigg[ \ln 
\left( \frac{(1 - \sigma)^2 m_A^2 + \sigma \mu^2}{\Lambda^2} \right) - 
\ln (1 - \sigma) \bigg] \Bigg\} \:, \label{Gformexp}
\eal
the higher terms in the expansion of the exponential of the finite terms
being suppressed by powers of $\Lambda$, as we shall see shortly. First, 
note that we can, by way of the substitutions
\be
y = \frac{1}{2 \sigma} \quad \text{and} \quad y = \frac{1}{2(1 - \sigma)}
\label{Ssubst}
\ee
in the first and second integral in Eq.\ \eqref{Gformexp}, respectively, 
express the contributions that contain the derivatives
of $\ln \sigma$ and $\ln (1 - \sigma)$ in terms of the exponential
integral functions \cite{AS65}
\be
E_n (x) = \int_1^\infty d t \, \frac{e^{- x t}}{t^n} \:, \quad x > 0 \:,
\quad n = 1, 2, 3 \:. \label{expintn}
\ee
For small $x$, we have the expansions \cite{AS65}
\bal
E_1 (x) &= -\gamma_E - \ln x + \cal{O} (x) \:, \n \\
E_2 (x) &= 1 + \cal{O} (x \ln x) \:, \n \\
E_3 (x) &= 1/2 + \cal{O} (x) \:, \label{expintnlim}
\eal
which lead to
\bal
&\phantom{=} - \frac{g^2}{(4 \pi)^2} \Bigg\{ - \int_0^{1/2} d \sigma 
\left[ \frac{\sigma^2}{2} \, p \cdot \gamma + \sigma m_A \right] \,
\frac{e^{- m_A^2/(\Lambda^2 \sigma)}}{\sigma} \n \\
&\hspace{6cm} {}+ \int_{1/2}^1 d \sigma 
\left[ \frac{\sigma^2}{2} \, p \cdot \gamma + \sigma m_A \right] \,
\frac{e^{- \mu^2/(\Lambda^2 (1 - \sigma))}}{1 - \sigma} \Bigg\} \n \\
&= - \frac{g^2}{(4 \pi)^2} \left[ \frac{1}{2} \, p \cdot \gamma + m_A \right]
\left[ \ln \left( \frac{\Lambda^2}{\mu^2} \right) - \gamma_E - \ln 2 - 1
\right] \label{GformSregdep}
\eal
in the limit of large $\Lambda$. In an analogous way, we can see that
the higher orders in the expansion of the exponential of the finite terms
(for $\sigma \to 0$ and $\sigma \to 1$) in Eq.\ \eqref{GformSibp} are 
suppressed by powers of $\Lambda$, taking into account that the
expressions resulting from this expansion and the first terms in the
integrals \eqref{Gformexp} are continuous bounded functions over the
intervals in question, and that furthermore the higher orders in the
expansion carry inverse powers of $\Lambda$.

As far as the first terms in the integrals \eqref{Gformexp} are concerned,
the limit $\Lambda \to \infty$ can be taken naively there (note that the
$\Lambda$-dependence in the argument of the logarithm is spurious since
its derivative with respect to $\sigma$ gives zero), as we will show in
a moment. If we suppose, for the time being, that this is indeed correct,
the exponentials can be replaced by one, and an integration by parts yields
\bal
&\phantom{=} - \frac{g^2}{(4 \pi)^2} \Bigg\{ \int_0^{1/2} d \sigma 
\left[ \frac{\sigma^2}{2} \, p \cdot \gamma + \sigma m_A \right] \,
e^{- m_A^2/(\Lambda^2 \sigma)} \frac{d}{d \sigma} \ln 
\left( \frac{(1 - \sigma)^2 m_A^2 + \sigma \mu^2}{\Lambda^2} \right) \n \\
&\hspace{2cm} {}+ \int_{1/2}^1 d \sigma 
\left[ \frac{\sigma^2}{2} \, p \cdot \gamma + \sigma m_A \right] \,
e^{- \mu^2/(\Lambda^2 (1 - \sigma))} \frac{d}{d \sigma} \ln \left( 
\frac{(1 - \sigma)^2 m_A^2 + \sigma \mu^2}{\Lambda^2} \right) \Bigg\} \n \\
&= \frac{g^2}{(4 \pi)^2} \Bigg\{ \left[ \frac{1}{2} \, p \cdot \gamma 
+ m_A \right] \ln \left( \frac{\Lambda^2}{\mu^2} \right)
+ \int_0^1 d \sigma \left[ \sigma \, p \cdot \gamma + m_A \right] \ln
\left( \frac{(1 - \sigma)^2 m_A^2 + \sigma \mu^2}{\Lambda^2} \right)
\Bigg\} \label{GformSlog}
\eal
for $\Lambda \to \infty$ (the $\Lambda$-dependence cancels between the
two terms). Using Eq.\ \eqref{GformSregdep} and the results \eqref{dimreG2a} 
and \eqref{dimreG2b} for the remaining integration, we finally arrive at
the expressions \eqref{SPTreG2a} and \eqref{SPTreG2b}, thus confirming
the approximation \eqref{Sapprox}.

Let us now show in detail that it is, indeed, correct to replace the
exponential on the left-hand side of Eq.\ \eqref{GformSlog} by one. To
this end, take the integrals on the left-hand side of Eq.\ 
\eqref{GformSlog} as they stand, perform the substitutions Eq.\ 
\eqref{Ssubst} and decompose the result in partial fractions to find,
after a somewhat lengthy calculation,
\bmu
- \frac{g^2}{(4 \pi)^2} \int_1^\infty d y \left\{ \frac{1}{2} \, 
p \cdot \gamma
\left[ \frac{1}{2 y^3} - \frac{M_+ + M_-}{2 y^2} + \frac{M_+^2 + M_-^2}{y}
- \frac{M_-^2}{y + M_+/2} - \frac{M_+^2}{y + M_-/2} \right] \right. \\
+ \left. m_A \left[ \frac{1}{y^2} - \frac{M_+ + M_-}{y} + \frac{M_-}{y + M_+/2}
+ \frac{M_+}{y + M_-/2} \right] \right\} e^{- 2 (m_A^2/\Lambda^2) y} \\
- \frac{g^2}{(4 \pi)^2} \int_1^\infty d y \left\{ \frac{1}{2} \, 
p \cdot \gamma \left[ -\frac{1}{2 y^3} - \frac{M_+ + M_- - 2}{2 y^2} \right. 
\right. \hspace{6cm} \\
\hspace{4cm} - \left. \frac{M_+^2 + M_-^2}{y} + 
\frac{M_+^2}{y - 1/(2 + 2 M_+)} + \frac{M_-^2}{y - 1/(2 + 2 M_-)} \right] \\
+ \left. m_A \left[ \frac{1}{y^2} + \frac{M_+ + M_-}{y} 
- \frac{M_+}{y - 1/(2 + 2 M_+)} - \frac{M_-}{y - 1/(2 + 2 M_-)} \right] 
\right\} e^{- 2 (\mu^2/\Lambda^2) y} \:, \label{GformSparfr}
\emu
where we have introduced the notations
\be
M_\pm = \frac{\mu^2}{2 m_A^2} - 1 \pm \sqrt{\frac{\mu^2}{2 m_A^2}
\left( \frac{\mu^2}{2 m_A^2} - 2 \right)} \:.
\ee
For simplicity, we have written this expression for the case 
$\mu^2 \ge 4 m_A^2$ where the square roots are real. To compare with Eqs.\
\eqref{SPTreG2a} and \eqref{SPTreG2b} in the end, one has to analytically 
continue the result to smaller values of $\mu^2$.

The integrals in Eq.\ \eqref{GformSparfr} are well-defined with and without
the exponentials. However, this is not true (in all cases) for the integrals 
over the partial fractions individually. It is clear then, that the
exponentials can be considered as regulating factors for the integrals over
the partial fractions. After summing up the individual results, the limit
$\Lambda \to \infty$ can safely be taken, thus removing the regulator. In
the end, this is equivalent to replacing the exponentials by one in Eq.\
\eqref{GformSlog}. The integrals in \eqref{GformSparfr} can also be explicitly 
calculated with the help of Eqs.\ \eqref{expintn} and \eqref{expintnlim} in 
the limit $\Lambda \to \infty$. As a result, Eq.\ \eqref{GformSlog} is 
recovered [using Eqs.\ \eqref{dimreG2a} and \eqref{dimreG2b} to evaluate
the integral in Eq.\ \eqref{GformSlog}].

These comments conclude the demonstration of the correctness of the
approximation \eqref{Sapprox} and hence of the results \eqref{SPTreG2a},
\eqref{SPTreG2b} and \eqref{SPTreGmA} for the Schwinger proper time 
regularization. It is clear from the above that the proper time regularization 
will not be the method of choice in a Hamiltonian (not explicitly covariant)
approach, considering the difficulties in establishing the regularized form
of the non-covariant expressions and the evaluation of the corresponding
(one-loop) integral as compared to the other regularizations discussed
before.

\subsection{Momentum cutoff regularization}

At last, we discuss the probably simplest regularization scheme available,
the use of a momentum cutoff. As it turns out, it does not have very
simple properties when used in the present context. We start with the
non-covariant expressions written in Euclidean space after a Wick rotation
in Eqs.\ \eqref{onepWa} and \eqref{onepWb}. The momentum cutoff is
implemented by restricting the integration domain (in Euclidean space) to
$(p^{\prime E})^2 \le \Lambda^2$, hence the complete regularized expression is
\bmu
G^{\text{MC}}_\Lambda (\bf{p}) = 
- g^2 \int^\Lambda \frac{d^4 p^{\prime E}}{(2 \pi)^4} 
\, \frac{p_0^{\prime E} \gamma_0^E + \bf{p}' \cdot \bm{\gamma}^E + m_A}
{(p_0^{\prime E})^2 + \bf{p}'{}^2 + m_A^2} \int \frac{d k_0^E}{2 \pi} 
\, \frac{1}{(k_0^E)^2 + (\bf{p} - \bf{p}')^2 + \mu^2} \\
\times \left[ \int_{-\infty}^0 d t^E \, e^{\epsilon^E t^E} 
e^{i (p_0^{\prime E} + k_0^E - p_0^E) t^E} + \int_{-\infty}^0 d t^E \, 
e^{\epsilon^E t^E} e^{-i (p_0^{\prime E} + k_0^E - p_0^E) t^E}
\right]_{p_0^E \to -i E_{\bf{p}}^A} \:. \label{onepMC}
\emu
We can perform the integrations over $k_0^E$ (with the
help of the residue theorem) and $t^E$ in Eq.\ \eqref{onepMC}, separately
for the two contributions, to obtain the result
\bal
G^{\text{MC}}_\Lambda (\bf{p}) &= 
\left. g^2 \int^\Lambda \frac{d^4 p^{\prime E}}{(2 \pi)^4} 
\, \frac{i}{2 \omega_{\bf{p} -\bf{p}'}} 
\frac{p_0^{\prime E} \gamma_0^E + \bf{p}' \cdot \bm{\gamma}^E + m_A}
{[p_0^{\prime E} - p_0^E - i \omega_{\bf{p} -\bf{p}'}] 
[(p_0^{\prime E})^2 + (E_{\bf{p}'}^A)^2 ]} 
\right|_{p_0^E \to -i E_{\bf{p}}^A} \n \\
&\phantom{=} {}- \left. g^2 \int^\Lambda \frac{d^4 p^{\prime E}}{(2 \pi)^4} 
\, \frac{i}{2 \omega_{\bf{p} -\bf{p}'}} 
\frac{p_0^{\prime E} \gamma_0^E + \bf{p}' \cdot \bm{\gamma}^E + m_A}
{[p_0^{\prime E} - p_0^E + i \omega_{\bf{p} -\bf{p}'}] 
[(p_0^{\prime E})^2 + (E_{\bf{p}'}^A)^2 ]} 
\right|_{p_0^E \to -i E_{\bf{p}}^A} \:. \label{onepMC2}
\eal

Adding up the integrands in Eq.\ \eqref{onepMC2} gives
\be
G^{\text{MC}}_\Lambda (\bf{p}) = \left. - g^2 \int^\Lambda 
\frac{d^4 p^{\prime E}}{(2 \pi)^4} 
\, \frac{p_0^{\prime E} \gamma_0^E + \bf{p}' \cdot \bm{\gamma}^E + m_A}
{[(p_0^E - p_0^{\prime E})^2 + (\bf{p} - \bf{p}')^2 + \mu^2] 
[(p_0^{\prime E})^2 + \bf{p}'{}^2 + m_A^2]} 
\right|_{p_0^E \to -i E_{\bf{p}}^A} \:, \label{onepMCov}
\ee
which obviously coincides with the cutoff regularization
of the Euclidean version \eqref{onepEucov} of the covariant form in Eq.\ 
\eqref{onepmom}, thus establishing the equivalence of the non-covariant
and covariant expressions in Eq.\ \eqref{onepmom} in the momentum cutoff
regularized form. One can use rotations in four-dimensional (Euclidean) 
space, taking into account the form of the integrand in Eq.\ \eqref{onepMCov}
as well as the invariance of the integration measure and the integration 
domain, to show that $G^{\text{MC}}_\Lambda (\bf{p})$ is in fact of the 
form of Eq.\ \eqref{Gform}. 

In order to obtain the three-dimensional form of the non-covariant
expression in cutoff regularization, we start with Eq.\
\eqref{onepMC2} and perform the $p_0^{\prime E}$-integration in
\be
\int^\Lambda \frac{d^4 p^{\prime E}}{(2 \pi)^4} = \int^\Lambda 
\frac{d^3 p'}{(2 \pi)^3} \, \int_{-\sqrt{\Lambda^2 - \bf{p}^{\prime 2}}}%
^{\sqrt{\Lambda^2 - \bf{p}^{\prime 2}}} \frac{d p_0^{\prime E}}{2 \pi}
\label{MCintegral}
\ee
(with the $d^3 p'$-integration restricted to $\bf{p}^{\prime 2}
\le \Lambda^2$). To this end, it is easiest to decompose the integrand 
in partial fractions with respect to $p_0^{\prime E}$, for the first
non-covariant contribution
\bmu
\frac{i}{2 \omega_{\bf{p} -\bf{p}'}} 
\frac{p_0^{\prime E} \gamma_0^E + \bf{p}' \cdot \bm{\gamma}^E + m_A}
{[p_0^{\prime E} - p_0^E - i \omega_{\bf{p} -\bf{p}'}] 
[(p_0^{\prime E})^2 + (E_{\bf{p}'}^A)^2 ]} =
\frac{i}{2 E_{\bf{p}'}^A \, 2 \omega_{\bf{p} -\bf{p}'}} \\
\times \left[ \frac{1}{E_{\bf{p}'}^A - \omega_{\bf{p} -\bf{p}'} + i p_0^E} 
\left( \frac{(p_0^E + i \omega_{\bf{p} -\bf{p}'}) \gamma_0^E 
+ \bf{p}' \cdot \bm{\gamma}^E + m_A}
{p_0^{\prime E} - p_0^E - i \omega_{\bf{p} -\bf{p}'}}
- \frac{i E_{\bf{p}'}^A \gamma_0^E + \bf{p}' \cdot \bm{\gamma}^E 
+ m_A}{p_0^{\prime E} - i E_{\bf{p}'}^A} \right) \right. \\
\left. + \frac{1}{E_{\bf{p}'}^A + \omega_{\bf{p} -\bf{p}'} - i p_0^E}
\left( \frac{(p_0^E + i \omega_{\bf{p} -\bf{p}'}) \gamma_0^E 
+ \bf{p}' \cdot \bm{\gamma}^E + m_A}
{p_0^{\prime E} - p_0^E - i \omega_{\bf{p} -\bf{p}'}}
- \frac{- i E_{\bf{p}'}^A \gamma_0^E + \bf{p}' \cdot \bm{\gamma}^E 
+ m_A}{p_0^{\prime E} + i E_{\bf{p}'}^A} \right) \right] \:.
\label{MCintegrand}
\emu
The decomposition in partial fractions for the second contribution in 
Eq.\ \eqref{onepMC2} can be obtained immediately from
the one above by realizing that the two contributions are complex conjugate 
to each other (if we consider $p_0^E$ and the Euclidean $\gamma$-matrices as
real, for example by using the Majorana representation). 

The $p_0^{\prime E}$-integration is then straightforward, although the result
is rather lengthy, indicating that the momentum cutoff would not usually
be the regularization of choice in the present not manifestly covariant
context, either. However, if we extended the $p_0^{\prime E}$-integration
to the whole real axis while sticking somewhat arbitrarily to 
$\bf{p}^{\prime 2} \le \Lambda^2$, we would obtain the following simple
result for the sum of the two contributions:
\bal
G^{\text{NC}}_\Lambda (\bf{p}) &= \left. - g^2 \int^\Lambda \frac{d^3 p'}
{(2 \pi)^3} \frac{1}{2 E_{\bf{p}'}^A \, 2 \omega_{\bf{p} - \bf{p}'}}
\frac{-i E_{\bf{p}'}^A \gamma_0^E + \bf{p}' \cdot \bm{\gamma}^E + m_A}
{E_{\bf{p}'}^A + \omega_{\bf{p} - \bf{p}'} - i p_0^E} 
\right|_{p_0^E \to -i E_{\bf{p}}^A} \n \\ 
&\phantom{=} \left. {}- g^2 \int^\Lambda \frac{d^3 p'}{(2 \pi)^3} 
\frac{1}{2 E_{\bf{p}'}^A \, 2 \omega_{\bf{p} - \bf{p}'}}
\frac{i E_{\bf{p}'}^A \gamma_0^E + \bf{p}' \cdot \bm{\gamma}^E + m_A}
{E_{\bf{p}'}^A + \omega_{\bf{p} - \bf{p}'} + i p_0^E} 
\right|_{p_0^E \to -i E_{\bf{p}}^A} \label{nococut} \:.
\eal
Eq.\ \eqref{nococut} results directly from the non-covariant expression in
Eq.\ \eqref{onepmom} by restricting the 3-momentum integration to 
$\bf{p}^{\prime 2} \le \Lambda^2$. It is hence, in a not manifestly
covariant approach, a very natural regularization scheme. Although it is 
apparent that  four-dimensional Euclidean rotational invariance is broken 
in Eq.\ \eqref{nococut} and that hence $G^{\text{NC}}_\Lambda (\bf{p})$ will 
\emph{not} be of the form \eqref{Gform} for any finite
value of $\Lambda$, it is not clear a priori whether the form \eqref{Gform} 
could not be recovered in the limit $\Lambda \to \infty$. In the special
case of a purely scalar theory, it was shown in Ref.\ \cite{KG00} that
covariance is indeed reestablished for $\Lambda \to \infty$ in the sense that
$G^{\text{NC}}_\Lambda (\bf{p})$ becomes independent of $\bf{p}$ in this
limit (and only depends on the square of the four-vector, $\left. p^2
\right|_{p_0 = E_{\bf{p}}^A} = m_A^2$).

In the following, we will consider the integration \eqref{MCintegral} of
the integrand \eqref{MCintegrand} and its complex conjugate counterpart
arising from the second non-covariant contribution in the limit of large
$\Lambda$ and compare it with $G^{\text{NC}}_\Lambda (\bf{p})$. To make
this comparison mathematically precise, we divide the integration over the
spatial momentum $\bf{p}'$ into the two regions $\bf{p}^{\prime 2} < K^2$ 
and $K^2 < \bf{p}^{\prime 2} < \Lambda^2$ with an intermediate scale $K$ 
which fulfills $\bf{p}^2, m_A^2, \mu^2 \ll K^2 \ll \Lambda^2$. Then, in
the limit $\Lambda \to \infty$, we can approximate the integral over the
first region,
\be
\int^K \frac{d^3 p'}{(2 \pi)^3} \, 
\int_{-\sqrt{\Lambda^2 - \bf{p}^{\prime 2}}}%
^{\sqrt{\Lambda^2 - \bf{p}^{\prime 2}}} \frac{d p_0^{\prime E}}{2 \pi}
\to \int^K \frac{d^3 p'}{(2 \pi)^3} \, \int_{-\infty}^\infty
\frac{d p_0^{\prime E}}{2 \pi} \:,
\ee
hence the integral over $\bf{p}^{\prime 2} < K^2$ of Eq.\ \eqref{MCintegrand}
and its complex conjugate, taking special care of the analytic continuation 
of the logarithms resulting from the $p_0^{\prime E}$-integration,
tend to
\bal
&- g^2 \left. \int^K \frac{d^3 p'}
{(2 \pi)^3} \frac{1}{2 E_{\bf{p}'}^A \, 2 \omega_{\bf{p} - \bf{p}'}}
\frac{-i E_{\bf{p}'}^A \gamma_0^E + \bf{p}' \cdot \bm{\gamma}^E + m_A}
{E_{\bf{p}'}^A + \omega_{\bf{p} - \bf{p}'} - i p_0^E} 
\right|_{p_0^E \to -i E_{\bf{p}}^A} \n \\ 
&{}- g^2 \left. \int^K \frac{d^3 p'}{(2 \pi)^3} 
\frac{1}{2 E_{\bf{p}'}^A \, 2 \omega_{\bf{p} - \bf{p}'}}
\frac{i E_{\bf{p}'}^A \gamma_0^E + \bf{p}' \cdot \bm{\gamma}^E + m_A}
{E_{\bf{p}'}^A + \omega_{\bf{p} - \bf{p}'} + i p_0^E} 
\right|_{p_0^E \to -i E_{\bf{p}}^A} \:. \label{MCfirstint}
\eal
The first corrections to this result, of relative order $(K/m_A) (K/\Lambda)$
[which can be suppressed in the limit of large $\Lambda$ through a suitable 
choice of $K$, e.g., $K = (m_A^3 \Lambda)^{1/4}$], cancel among the two 
contributions. Incidentally, the second corrections, of order $(K/m_A) 
(K/\Lambda)^2$, also vanish.

Now, for the other integration region $K^2 < \bf{p}^{\prime 2} < \Lambda^2$,
we have $\bf{p}^2, m_A^2, \mu^2 \ll \bf{p}^{\prime 2}, \Lambda^2$.
A lengthy calculation leads to the following result: the leading terms in a 
systematic expansion vanish for both contributions, the subleading terms 
cancel among the two contributions for the $\gamma_0^E$-coefficients or as a 
result of the $(\bf{p}' \to - \bf{p}')$-symmetry of integrand, integration 
measure and domain for the $\bm{\gamma}^E$-coefficient, and it is hence the 
subsubleading terms that give the dominant contributions which turn out to
be
\bmu
- g^2 \int_K^\Lambda \frac{d^3 p'}{(2 \pi)^3} \,
\frac{p_0^E \gamma_0^E + \bf{p} \cdot \bm{\gamma}^E + 2 m_A}
{4 \pi \bf{p}^{\prime 2}}
\left[ \frac{\pi}{2 \abs{\bf{p}'}} - \frac{\arcsin (\abs{\bf{p}'}/
\Lambda)}{\abs{\bf{p}'}} + \frac{\sqrt{1 - \bf{p}^{\prime 2}/\Lambda^2}}
{\Lambda} \right] \\
+ g^2 \int_K^\Lambda \frac{d^3 p'}{(2 \pi)^3} \,
\frac{3 p_0^E \gamma_0^E - \bf{p} \cdot \bm{\gamma}^E}{6 \pi \Lambda^2} 
\, \frac{\sqrt{1 - \bf{p}^{\prime 2}/\Lambda^2}}{\Lambda} \:.
\label{intexpand}
\emu

The first term in square brackets in Eq.\ \eqref{intexpand} coincides with
the dominant contribution from the part of $G^{\text{NC}}_\Lambda (\bf{p})$
that stems from the second integration region $K^2 < \bf{p}^{\prime 2} 
< \Lambda^2$ [cf.\ Eq.\ \eqref{nococut}], and hence, when added to Eq.\
\eqref{MCfirstint}, combines to $G^{\text{NC}}_\Lambda (\bf{p})$. The
other terms in Eq.\ \eqref{intexpand} are readily integrated \cite{GR00}
in the limit $K/\Lambda \to 0$ through the substitution 
$y = \abs{\bf{p}'}/\Lambda$. The final result is 
\bmu
G^{\text{NC}}_\Lambda (\bf{p}) = \left. - g^2 \int^\Lambda 
\frac{d^4 p^{\prime E}}{(2 \pi)^4} 
\, \frac{p_0^{\prime E} \gamma_0^E + \bf{p}' \cdot \bm{\gamma}^E + m_A}
{[(p_0^E - p_0^{\prime E})^2 + (\bf{p} - \bf{p}')^2 + \mu^2] 
[(p_0^{\prime E})^2 + \bf{p}'{}^2 + m_A^2]} 
\right|_{p_0^E \to -i E_{\bf{p}}^A} \\
+ \frac{g^2}{(4 \pi)^2} \left( \frac{1}{2} - \ln 2 \right) 
\left( E_{\bf{p}}^A \gamma_0 - \bf{p} \cdot \bm{\gamma} + 2 m_A \right) -
\frac{g^2}{(4 \pi)^2} \, \frac{1}{12} \left( 3 E_{\bf{p}}^A \gamma_0 + 
\bf{p} \cdot \bm{\gamma} \right) \label{NCresult}
\emu
in the limit $\Lambda \to \infty$.

The result \eqref{NCresult} has two important consequences: first,
together with the expressions \eqref{MC0expl} and \eqref{MC1expl} for 
$G^{\text{MC}}_\Lambda (\bf{p})$ below, it provides an explicit expression 
for $G^{\text{NC}}_\Lambda (\bf{p})$ in the limit of large $\Lambda$.
Second, and more importantly, it shows that Lorentz invariance \emph{is}
broken, even in the limit $\Lambda \to \infty$, through the last term in
Eq.\ \eqref{NCresult}. For example, it can explicitly be shown that this
term leads to a non-covariant contribution to the energies of the 
one-particle states by following the procedure detailed in Eqs.\
\eqref{onepform2}--\eqref{onepformren}. It is then clear that the
non-covariant cutoff regularization is not suitable for a fermionic 
theory. On the other hand, in a purely scalar theory, given by the 
$m_A$-coefficient in Eq.\ \eqref{NCresult}, Lorentz invariance is recovered 
in the limit $\Lambda \to \infty$, in agreement with
the result in Ref.\ \cite{KG00} which was obtained by a different method.

Let us now return to the expression for $G^{\text{MC}}_\Lambda (\bf{p})$
given in Eq.\ \eqref{onepMCov}. In order to calculate $G_0 (m_A^2)$ and
$G_1 (m_A^2)$ in cutoff regularization, one could use four-dimensional
spherical coordinates for the $d^4 p^{\prime E}$-integration (see also
below), which would also explicitly confirm that $G^{\text{MC}}_\Lambda 
(\bf{p})$ is of the form of Eq.\ \eqref{Gform} for any finite value of
$\Lambda$. However, as long as one is only interested in the result for 
large enough $\Lambda$, it is much quicker in the present situation to 
introduce Feynman parameters and proceed in analogy with Eq.\ \eqref{Gform2}
[in $(D=4)$-dimensional Euclidean space and for the integration domain
$(p^{\prime E})^2 \le \Lambda^2$]. We thus arrive at
\be
G^{\text{MC}}_\Lambda (\bf{p}) = \left. - g^2 \int_0^1 dx 
\int^{\Lambda'} \frac{d^4 q^E}{(2 \pi)^4} \frac{(q^E + x p^E) \cdot
\gamma^E + m_A}{\left[ (q^E)^2 + x (1 - x) (p^E)^2 + x \mu^2 + (1 - x)
m_A^2 \right]^2} \right|_{p_0^E \to -i E_{\bf{p}}^A} \label{onepMCov2}
\ee
with $q^E$ to be integrated over the four-dimensional Euclidean domain
$(q^E + x p^E)^2 \le \Lambda^2$ (as a result of the shift of the
integration variable to $q^E = p^{\prime E} - x p^E$). To evaluate this
expression it is easiest to calculate the difference
\be
\Delta G^{\text{MC}}_{\Lambda, \Lambda'} (\bf{p}) = 
G^{\text{MC}}_\Lambda (\bf{p}) - G^{\text{MC}}_{\Lambda'} (\bf{p})
\label{Gdifdef}
\ee
between $G^{\text{MC}}_\Lambda (\bf{p})$ and
\be
G^{\text{MC}}_{\Lambda'} (\bf{p}) = \left. - g^2 \int_0^1 dx 
\int^\Lambda \frac{d^4 q^E}{(2 \pi)^4} \frac{x p^E \cdot
\gamma^E + m_A}{\left[ (q^E)^2 + x (1 - x) (p^E)^2 + x \mu^2 + (1 - x)
m_A^2 \right]^2} \right|_{p_0^E \to -i E_{\bf{p}}^A} \label{onepMCov3}
\ee
with the integration domain restricted to $(q^E)^2 \le \Lambda^2$
[observe that in this case the first term in the numerator in Eq.\ 
\eqref{onepMCov2} does not contribute because of the symmetry of the 
integration domain]. The $d^4 q^E$-integral in $G^{\text{MC}}_{\Lambda'} 
(\bf{p})$ can be evaluated by standard methods. 

Incidentally, it would in principle be possible to define cutoff 
regularization via Eq.\ \eqref{onepMCov3} [instead of Eq.\ \eqref{onepMCov2}], 
a choice that would obviously fulfill Eq.\ \eqref{Gform}, too. The 
non-covariant expression corresponding to Eq.\ \eqref{onepMC} in this 
alternative regularization is
\bmu
G^{\text{MC}}_{\Lambda'} (\bf{p}) = \\
- g^2 \int_0^1 d x \int^{\Lambda'} \frac{d^4 p^{\prime E}}{(2 \pi)^4} 
\int \frac{d k_0^E}{2 \pi} \, \frac{p_0^{\prime E} \gamma_0^E + 
\bf{p}' \cdot \bm{\gamma}^E + m_A}
{\left[ \left( k_0^E \right)^2 x + \omega_{\bf{p} - \bf{p}'}^2 \, x + 
\left( p_0^{\prime E} \right)^2 (1 - x) + 
\left( E_{\bf{p}'}^A \right)^2 (1 - x) \right]^2} \\
\times \left[ \int_{-\infty}^0 d t^E \, e^{\epsilon^E t^E} 
e^{i (p_0^{\prime E} + k_0^E - p_0^E) t^E} + \int_{-\infty}^0 d t^E \, 
e^{\epsilon^E t^E} e^{-i (p_0^{\prime E} + k_0^E - p_0^E) t^E}
\right]_{p_0^E \to -i E_{\bf{p}}^A} \:, \label{onepMCvar}
\emu
where the $d^4 p^{\prime E}$-integration is now over $(p^{\prime E} - 
x p^E)^2 \le \Lambda^2$. When comparing this expression with Eq.\ 
\eqref{onepMC}, it becomes clear that this regularization is not too natural 
in the present non-covariant context. Still, we can integrate over $k_0^E$
(again with the help of the residue theorem) and $t_E$ in Eq.\ 
\eqref{onepMCvar}. The result for the first contribution is
\bmu
- g^2 \int_0^1 d x \int^{\Lambda'} \frac{d^4 p^{\prime E}}{(2 \pi)^4} 
\, \frac{p_0^{\prime E} \gamma_0^E + \bf{p}' \cdot \bm{\gamma}^E + m_A}
{\left[ \left( p_0^E - p_0^{\prime E} \right)^2 x + 
\omega_{\bf{p} - \bf{p}'}^2 \, x + \left( p_0^{\prime E} \right)^2 (1 - x) + 
\left( E_{\bf{p}'}^A \right)^2 (1 - x) \right]^2} \\
\times \left\{ \frac{1}{2} + i \, \frac{(p_0^E - p_0^{\prime E}) 
\sqrt{x} \left[ \left( p_0^E - p_0^{\prime E} \right)^2 x + 
3 \, \omega_{\bf{p} - \bf{p}'}^2 \, x + 
3 \left( p_0^{\prime E} \right)^2 (1 - x) 
+ 3 \left( E_{\bf{p}'}^A \right)^2 (1 - x) \right]}
{4 \left[ \omega_{\bf{p} - \bf{p}'}^2 \, x + 
\left( p_0^{\prime E} \right)^2 (1 - x) + 
\left( E_{\bf{p}'}^A \right)^2 (1 - x) \right]^{3/2}} \right\} \:,
\label{onepMCvar2}
\emu
to be analytically continued to $p_0^E \to -i E_{\bf{p}}^A$.
The result for the second contribution can again be obtained by
complex conjugation from the above (considering $p_0^E$ and the Euclidean
$\gamma$-matrices as real). The integrations over both $x$ and 
$p_0^{\prime E}$ in Eq.\ \eqref{onepMCvar2} look forbidding. However, the
sum of the two non-covariant contributions is readily seen to give 
$G^{\text{MC}}_{\Lambda'} (\bf{p})$ as defined in Eq.\ \eqref{onepMCov3},
after shifting the four-momentum integration variable to $q^E = p^{\prime E}
- x p^E$.

Returning to the explicit calculation of $G^{\text{MC}}_{\Lambda} (\bf{p})$,
we note that the difference $\Delta G^{\text{MC}}_{\Lambda, \Lambda'} 
(\bf{p})$ is an integral over the difference of the four-balls 
$(q^E + x p^E)^2 \le \Lambda^2$ and $(q^E)^2 \le \Lambda^2$. We introduce 
four-dimensional (Euclidean) spherical coordinates with the fourth axis 
oriented in direction of $p^E$ and the corresponding polar angle denoted 
as $\chi$. Then for the four-ball $(q^E + x p^E)^2 \le \Lambda^2$ the 
integration along the radius runs up to $R (\chi)$,
\be
R (\chi) = \Lambda - x \abs{p^E} \cos \chi + \cal{O} (1/\Lambda)
\ee
in the limit of large $\Lambda$. Furthermore, the denominator of the 
integrand in the region between the two four-balls can be approximated 
by $\Lambda^4$ in this limit. We then have
\be
\Delta G^{\text{MC}}_{\Lambda, \Lambda'} (\bf{p}) = -g^2 \int_0^1 d x \,
4 \pi \int_0^\pi d \chi \, \sin^2 \chi \int_\Lambda^{R (\chi)} dr \, r^3
\frac{r \cos \chi \, \gamma^E_4 + x (p^E \cdot \gamma^E) + m_A}
{(2 \pi)^4 \Lambda^4} \:, \label{fourbdif}
\ee
while the other components of $q^E$ in the numerator do not contribute for
symmetry reasons (integration over the other angular coordinates). The
integrations in Eq.\ \eqref{fourbdif} yield, in the
limit of large $\Lambda$,
\be
\Delta G^{\text{MC}}_{\Lambda, \Lambda'} (\bf{p}) = \frac{g^2}{(4 \pi)^2}
\frac{\abs{p^E} \gamma^E_4}{4} = \frac{g^2}{(4 \pi)^2}
\frac{p \cdot \gamma}{4} \:, \label{Gdifres}
\ee
where we have used $\abs{p^E} \gamma^E_4 = p^E \cdot \gamma^E$ and replaced 
$p_0^E \to -i E_{\bf{p}}^A \equiv -i p^0$. In particular, for a purely
scalar theory, the difference $\Delta G^{\text{MC}}_{\Lambda, \Lambda'} 
(\bf{p})$ tends to zero for $\Lambda \to \infty$.

Now from Eqs.\ \eqref{Gdifdef} and \eqref{onepMCov3}, the above result
\eqref{Gdifres} and the result of the integration over the Feynman parameter
$x$ in Eqs.\ \eqref{dimreG2a} and \eqref{dimreG2b}, we finally get the
explicit expressions
\bal
G^{\text{MC}}_{0, \Lambda} (m_A^2) &= - \frac{g^2}{(4 \pi)^2} \Bigg\{ 
\ln \frac{\Lambda^2}{m_A^2} + 1 - \frac{1}{2} 
\frac{\mu^2}{m_A^2} \ln \frac{\mu^2}{m_A^2} \n \\
&\phantom{=} {}- 2
\sqrt{\frac{\mu^2}{m_A^2} \left( 1 - \frac{1}{4} \frac{\mu^2}{m_A^2} \right)} 
\, \text{arcctg} \sqrt{\frac{\mu^2/(4 m_A^2)}{1 - \mu^2/(4 m_A^2)}} \Bigg\}
\:, \label{MC0expl} \\
G^{\text{MC}}_{1, \Lambda} (m_A^2) &= - \frac{g^2}{(4 \pi)^2} \Bigg\{ 
\frac{1}{2} \ln \frac{\Lambda^2}{m_A^2} + \frac{3}{4} - \frac{1}{2} 
\frac{\mu^2}{m_A^2} - \left( \frac{\mu^2}{m_A^2} -
\frac{1}{4} \frac{\mu^4}{m_A^4} \right) \ln \frac{\mu^2}{m_A^2} \n \\
&\phantom{=} {}- \left( 2 - \frac{\mu^2}{m_A^2} \right) 
\sqrt{\frac{\mu^2}{m_A^2} \left( 1 - \frac{1}{4} \frac{\mu^2}{m_A^2} \right)} 
\, \text{arcctg} \sqrt{\frac{\mu^2/(4 m_A^2)}{1 - \mu^2/(4 m_A^2)}} \Bigg\}
\label{MC1expl}
\eal
($\Lambda^2 \gg m_A^2$). Again, the result for 
$G^{\text{MC}}_{0, \Lambda} (m_A^2)$ gives directly the 
expression for $\Delta m_A^2$ in momentum cutoff regularization for the purely 
scalar case, while for Yukawa theory we have
\bal
\frac{\Delta m_A^2}{2 m_A^2} &= - \frac{g^2}{(4 \pi)^2} \Bigg\{ \frac{3}{2} 
\ln \frac{\Lambda^2}{m_A^2} + \frac{7}{4} - \frac{1}{2} 
\frac{\mu^2}{m_A^2} - \left( \frac{3}{2} \frac{\mu^2}{m_A^2} -
\frac{1}{4} \frac{\mu^4}{m_A^4} \right) \ln \frac{\mu^2}{m_A^2} \\
&\phantom{=} {}- \left( 4 - \frac{\mu^2}{m_A^2} \right) 
\sqrt{\frac{\mu^2}{m_A^2} \left( 1 - \frac{1}{4} \frac{\mu^2}{m_A^2} \right)} 
\, \text{arcctg} \sqrt{\frac{\mu^2/(4 m_A^2)}{1 - \mu^2/(4 m_A^2)}} \Bigg\} 
\:. 
\eal

\section{Separation of angular variables and spin \label{appang}}

In terms of the well--known eigenstates $\chi_{S, m_S}$ of total spin 
$\bf{S} = \bf{s}_A + \bf{s}_B$ ($S = 0$ or $S = 1$), one has the following 
explicit expressions for the eigenstates of ${\bf J}^2$ and $J_z$ with 
eigenvalues $J (J + 1)$ and $M$:
\bal
\supfi{1} \cal{Y}_{J M}^J (\hat{\bf{p}}) &= Y_{J M} (\hat{\bf{p}}) \, 
\chi_{00} \:, \n \\[1mm]
\supfi{3} \cal{Y}_{J-1, M}^J (\hat{\bf{p}}) &= \frac{1}{\sqrt{2 J
(2J - 1)}} \left[ \sqrt{(J - M - 1)(J - M)} \, Y_{J - 1, M + 1} 
(\hat{\bf{p}}) \, \chi_{1, -1} \right. \n \\
&\phantom{=} {}+ \sqrt{2 (J - M)(J + M)} \, Y_{J - 1, M} (\hat{\bf{p}}) \,
\chi_{10} \n \\
&\phantom{=} \left. {}+ \sqrt{(J + M - 1)(J + M)} \, Y_{J - 1, M - 1}
(\hat{\bf{p}}) \, \chi_{11} \right]  \quad (J \ge 1) \:, \n \\[1mm]
\supfi{3} \cal{Y}_{J M}^J (\hat{\bf{p}}) &= \frac{1}{\sqrt{2 J (J + 1)}}
\left[ \sqrt{(J - M)(J + M + 1)} \, Y_{J, M + 1} (\hat{\bf{p}}) \, 
\chi_{1, -1} \right. \n \\
&\phantom{=} {}+ \sqrt{2} \, M \, Y_{J M} (\hat{\bf{p}}) \, \chi_{10} \n \\
&\phantom{=} \left. {}- \sqrt{(J - M + 1)(J + M)} \, Y_{J, M - 1} 
(\hat{\bf{p}}) \, \chi_{11} \right] \quad (J \ge 1) \:, \n \\[1mm]
\supfi{3} \cal{Y}_{J + 1, M}^J (\hat{\bf{p}}) &= \frac{1}{\sqrt{2(J + 1)
(2 J + 3)}} \left[ \sqrt{(J + M + 1)(J + M + 2)} \, Y_{J + 1, M + 1} 
(\hat{\bf{p}}) \, \chi_{1, -1} \right. \n \\
&\phantom{=} {}- \sqrt{2 (J - M + 1)(J + M + 1)} \, Y_{J + 1, M} (\hat{\bf p}) 
\, \chi_{10} \n \\
&\phantom{=} \left. {}+ \sqrt{(J - M + 1)(J - M + 2)} \, Y_{J + 1, M - 1} 
(\hat{\bf{p}}) \, \chi_{11} \right] \:.
\eal

To determine the action of the helicity operators on these eigenfunctions,
the explicit spinor representation
\be
\hat{\bf{p}} \cdot \bm{\sigma} = \left( \ba{cc}
\cos \vartheta & \sin \vartheta e^{-i \varphi} \\
\sin \vartheta e^{i \varphi} & - \cos \vartheta
\ea \right) \label{helexplicit}
\ee 
can be used, together with the following special instances of the spherical
harmonics addition relation \cite{CDL77}:
\bal
\cos \vartheta \, Y_{lm} (\hat{\bf{p}}) &= \sqrt{\frac{(l - m)(l + m)}
{(2l - 1)(2l + 1)}} \, Y_{l - 1, m} (\hat{\bf{p}}) \n \\
&\phantom{=} {}+ \sqrt{\frac{(l - m + 1)(l + m + 1)}{(2l + 1)(2l + 3)}} \,
Y_{l + 1, m} (\hat{\bf{p}}) \:, \n \\[2mm]
\sin \vartheta e^{i \varphi} Y_{lm} (\hat{\bf{p}}) &= 
\sqrt{\frac{(l - m - 1)(l - m)}
{(2l - 1)(2l + 1)}} \, Y_{l - 1, m + 1} (\hat{\bf{p}}) \n \\
&\phantom{=} {}- \sqrt{\frac{(l + m + 1)(l + m + 2)}{(2l + 1)(2l + 3)}} \,
Y_{l + 1, m + 1} (\hat{\bf{p}}) \:, \n \\[2mm]
\sin \vartheta e^{-i \varphi} Y_{lm} (\hat{\bf{p}}) &= 
- \sqrt{\frac{(l + m - 1)(l + m)}
{(2l - 1)(2l + 1)}} \, Y_{l - 1, m - 1} (\hat{\bf{p}}) \n \\
&\phantom{=} {}+ \sqrt{\frac{(l - m + 1)(l - m + 2)}{(2l + 1)(2l + 3)}} \,
Y_{l + 1, m - 1} (\hat{\bf{p}}) \:. \label{SHAR}
\eal

It is then a straightforward, though somewhat lengthy exercise to derive
the following formulas for the application of the helicity operators to
the total angular momentum eigenstates:
\bal
(\hat{\bf{p}} \cdot \bm{\sigma}_A) \, \supfi{1} \cal{Y}_{J M}^J
(\hat{\bf{p}}) &= \sqrt{\frac{J}{2 J + 1}} \, \supfi{3}{\cal Y}_{J - 1, M}^J
(\hat{\bf{p}}) - \sqrt{\frac{J + 1}{2 J + 1}} \, 
\supfi{3} {\cal Y}_{J + 1, M}^J (\hat{\bf{p}}) \:, \n \\
(\hat{\bf{p}} \cdot \bm{\sigma}_A) \, \supfi{3} \cal{Y}_{J - 1, M}^J
(\hat{\bf{p}}) &= \sqrt{\frac{J}{2 J + 1}} \, \supfi{1} \cal{Y}_{J M}^J
(\hat{\bf{p}}) - \sqrt{\frac{J + 1}{2 J + 1}} \, 
\supfi{3} \cal{Y}_{J M}^J (\hat{\bf{p}}) \:, \n \\
(\hat{\bf{p}} \cdot \bm{\sigma}_A) \, \supfi{3} \cal{Y}_{J M}^J
(\hat{\bf{p}}) &= - \sqrt{\frac{J + 1}{2 J + 1}} \, 
\supfi{3} \cal{Y}_{J - 1, M}^J (\hat{\bf{p}}) - \sqrt{\frac{J}{2 J + 1}} \, 
\supfi{3} \cal{Y}_{J + 1, M}^J (\hat{\bf{p}}) \:, \n \\
(\hat{\bf{p}} \cdot \bm{\sigma}_A) \, \supfi{3} \cal{Y}_{J + 1, M}^J
(\hat{\bf{p}}) &= - \sqrt{\frac{J + 1}{2 J + 1}} \, 
\supfi{1} {\cal Y}_{J M}^J (\hat{\bf{p}}) - \sqrt{\frac{J}{2 J + 1}} \, 
\supfi{3} \cal{Y}_{J M}^J (\hat{\bf{p}}) \:, 
\eal 
and
\bal
(\hat{\bf{p}} \cdot \bm{\sigma}_B) \, \supfi{1} \cal{Y}_{J M}^J
(\hat{\bf{p}}) &= - \sqrt{\frac{J}{2 J + 1}} \, \supfi{3}{\cal Y}_{J - 1, M}^J
(\hat{\bf{p}}) + \sqrt{\frac{J + 1}{2 J + 1}} \, 
\supfi{3} {\cal Y}_{J + 1, M}^J (\hat{\bf{p}}) \:, \n \\
(\hat{\bf{p}} \cdot \bm{\sigma}_B) \, \supfi{3} \cal{Y}_{J - 1, M}^J
(\hat{\bf{p}}) &= - \sqrt{\frac{J}{2 J + 1}} \, \supfi{1} \cal{Y}_{J M}^J
(\hat{\bf{p}}) - \sqrt{\frac{J + 1}{2 J + 1}} \, 
\supfi{3} \cal{Y}_{J M}^J (\hat{\bf{p}}) \:, \n \\
(\hat{\bf{p}} \cdot \bm{\sigma}_B) \, \supfi{3} \cal{Y}_{J M}^J
(\hat{\bf{p}}) &= - \sqrt{\frac{J + 1}{2 J + 1}} \, 
\supfi{3} \cal{Y}_{J - 1, M}^J (\hat{\bf{p}}) - \sqrt{\frac{J}{2 J + 1}} \, 
\supfi{3} \cal{Y}_{J + 1, M}^J (\hat{\bf{p}}) \:, \n \\
(\hat{\bf{p}} \cdot \bm{\sigma}_B) \, \supfi{3} \cal{Y}_{J + 1, M}^J
(\hat{\bf{p}}) &= \sqrt{\frac{J + 1}{2 J + 1}} \, 
\supfi{1} {\cal Y}_{J M}^J (\hat{\bf{p}}) - \sqrt{\frac{J}{2 J + 1}} \, 
\supfi{3} \cal{Y}_{J M}^J (\hat{\bf{p}}) \:.
\eal 
In the special case $J=0$, the states $\supfi{3} \cal{Y}_{J - 1, M}^J 
(\hat{\bf p})$ and $\supfi{3} \cal{Y}_{J M}^J (\hat{\bf p})$ do not exist, 
and on the right--hand sides for the application of one of the helicity 
operators to $\supfi{1} \cal{Y}_{J M}^J (\hat{\bf p})$ and
$\supfi{3} \cal{Y}_{J + 1, M}^J (\hat{\bf p})$, only one term remains.

By use of Eq.\ \eqref{addtheorem}, where now
\bmu
a_l (p, p') = \frac{2l + 1}{2} \int_{-1}^1 d \cos \theta \, P_l (\cos \theta)
\, \frac{1}{2 \abs{\bf{p} - \bf{p}'}} \\
\times \left( \frac{1}{E_p^A + \abs{\bf{p} - \bf{p}'} - E_{p'}^A} + 
\frac{1}{E_p^B + \abs{\bf{p} - \bf{p}'} - E_{p'}^B} \right) \label{defal}
\emu
(for $\mu = 0$), the effective Schr\"odinger equation Eq.\ \eqref{spinrep}
decouples into pairs of coupled one-dimensional integral equations. For the 
S-coupled states, we introduce the wave function
\be
\phi (\bf{p}) = \supfi{S} \phi_0^J (p) \supfi{1} \cal{Y}_{J M}^J
(\hat{\bf{p}}) + \supfi{S} \phi_1^J (p) \supfi{3} \cal{Y}_{J M}^J
(\hat{\bf{p}}) \:.
\ee
The effective Schr\"odinger equation for the coefficient functions becomes
\bmu
\left( \sqrt{M_A^2 + p^2} + \sqrt{M_B^2 + p^2} \right)
\left( \ba{c} \supfi{S} \phi_0^J (p) \\
\supfi{S} \phi_1^J (p) \ea \right) \\
{}- \frac{g^2}{2 \pi^2} \int_0^\infty d p' \, {p'}^2 \sqrt{
\frac{E_p^A + M_A}{2 E_p^A} \, \frac{E_p^B + M_B}{2 E_p^B} \, 
\frac{E_{p'}^A + M_A}{2 E_{p'}^A} \, \frac{E_{p'}^B + M_B}{2 E_{p'}^B}} \\
{}\times \frac{1}{2 J + 1} \left( \ba{cc}
\supfi{S} V_{0 0}^J (p, p') & \supfi{S} V_{0 1}^J (p, p') \\
\supfi{S} V_{1 0}^J (p, p') & \supfi{S} V_{1 1}^J (p, p') \ea \right)
\left( \ba{c} \supfi{S} \phi_0^J (p') \\
\supfi{S} \phi_1^J (p') \ea \right) \\[1mm]
= \left( E - E_V \right) \left( \ba{c} \supfi{S} \phi_0^J (p) \\
\supfi{S} \phi_1 (p) \ea \right) \:, \label{Scoupleff}
\emu
with
\bal
\supfi{S} V_{0 0}^J (p, p') &= \left[ 1 + \frac{p}{E_p^A + M_A} \, 
\frac{p}{E_p^B + M_B} \, \frac{p'}{E_{p'}^A + M_A} \,
\frac{p'}{E_{p'}^B + M_B} \right] a_J (p, p') \n \\
&\phantom{=} {}- \left[ \frac{p}{E_p^A + M_A} \, \frac{p'}{E_{p'}^A + M_A}
+ \frac{p}{E_p^B + M_B} \, \frac{p'}{E_{p'}^B + M_B} \right] \n \\
&\phantom{=} {}\times \left[ J \, \frac{a_{J - 1} (p, p')}{2 J - 1} + 
(J + 1) \, \frac{a_{J + 1} (p, p')}{2 J + 3} \right] \:, \n \\[1mm]
\supfi{S} V_{1 1}^J (p, p') &= \left[ 1 + \frac{p}{E_p^A + M_A} \, 
\frac{p}{E_p^B + M_B} \, \frac{p'}{E_{p'}^A + M_A} \,
\frac{p'}{E_{p'}^B + M_B} \right] a_J (p, p') \n \\
&\phantom{=} {}- \left[ \frac{p}{E_p^A + M_A} \, \frac{p'}{E_{p'}^A + M_A}
+ \frac{p}{E_p^B + M_B} \, \frac{p'}{E_{p'}^B + M_B} \right] \n \\
&\phantom{=} {}\times \left[ (J + 1) \, \frac{a_{J - 1} (p, p')}{2 J - 1} + 
J \, \frac{a_{J + 1} (p, p')}{2 J + 3} \right] \:, \n \\[1mm]
\supfi{S} V_{0 1}^J (p, p') &= \supfi{S} V_{1 0}^J (p, p') \n \\
&= \left[ \frac{p}{E_p^A + M_A} \, \frac{p'}{E_{p'}^A + M_A}
- \frac{p}{E_p^B + M_B} \, \frac{p'}{E_{p'}^B + M_B} \right] \n \\
&\phantom{=} {}\times \sqrt{J (J + 1)} \left[ 
\frac{a_{J - 1} (p, p')}{2 J - 1} - \frac{a_{J + 1} (p, p')}{2 J + 3} \right] 
\:. \label{Scouplepot}
\eal

On the other hand, with the wave function
\be
\phi (\bf{p}) = \supfi{L} \phi_{J - 1}^J (p) \supfi{3} \cal{Y}_{J - 1, M}^J
(\hat{\bf{p}}) + \supfi{L} \phi_{J + 1}^J (p) \supfi{3} \cal{Y}_{J + 1, M}^J
(\hat{\bf{p}}) \:, \label{Lcouplewavef}
\ee
the effective Schr\"odinger equation for the L-coupled states becomes
\bmu
\left( \sqrt{M_A^2 + p^2} + \sqrt{M_B^2 + p^2} \right)
\left( \ba{c} \supfi{L} \phi_{J - 1}^J (p) \\
\supfi{L} \phi_{J + 1}^J (p) \ea \right) \\
{}- \frac{g^2}{2 \pi^2} \int_0^\infty d p' \, {p'}^2 \sqrt{
\frac{E_p^A + M_A}{2 E_p^A} \, \frac{E_p^B + M_B}{2 E_p^B} \, 
\frac{E_{p'}^A + M_A}{2 E_{p'}^A} \, \frac{E_{p'}^B + M_B}{2 E_{p'}^B}} \\
{}\times \frac{1}{2 J + 1} \left( \ba{cc}
\supfi{L} V_{J - 1, J - 1}^J (p, p') & \supfi{L} V_{J - 1, J + 1}^J (p, p') \\
\supfi{L} V_{J + 1, J - 1}^J (p, p') & \supfi{L} V_{J + 1, J + 1}^J (p, p') 
\ea \right) \left( \ba{c} \supfi{L} \phi_{J - 1}^J (p') \\
\supfi{L} \phi_{J + 1}^J (p') \ea \right) \\[1mm]
= \left( E - E_V \right) \left( \ba{c} \supfi{L} \phi_{J - 1}^J (p) \\
\supfi{L} \phi_{J + 1}^J (p) \ea \right) \:, \label{Lcoupleff}
\emu
where
\bal
\supfi{L} V_{J - 1, J - 1}^J (p, p') &= (2 J + 1) \, \frac{a_{J - 1} (p, p')}
{2 J - 1} \n \\
&\phantom{=} {}- \left[ \frac{p}{E_p^A + M_A} \, \frac{p'}{E_{p'}^A + M_A}
+ \frac{p}{E_p^B + M_B} \, \frac{p'}{E_{p'}^B + M_B} \right] 
a_J (p, p') \n \\
&\phantom{=} {}+ \frac{p}{E_p^A + M_A} \, 
\frac{p}{E_p^B + M_B} \, \frac{p'}{E_{p'}^A + M_A} \,
\frac{p'}{E_{p'}^B + M_B} \n \\
&\phantom{=} {}\times \frac{1}{2 J + 1} \left[ \frac{a_{J - 1} (p, p')}
{2 J - 1} + 4 J (J + 1) \, \frac{a_{J + 1} (p, p')}{2 J + 3} \right] \:, 
\n \\[1mm]
\supfi{L} V_{J + 1, J + 1}^J (p, p') &= (2 J + 1) \, \frac{a_{J + 1} (p, p')}
{2 J + 3} \n \\
&\phantom{=} {}- \left[ \frac{p}{E_p^A + M_A} \, \frac{p'}{E_{p'}^A + M_A}
+ \frac{p}{E_p^B + M_B} \, \frac{p'}{E_{p'}^B + M_B} \right] 
a_J (p, p') \n \\
&\phantom{=} {}+ \frac{p}{E_p^A + M_A} \, 
\frac{p}{E_p^B + M_B} \, \frac{p'}{E_{p'}^A + M_A} \,
\frac{p'}{E_{p'}^B + M_B} \n \\
&\phantom{=} {}\times \frac{1}{2 J + 1} \left[ 4 J (J + 1) \, \frac{a_{J - 1} 
(p, p')}{2J - 1} + \frac{a_{J + 1} (p, p')}{2 J + 3} \right] \:, \n \\[1mm]
\supfi{L} V_{J - 1, J + 1} (p, p') &= \supfi{L} V_{J + 1, J - 1} (p, p') \n \\
&= \frac{p}{E_p^A + M_A} \, 
\frac{p}{E_p^B + M_B} \, \frac{p'}{E_{p'}^A + M_A} \,
\frac{p'}{E_{p'}^B + M_B} \n \\
&\phantom{=} {}\times \frac{2 \sqrt{J (J + 1)}}{2 J + 1} \left[ 
\frac{a_{J - 1} (p, p')}{2 J - 1} - \frac{a_{J + 1} (p, p')}{2 J + 3} \right] 
\:. \label{Lcouplepot}
\eal
In the special case $J = 0$, $\supfi{S} \phi_1^J (p) \equiv
\supfi{L} \phi_{J - 1}^J (p) \equiv 0$, and the equations for
$\supfi{S} \phi_0^J (p)$ and $\supfi{L} \phi_{J + 1}^J (p)$ 
decouple. 

The functions $a_l (p, p')$ are quite straightforwardly calculated
from Eq.\ \eqref{defal}, for any given value of $l$. The results for the 
functions which are relevant for the numerical calculations presented here 
($l \le 3$) are:
\bal
a_0 (p, p') &= \frac{1}{4 p p'} \left\{ 
\ln \left( \frac{E^A_p + (p + p') - E^A_{p'}}
{E^A_p + |p - p'| - E^A_{p'}} \right) +
\ln \left( \frac{E^B_p + (p + p') - E^B_{p'}}
{E^B_p + |p - p'| - E^B_{p'}} \right) \right\} \:, \n \\[2mm]
a_1 (p, p') &= \frac{3}{4 p p'} \left\{
\frac{(E^A_p - E^A_{p'} + E^B_p - E^B_{p'}) (p + p' - |p - p'|)}{2 p p'} 
- 2 \right. \n \\
&\phantom{=} {}+ \frac{p^2 + {p'}^2 - (E^A_p - E^A_{p'})^2}{2 p p'} 
\ln \left( \frac{E^A_p + (p + p') - E^A_{p'}}{E^A_p + |p - p'| - E^A_{p'}} 
\right) \n \\
&\phantom{=} \left. {}+ \frac{p^2 + {p'}^2 - (E^B_p - E^B_{p'})^2}{2 p p'} 
\ln \left( \frac{E^B_p + (p + p') - E^B_{p'}}{E^B_p + |p - p'| - E^B_{p'}} 
\right) \right\} \:, \n \\[2mm]
a_2 (p, p') &= \frac{5}{8 p p'} \left\{ \frac{ 3 [2 (p^2 + {p'}^2) 
- (E_p^A - E_{p'}^A)^2] (E_p^A - E_{p'}^A) ( p + p'
- |p - p'|)}{4 p^2 {p'}^2} \right. \n \\
&\phantom{=} {}+ \frac{ 3 [2 (p^2 + {p'}^2) - (E_p^B - E_{p'}^B)^2] 
(E_p^B - E_{p'}^B) ( p + p' - |p - p'|)}{4 p^2 {p'}^2} \n \\
&\phantom{=} {}- \frac{ (E_p^A - E_{p'}^A + E_p^B - E_{p'}^B) 
[(p + p')^3 - |p - p'|^3]}{4 p^2 {p'}^2} \n \\
&\phantom{=} {}- \frac{3 [2 (p^2 + {p'}^2) -
(E_p^A - E_{p'}^A)^2 - (E_p^B - E_{p'}^B)^2]}{2 p p'} \n \\
&\phantom{=} {}+ \left[ \frac{3 [p^2 + {p'}^2 -
(E_p^A - E_{p'}^A)^2]^2}{4 p^2 {p'}^2} - 1 \right]
\ln \left( \frac{E^A_p + (p + p') - E^A_{p'}}
{E^A_p + |p - p'| - E^A_{p'}} \right) \n \\
&\phantom{=} \left. {}+ \left[ \frac{3 [p^2 + {p'}^2 -
(E_p^B - E_{p'}^B)^2]^2}{4 p^2 {p'}^2} - 1 \right]
\ln \left( \frac{E^B_p + (p + p') - E^B_{p'}}
{E^B_p + |p - p'| - E^B_{p'}} \right) \right\} \:, \n \\[2mm]
a_3 (p, p') &= \frac{7}{8 p p'} \left\{ \left[
\frac{5 [3 (p^2 + {p'}^2)^2 - 3 (p^2 + {p'}^2) (E_p^A - E_{p'}^A)^2 
+ (E_p^A - E_{p'}^A)^4] (E_p^A - E_{p'}^A)}{8 p^3 {p'}^3} \right. 
\right. \n \\
&\phantom{=} {}+ \frac{5 [3 (p^2 + {p'}^2)^2 - 3 (p^2 + {p'}^2) 
(E_p^B - E_{p'}^B)^2 + (E_p^B - E_{p'}^B)^4] (E_p^B - E_{p'}^B)}
{8 p^3 {p'}^3} \n \\
&\phantom{=} \left. {}- \frac{3 (E_p^A - E_{p'}^A + E_p^B - E_{p'}^B)}
{2 p p'} \right] (p + p' - |p - p'|) \n \\
&\phantom{=} {}- \left[ \frac{5 (E_p^A - E_{p'}^A) [3(p^2+ {p'}^2) -
(E_p^A - E_{p'}^A)^2]}{24 p^3 {p'}^3} \right. \n \\
&\phantom{=} \left. {}+ \frac{5 (E_p^B - E_{p'}^B) [3(p^2+ {p'}^2) -
(E_p^B - E_{p'}^B)^2]}{24 p^3 {p'}^3} \right] [ (p + p')^3 - |p - p'|^3 ]
\n \\
&\phantom{=} {}+ \frac{(E_p^A - E_{p'}^A + E_p^B - E_{p'}^B) [ (p + p')^5
- |p - p'|^5 ]}{8 p^3 {p'}^3} \n \\
&\phantom{=} {}- \frac{5 [p^2 + {p'}^2 - (E_p^A - E_{p'}^A)^2]^2
+ 5 [p^2 + {p'}^2 - (E_p^B - E_{p'}^B)^2]^2}{4 p^2 {p'}^2} + \frac{8}{3}
\n \\
&\phantom{=} {}+ \left[ \frac{5 [p^2 + {p'}^2 - (E_p^A - E_{p'}^A)^2]^3}
{8 p^3 {p'}^3} - \frac{3 [p^2 + {p'}^2 - (E_p^A - E_{p'}^A)^2]}{2 p p'}
\right] \n \\
&\phantom{=} {}\times \ln \left( \frac{E^A_p + (p + p') - E^A_{p'}}
{E^A_p + |p - p'| - E^A_{p'}} \right) \n \\
&\phantom{=} {}+ \left[ \frac{5 [p^2 + {p'}^2 - 
(E_p^B - E_{p'}^B)^2]^3}{8 p^3 {p'}^3} - 
\frac{3 [p^2 + {p'}^2 - (E_p^B - E_{p'}^B)^2]}{2 p p'}
\right] \n \\
&\phantom{=} \left. {}\times \ln \left( \frac{E^B_p + (p + p') - E^B_{p'}}
{E^B_p + |p - p'| - E^B_{p'}} \right) \right\} \:. \label{pwcoeff}
\eal
In the actual numerical calculations we did not rely on these
explicit expressions, but rather used a quasi-algebraic method which
increases speed and accuracy. To this end, the integrand in Eq.\ 
\eqref{defal} is written, after changing variables from $\cos \theta$ to
$x = \abs{\bf{p} - \bf{p}'}$, as a (finite) Laurent series in $x$ around 
$E_{p'}^A - E_p^A$ (or $E_{p'}^B - E_p^B$ for the second term). This requires 
expressing $P_l (\cos \theta)$ as a polynomial in $x$, a task left to the 
computer. Each term in the Laurent series has a known integral which is simply
inserted. Any value of $l$ can be handled easily by this method.

In the one-body limit $M_B \to \infty$, the matrix for the effective
potential in the S-coupled sector, Eq.\ \eqref{Scouplepot}, tends to
\bmu
\left( \ba{cc}
\supfi{S} V_{0 0}^J (p, p') & \supfi{S} V_{0 1}^J (p, p') \\
v\supfi{S} V_{1 0}^J (p, p') & \supfi{S} V_{1 1}^J (p, p')
\ea \right)
= \left( \ba{cc}
1 & 0 \\
0 & 1 
\ea \right) a_J (p, p') \\
- \frac{p}{E_p^A + M_A} \, \frac{p'}{E_{p'}^A + M_A}
\left[ \left( \ba{cc}
J & - \sqrt{J (J + 1)} \\[1mm]
- \sqrt{J (J + 1)} & J + 1
\ea \right) \frac{a_{J - 1} (p, p')}{2 J - 1} \right. \\
+ \left. \left( \ba{cc}
J + 1 & \sqrt{J (J + 1)} \\[1mm]
\sqrt{J (J + 1)} & J
\ea \right) \frac{a_{J + 1} (p, p')}{2 J + 3} \right] \:.
\emu
This matrix is diagonalized by the orthogonal linear combinations
\be
\left( \ba{c}
\supfi{S} \phi^J_0 (p) \\ \supfi{S} \phi^J_1 (p)
\ea \right)
= \left( \ba{c}
- \sqrt{J} \\[1mm] \sqrt{J + 1}
\ea \right) 
\frac{\supfi{S} \phi^J_{J - 1/2} (p)}{\sqrt{2 J + 1}} \label{onebBwavef1}
\ee
and
\be
\left( \ba{c}
\supfi{S} \phi^J_0 (p) \\ \supfi{S} \phi^J_1 (p)
\ea \right)
= \left( \ba{c}
\sqrt{J + 1} \\[1mm] \sqrt{J}
\ea \right) 
\frac{\supfi{S} \phi^J_{J + 1/2} (p)}{\sqrt{2 J + 1}} \:,
 \label{onebBwavef2}
\ee
with the respective eigenvalues
\bal
&{} (2 J + 1) \left( \frac{a_J (p, p')}{2 J + 1} - \frac{p}{E_p^A + M_A} \, 
\frac{p'}{E_{p'}^A + M_A} \, \frac{a_{J - 1} (p, p')}{2 J - 1} \right)
\:, \n \\
&{} (2 J + 1) \left( \frac{a_J (p, p')}{2 J + 1} - \frac{p}{E_p^A + M_A} \, 
\frac{p'}{E_{p'}^A + M_A} \, \frac{a_{J + 1} (p, p')}{2 J + 3} \right) \:.
\eal
Alternatively, we can write the wave function as
\be
\phi (\bf{p}) = \supfi{S} \phi^J_{J - 1/2} (p) 
\subfi{J - 1/2} \cal{Y}^J_{J M} (\hat{\bf{p}}) \label{jAwavef1}
\ee
or
\be
\phi (\bf{p}) = \supfi{S} \phi^J_{J + 1/2} (p) 
\subfi{J + 1/2} \cal{Y}^J_{J M} (\hat{\bf{p}}) \label{jAwavef2}
\ee
with
\bal
\subfi{J - 1/2} \cal{Y}^J_{J M} (\hat{\bf{p}}) &= - \sqrt{\frac{J}{2 J + 1}}
\supfi{1} \cal{Y}^J_{J M} (\hat{\bf{p}}) + \sqrt{\frac{J + 1}{2 J + 1}}
\supfi{3} \cal{Y}^J_{J M} (\hat{\bf{p}}) \:, \n \\
\subfi{J + 1/2} \cal{Y}^J_{J M} (\hat{\bf{p}}) &= \sqrt{\frac{J + 1}{2 J + 1}}
\supfi{1} \cal{Y}^J_{J M} (\hat{\bf{p}}) + \sqrt{\frac{J}{2 J + 1}}
\supfi{3} \cal{Y}^J_{J M} (\hat{\bf{p}}) \:, \label{jAcouple}
\eal
anticipating the notation $\subfi{j_A} \cal{Y}^J_{l M} (\hat{\bf{p}})$.
In the special case $J=0$, of course, $\supfi{S} \phi_1^J (p) \equiv 0$,
$\supfi{S} \phi_{J - 1/2}^J (p) \equiv 0$, and
$\subfi{J - 1/2} \cal{Y}_{J M}^J (\hat{\bf{p}})$ does not exist, since there
is only one state in the S-coupled $J=0$ sector.

The matrix \eqref{Lcouplepot} of the effective potential for L-coupled 
states, on the other hand, becomes diagonal in the limit $M_B \to \infty$,
hence there is no L-coupling in this limit. The diagonal matrix elements 
tend to
\bal
\supfi{L} V_{J - 1, J - 1}^J (p, p') &= (2 J + 1) \left( 
\frac{a_{J - 1} (p, p')}{2 J - 1} - \frac{p}{E_p^A + M_A} \, 
\frac{p'}{E_{p'}^A + M_A} \, \frac{a_J (p, p')}{2 J + 1} \right) \:, \n \\
\supfi{L} V_{J + 1, J + 1}^J (p, p') &= (2 J + 1) \left( 
\frac{a_{J + 1} (p, p')}{2 J + 3} - \frac{p}{E_p^A + M_A} \, 
\frac{p'}{E_{p'}^A + M_A} \, \frac{a_J (p, p')}{2 J + 1} \right) \:.
\eal
Then, from Eqs.\ \eqref{Scoupleff} and \eqref{Lcoupleff}, 
$\supfi{S} \phi^J_{J - 1/2} (p)$ in Eq.\ \eqref{onebBwavef1} and
$\supfi{L} \phi^{J - 1}_{J - 1/2} (p) \equiv \supfi{L} \phi^{J - 1}_J (p)$ 
in Eq.\ \eqref{Lcouplewavef},
\be
\phi (\bf{p}) = \supfi{L} \phi^{J - 1}_{J - 1/2} (p) 
\supfi{3} \cal{Y}_{J M}^{J - 1} (\hat{\bf{p}}) \:, \label{jAwavef3}
\ee
fulfill exactly the same one-dimensional equation in the limit 
$M_B \to \infty$ (for $J \ge 1$),
\bmu
\sqrt{M_A^2 + p^2} \left( \ba{c} \supfi{S} \phi_{J - 1/2}^J (p) \\
\supfi{L} \phi_{J - 1/2}^{J - 1} (p) \ea \right)
- \frac{g^2}{2 \pi^2} \int_0^\infty d p' \, {p'}^2 \sqrt{
\frac{E_p^A + M_A}{2 E_p^A} \, \frac{E_{p'}^A + M_A}{2 E_{p'}^A}} \\
\times \left( \frac{a_J (p, p')}{2 J + 1} - \frac{p}{E_p^A + M_A} \, 
\frac{p'}{E_{p'}^A + M_A} \, \frac{a_{J - 1} (p, p')}{2 J - 1} \right)
\left( \ba{c} \supfi{S} \phi_{J - 1/2}^J (p') \\
\supfi{L} \phi_{J - 1/2}^{J - 1} (p') \ea \right) \\
= \left( E - E_V - M_B \right) \left( \ba{c} \supfi{S} \phi_{J - 1/2}^J (p) \\
\supfi{L} \phi_{J - 1/2}^{J - 1} (p) \ea \right) \:. \label{onebBeffseq1}
\emu
Likewise, 
$\supfi{S} \phi^J_{J + 1/2} (p)$ in Eq.\ \eqref{onebBwavef2} and
$\supfi{L} \phi^{J + 1}_{J + 1/2} (p) \equiv \supfi{L} \phi^{J + 1}_J (p)$ 
in Eq.\ \eqref{Lcouplewavef},
\be
\phi (\bf{p}) = \supfi{L} \phi^{J + 1}_{J + 1/2} (p) 
\supfi{3} \cal{Y}_{J M}^{J + 1} (\hat{\bf{p}}) \:, \label{jAwavef4}
\ee
both fulfill the equation
\bmu
\sqrt{M_A^2 + p^2} \left( \ba{c} \supfi{S} \phi_{J + 1/2}^J (p) \\
\supfi{L} \phi_{J + 1/2}^{J + 1} (p) \ea \right)
- \frac{g^2}{2 \pi^2} \int_0^\infty d p' \, {p'}^2 \sqrt{
\frac{E_p^A + M_A}{2 E_p^A} \, \frac{E_{p'}^A + M_A}{2 E_{p'}^A}} \\
\times \left( \frac{a_J (p, p')}{2 J + 1} - \frac{p}{E_p^A + M_A} \, 
\frac{p'}{E_{p'}^A + M_A} \, \frac{a_{J + 1} (p, p')}{2 J + 3} \right)
\left( \ba{c} \supfi{S} \phi_{J + 1/2}^J (p') \\
\supfi{L} \phi_{J + 1/2}^{J + 1} (p') \ea \right) \\
= \left( E - E_V - M_B \right) \left( \ba{c} \supfi{S} \phi_{J + 1/2}^J (p) \\
\supfi{L} \phi_{J + 1/2}^{J + 1} (p) \ea \right) \:. \label{onebBeffseq2}
\emu
As a result, every state is (at least) twofold degenerate in the one-body
limit.

The states $\subfi{J - 1/2} \cal{Y}^J_{J M} (\hat{\bf{p}})$ and
$\subfi{J + 1/2} \cal{Y}^J_{J M} (\hat{\bf{p}})$ defined in Eq.\ 
\eqref{jAcouple}, as well as $\subfi{J - 1/2} \cal{Y}^{J - 1}_{J M} 
(\hat{\bf{p}}) \equiv \supfi{3} \cal{Y}^{J - 1}_{J M} (\hat{\bf{p}})$ and
$\subfi{J + 1/2} \cal{Y}^{J + 1}_{J M} (\hat{\bf{p}}) \equiv
\supfi{3} \cal{Y}^{J + 1}_{J M} (\hat{\bf{p}})$, can be rewritten in
terms of the eigenstates $Y^{j_A}_{l m} (\hat{\bf{p}})$ of $\bf{j}_A^2$,
$j_{A, z}$, and $\bf{L}^2$ (where $\bf{j}_A = \bf{L} + \bf{s}_A$),
\bal
Y^{l - 1/2}_{l m} (\hat{\bf{p}}) &= \frac{1}{\sqrt{2 l + 1}}
\left( \ba{c}
\ds - \sqrt{l - m + 1/2} \, Y_{l, m - 1/2} (\hat{\bf{p}}) \\[2mm] 
\ds \sqrt{l + m + 1/2} \, Y_{l, m + 1/2} (\hat{\bf{p}})
\ea \right) \quad (l \ge 1) \:, \n \\[2mm]
Y^{l + 1/2}_{l m} (\hat{\bf{p}}) &= \frac{1}{\sqrt{2 l + 1}}
\left( \ba{c}
\ds \sqrt{l + m + 1/2} \, Y_{l, m - 1/2} (\hat{\bf{p}}) \\[2mm] 
\ds \sqrt{l - m + 1/2} \, Y_{l, m + 1/2} (\hat{\bf{p}})
\ea \right) \:,
\eal
as
\bal
\subfi{J - 1/2} \cal{Y}_{J M}^J (\hat{\bf{p}}) &=
\sqrt{\frac{J - M}{2 J}} \, Y_{J, M + 1/2}^{J - 1/2} (\hat{\bf{p}})
\otimes \left( \ba{c} 0 \\ 1 \ea \right)
+ \sqrt{\frac{J + M}{2 J}} \, Y_{J, M - 1/2}^{J - 1/2} 
(\hat{\bf{p}}) \otimes \left( \ba{c} 1 \\ 0 \ea \right) \:, \n \\[2mm]
\subfi{J - 1/2} \cal{Y}_{J M}^{J - 1} (\hat{\bf{p}}) &= 
\sqrt{\frac{J + M}{2 J}} \, Y_{J, M + 1/2}^{J - 1/2}
(\hat{\bf{p}}) \otimes \left( \ba{c} 0 \\ 1 \ea \right)
- \sqrt{\frac{J - M}{2 J}} \, Y_{J, M - 1/2}^{J - 1/2} 
(\hat{\bf{p}}) \otimes \left( \ba{c} 1 \\ 0 \ea \right) \:, \n \\[2mm]
\subfi{J + 1/2} \cal{Y}_{J M}^J (\hat{\bf{p}}) &=
\sqrt{\frac{J + M + 1}{2 (J + 1)}} \, Y_{J, M + 1/2}^{J + 1/2} 
(\hat{\bf p}) \otimes \left( \ba{c} 0 \\ 1 \ea \right)
- \sqrt{\frac{J - M + 1}{2 (J + 1)}} \, Y_{J, M - 1/2}^{J + 1/2} 
(\hat{\bf{p}}) \otimes \left( \ba{c} 1 \\ 0 \ea \right) \:, \n \\[2mm]
\subfi{J + 1/2} \cal{Y}_{J M}^{J + 1} (\hat{\bf{p}}) &= 
\sqrt{\frac{J - M + 1}{2 (J + 1)}} \, Y_{J, M + 1/2}^{J + 1/2} (\hat{\bf{p}})
\otimes \left( \ba{c} 0 \\ 1 \ea \right)
+ \sqrt{\frac{J + M + 1}{2 (J + 1)}} \, Y_{J, M - 1/2}^{J + 1/2}
(\hat{\bf{p}}) \otimes \left( \ba{c} 1 \\ 0 \ea \right) \:,
\eal
hence they are simultaneous eigenstates of $\bf{J}^2$, $J_z$, $\bf{j}_A^2$,
and $\bf{L}^2$. 

Using the formulas
\bal
(\hat{\bf{p}} \cdot \bm{\sigma}) Y_{l m}^{l - 1/2} (\hat{\bf{p}}) &= 
- Y_{l - 1, m}^{l - 1/2} (\hat{\bf{p}}) \:, \n \\
(\hat{\bf{p}} \cdot \bm{\sigma}) Y_{l m}^{l + 1/2} (\hat{\bf{p}}) &= 
- Y_{l + 1, m}^{l + 1/2} (\hat{\bf{p}}) \:,
\eal
proven, e.g., with the help of Eqs.\ \eqref{helexplicit} and \eqref{SHAR},
and consistent with the fact that the application of helicity operators
conserves angular momentum $\bf{j}_A = \bf{L} + \bf{s}_A$ and changes
spatial parity $(-1)^l$, it is easily seen that for the wave functions
\eqref{jAwavef1} and \eqref{jAwavef3} the effective Schr\"odinger
equation in the one-body limit, Eq.\ \eqref{1Blimit}, is equivalent to
Eq.\ \eqref{onebBeffseq1}, while for the wave functions \eqref{jAwavef2}
and \eqref{jAwavef4}, Eq.\ \eqref{1Blimit} is equivalent to Eq.\ 
\eqref{onebBeffseq2}.

The coefficient functions $a_l (p, p')$ of the partial wave expansion
in the one-body limit $M_B \to \infty$ are defined by
\bmu
a_l (p, p') = \frac{2l + 1}{2} \int_{-1}^1 d \cos \theta \, P_l (\cos \theta)
\, \frac{1}{2 \abs{\bf{p} - \bf{p}'}} 
\left( \frac{1}{\abs{\bf{p} - \bf{p}'}} +
\frac{1}{E_p^A + \abs{\bf{p} - \bf{p}'} - E_{p'}^A} \right) \:,
\emu
see Eq.\ \eqref{1Blimit}. The explicit expressions in the cases 
$l = 0, 1, 2$, which are the ones relevant for the numerical results 
presented here, are
\bal
a_0 (p, p') &= \frac{1}{4 p p'} \left\{ \ln \frac{p + p'}{|p - p'|} + 
\ln \left( \frac{E^A_p + (p + p') - E^A_{p'}}
{E^A_p + |p - p'| - E^A_{p'}} \right) \right\} \:, \n \\[2mm]
a_1 (p, p') &= \frac{3}{4 p p'} \left\{
\frac{(E^A_p - E^A_{p'}) (p + p' - |p - p'|)}{2 p p'} - 2 \right. \n \\
&\phantom{=} \left. {}+ \frac{p^2 + {p'}^2}{2 p p'}
\ln \frac{p + p'}{|p - p'|} +
\frac{p^2 + {p'}^2 - (E^A_p - E^A_{p'})^2}{2 p p'} \ln
\left( \frac{E^A_p + (p + p') - E^A_{p'}}{E^A_p + |p - p'| - E^A_{p'}} \right)
\right\} \:, \n \\[2mm]
a_2 (p, p') &= \frac{5}{8 p p'} \left\{ \frac{ 3 [2 (p^2 + {p'}^2) 
- (E_p^A - E_{p'}^A)^2] (E_p^A - E_{p'}^A) ( p + p'
- |p - p'|)}{4 p^2 {p'}^2} \right. \n \\
&\phantom{=} {}- \frac{ (E_p^A - E_{p'}^A) [(p + p')^3 - |p - p'|^3
]}{4 p^2 {p'}^2} - \frac{3 [2 (p^2 + {p'}^2) -
(E_p^A - E_{p'}^A)^2]}{2 p p'} \n \\
&\phantom{=} {}+ \left[ \frac{3 (p^2 + {p'}^2)^2}{4 p^2 {p'}^2} - 1 
\right] \ln \frac{p + p'}{|p - p'|} \n \\
&\phantom{=} \left. {}+ \left[ \frac{3 [p^2 + {p'}^2 -
(E_p^A - E_{p'}^A)^2]^2}{4 p^2 {p'}^2} - 1 \right]
\ln \left( \frac{E^A_p + (p + p') - E^A_{p'}}
{E^A_p + |p - p'| - E^A_{p'}} \right) \right\} \:.
\eal
They coincide with the limit $M_B \to \infty$ of the
corresponding expressions in Eq.\ \eqref{pwcoeff}.

The consistency of the one-body limit is hence fully
established at the level of the effective Schr\"odinger equation
after the separation of angular and spin variables.

\section{Approximate diagonalization of the effective potential matrices
\label{appert}}

To order $\alpha^4$, i.e., for the lowest-order relativistic corrections, 
we can approximate
\be
\frac{p}{E_p^A + M_A} = \frac{p}{2 M_A} \:, \quad
\frac{p}{E_p^B + M_B} = \frac{p}{2 M_B} \:,
\ee
and analogously for $p'$ instead of $p$, in the potential terms of the 
effective Schr\"odinger equations \eqref{Scoupleff} and 
\eqref{Lcoupleff}. Furthermore, terms containing
\be
\frac{p}{E_p^A + mM_A} \, \frac{p}{E_p^B + M_B} \, \frac{p'}{E_{p'}^A + M_A}
\, \frac{p'}{E_{p'}^B + M_B}
\ee
can be neglected. The matrix \eqref{Lcouplepot} in the 
L-coupled sector can then be approximated by the following diagonal matrix
\bal
&\phantom{=} \frac{1}{2 J + 1} \left( \ba{cc}
\supfi{L} V_{J - 1, J - 1}^J (p, p') & \supfi{L} V_{J - 1, J + 1}^J (p, p') \\
\supfi{L} V_{J + 1, J - 1}^J (p, p') & \supfi{L} V_{J + 1, J + 1}^J (p, p') 
\ea \right) \n \\[2mm]
&= \left( \ba{cc}
\ds \frac{a_{J - 1} (p, p')}{2 J - 1} & 0 \\
0 & \ds \frac{a_{J + 1} (p, p')}{2 J + 3} \ea \right)
- \left( \frac{p p'}{4 M_A^2} + \frac{p p'}{4 M_B^2} \right)
\left( \ba{cc}
\ds \frac{a_J (p, p')}{2 J + 1} & 0 \\
0 & \ds \frac{a_J (p, p')}{2 J + 1} \ea \right) \:.
\eal
In particular, to order $\alpha^4$ there is \emph{no} L-coupling.
We hence expect, at least for moderate values of $\alpha$, the mixing for 
the solutions in the L-coupled sector (with $J \ge 1$) to be rather small.

The matrix elements \eqref{Scouplepot} in the S-coupled sector can be 
approximated by
\bal
\supfi{S} V_{0 0}^J (p, p') &=
a_J (p, p') - \left[ \frac{p p'}{4 M_A^2} + \frac{p p'}{4 M_B^2} \right]
\left[ J \, \frac{a_{J - 1} (p, p')}{2 J - 1} + 
(J + 1) \, \frac{a_{J + 1} (p, p')}{2 J + 3} \right] \:, \n \\
\supfi{S} V_{1 1}^J (p, p') &=
a_J (p, p') - \left[ \frac{p p'}{4 M_A^2} + \frac{p p'}{4 M_B^2} \right]
\left[ (J + 1) \, \frac{a_{J - 1} (p, p')}{2 J - 1} + 
J \, \frac{a_{J + 1} (p, p')}{2 J + 3} \right] \:, \n \\[2mm]
\supfi{S} V_{0 1}^J (p, p') &= \supfi{S} V_{1 0}^J (p, p')
= \left[ \frac{p p'}{4 M_A^2} - \frac{p p'}{4 M_B^2} \right]
\sqrt{J (J + 1)} \left[ \frac{a_{J - 1} (p, p')}{2 J - 1} - 
\frac{a_{J + 1} (p, p')}{2 J + 3} \right] \:.
\eal
This approximate matrix, somewhat fortunately, can be diagonalized through
a $p$- and $p'$-independent transformation. After some algebraic labor,
one obtains the following eigenvectors:
\bal
v^{(+)}_{J, x} &= \frac{1}{\sqrt{2}} \left( \ba{c}
\ds - \left( 1 - \frac{1}{\sqrt{1 + 4 J (J + 1) x^2}} \right)^{1/2} \\[6mm]
\ds \left( 1 + \frac{1}{\sqrt{1 + 4 J (J + 1) x^2}} \right)^{1/2}
\ea \right) \:, \n \\[2mm]
v^{(-)}_{J, x} &= \frac{1}{\sqrt{2}} \left( \ba{c}
\ds \left( 1 + \frac{1}{\sqrt{1 + 4 J (J + 1) x^2}} \right)^{1/2} \\[6mm]
\ds \left( 1 - \frac{1}{\sqrt{1 + 4 J (J + 1) x^2}} \right)^{1/2}
\ea \right)
\eal
($J \ge 1$), where we have expressed the mass dependence through the parameter
\be
x = \frac{M_B^2 - M_A^2}{M_B^2 + M_A^2} \:.
\ee
The corresponding eigenvalues are
\bal
\frac{1}{2 J + 1} \, \supfi{S} V^{(+)}_{J, x} (p, p') &= 
\frac{a_J (p, p')}{2 J + 1} -
\frac{p p'}{4 M_A^2} \, \frac{1}{1 + x} \left[ \left( 1 +
\frac{\sqrt{1 + 4 J (J + 1) x^2}}{2 J + 1} \right) \frac{a_{J - 1} (p, p')}
{2 J - 1} \right. \n \\[1mm]
& \hspace{4cm} \left. {}+ \left( 1 - \frac{\sqrt{1 + 4 J (J + 1) x^2}}
{2 J + 1} \right) \frac{a_{J + 1} (p, p')}{2 J + 3} \right] \:, \n \\[2mm]
\frac{1}{2 J + 1} \, \supfi{S} V^{(-)}_{J, x} (p, p') &= 
\frac{a_J (p, p')}{2 J + 1} -
\frac{p p'}{4 M_A^2} \, \frac{1}{1 + x} \left[ \left( 1 -
\frac{\sqrt{1 + 4 J (J + 1) x^2}}{2 J + 1} \right) \frac{a_{J - 1} (p, p')}
{2 J - 1} \right. \n \\[1mm]
& \hspace{4cm} \left. {}+ \left( 1 + \frac{\sqrt{1 + 4 J (J + 1) x^2}}
{2 J + 1} \right) \frac{a_{J + 1} (p, p')}{2 J + 3} \right] \:. 
\label{Sapproxdiag}
\eal
These formulas are valid for $x \ge 0$ or $M_B \ge M_A$.
Comparing these results with the approximate diagonal matrix elements
for L-coupling,
\bal
\frac{1}{2 J - 1} \, \supfi{L} V_{J J}^{J - 1} (p, p') &= 
\frac{a_J (p, p')}{2 J + 1} - \frac{p p'}{4 M_A^2} \, \frac{2}{1 + x}
\, \frac{a_{J - 1} (p, p')}{2 J - 1} \:, \n \\
\frac{1}{2 J + 3} \, \supfi{L} V_{J J}^{J + 1} (p, p') &= 
\frac{a_J (p, p')}{2 J + 1} - \frac{p p'}{4 M_A^2} \, \frac{2}{1 + x}
\, \frac{a_{J + 1} (p, p')}{2 J + 3} \:,
\eal
one notes the coincidence in the one-body limit $x \to 1$. For $x$ close to
one, Eq.\ \eqref{Sapproxdiag} can be expanded in $x$ around one. Consequently,
the energy levels that are degenerate at $x=1$ split for $M_A \ll M_B$
by terms of the order of (at least) $(M_A/M_B)^2$, which points to the fact 
that there is \emph{no hyperfine splitting} of the levels in the strict sense,
i.e., to orders $\alpha^4$ and $M_A/M_B$.

\end{appendix}


\begin{thebibliography}{99}
\bibitem{BS51} E.E. Salpeter and H.A. Bethe, Phys. Rev. \textbf{84},
1232 (1951).
\bibitem{GL51} M. Gell-Mann and F. Low, Phys. Rev. \textbf{84},
350 (1951).
\bibitem{WC54} G.C. Wick, Phys. Rev. \textbf{96}, 1124 (1954);
R.E. Cutkosky, \textit{ibid.} \textbf{96}, 1135 (1954).
\bibitem{Nak69} for a review of early developments, see N. Nakanishi, 
Prog. Theor. Phys. (Suppl.) \textbf{43}, 1 (1969).
\bibitem{Gol53} J.S. Goldstein, Phys. Rev. \textbf{91},
1516 (1953).
\bibitem{Gre55} H.S. Green, Phys. Rev. \textbf{97}, 540 (1955).
\bibitem{Set88} N. Set\^o, Prog. Theor. Phys. (Suppl.) \textbf{95},
25 (1988).
\bibitem{BSL63} A.A. Logunov and A.N. Tavkhelidze, Nuovo Cim. 
\textbf{29}, 380 (1963); R. Blankenbecler and R. Sugar, Phys. Rev. 
\textbf{142}, 1051 (1966). 
\bibitem{Gro69} F. Gross, Phys. Rev. \textbf{186}, 1448 (1969).
\bibitem{Gro82} F. Gross, Phys. Rev. C \textbf{26}, 2203 (1982).
\bibitem{Pas97}V. Pascalutsa and J.A. Tjon, Phys. Lett. B \textbf{435},
245 (1998); Phys. Rev. C \textbf{60}, 034005 (1999).
\bibitem{ZT80} J. Fleischer and J.A. Tjon, Nucl. Phys. \textbf{B84},
375 (1975); M.J. Zuilhof and J.A. Tjon, Phys. Rev. C \textbf{22},
2369 (1980).
\bibitem{Gro74} F. Gross, Phys. Rev. D \textbf{10}, 223 (1974);
W.W. Buck and F. Gross, \textit{ibid.} \textbf{20}, 2361 (1979);
F. Gross, J.W. Van Orden, and K. Holinde, Phys. Rev. C \textbf{45},
2094 (1992).
\bibitem{Mac87} R. Machleidt, K. Holinde, and C. Elster, Phys. Rep.
\textbf{149}, 1 (1987).
\bibitem{BH58} C. Bloch and J. Horowitz, Nucl. Phys. \textbf{8},
91 (1958).
\bibitem{KG83} U. Kaulfuss and M. Gari, Nucl. Phys. \textbf{A408},
507 (1983); J. Flender and M.F. Gari, Phys. Rev. C \textbf{51},
R1 (1995).
\bibitem{Oku54} S. Okubo, Prog. Theor. Phys. \textbf{12}, 603 (1954).
\bibitem{GW93} S. G{\l}azek, A. Harindranath, S. Pinsky, J. Shigemutsu,
and K. Wilson, Phys. Rev. D \textbf{47}, 1599 (1993).
\bibitem{MCK03} M. Mangin-Brinet, J. Carbonell, and V.A. Karmanov,
Phys. Rev. C \textbf{68}, 055203 (2003).
\bibitem{IB04} M. van Iersel and B.L.G. Bakker, hep-ph/0407318.
\bibitem{SFC01} J.H.O. Sales, T. Frederico, B.V. Carlson, and P.U. Sauer,
Phys. Rev. C \textbf{63}, 064003 (2001).
\bibitem{Web00} A. Weber, in \textit{Particles and Fields --- Seventh Mexican 
Workshop}, edited by A. Ayala, G. Contreras, and G. Herrera, AIP Conf.
Proc. No. 531 (AIP, New York, 2000), p. 305, hep-th/9911198.
\bibitem{WL02} A. Weber and N.E. Ligterink, Phys. Rev. D \textbf{65},
025009 (2002).
\bibitem{FW71} A.L. Fetter and J.D. Walecka, \textit{Quantum Theory of
Many-Particle Systems} (McGraw-Hill, New York, 1971).
\bibitem{PS95} see, e.g., M.E. Peskin and D.V. Schroeder, \textit{An 
Introduction to Quantum Field Theory} (Perseus Books, Cambridge, MA, 1995).
\bibitem{Est03} F. Estrada Ch\'avez, M.Sc. thesis, Universidad Michoacana
de San Nicol\'as de Hidalgo, 2003.
\bibitem{KG00} A. Kr\"uger and W. Gl\"ockle, Phys. Rev. C
\textbf{60}, 024004 (1999).
\bibitem{GR00} see, e.g., I.S. Gradshteyn and I.M. Ryzhik, \textit{Table of 
Integrals, Series, and Products} (Academic Press, San Diego, 
2000), sixth edition.
\bibitem{AS65} M. Abramowitz and I.A. Stegun, \textit{Handbook of
Mathematical Functions} (Dover Publications, New York, 1965).
\bibitem{CDL77} see, e.g., C. Cohen-Tannoudji, B. Diu, and F. Lalo\"e, 
\textit{Quantum Mechanics} (Wiley-Interscience, Paris, 1977).
\end{thebibliography}
\end{document}